\documentclass{aa}  

\usepackage{graphicx}
\usepackage{txfonts}
\usepackage{subfigure}
\usepackage{threeparttable}
\bibpunct{(}{)}{;}{a}{}{,} 
\usepackage{gensymb}
\usepackage{natbib,twoopt}
\usepackage{footnote}
\usepackage[utf8]{inputenc}
\usepackage{array}
\usepackage{color}
\usepackage{mathtools}
\usepackage[export]{adjustbox}
\usepackage{amsmath,amssymb}
\usepackage{sidecap}
\sidecaptionvpos{figure}{c}
 
\begin{document} 

\def\mean#1{\left< #1 \right>}

   \title{Iron abundance distribution in the hot gas of merging galaxy clusters}

   \author{I. Urdampilleta \inst{1,2}
          \and F. Mernier \inst{1,3,4}
          \and J. S. Kaastra\inst{1,2}
          \and A. Simionescu\inst{1,2,5}
          \and J. de Plaa\inst{1}
          \and S. Kara\inst{6}
          \and E. N. Ercan\inst{6}
      }
             
   \institute{SRON Netherlands Institute for Space Research, Sorbonnelaan 2, 3584 CA Utrecht, The Netherlands\\
              \email{i.urdampilleta@sron.nl}
         \and
            Leiden Observatory, Leiden University, PO Box 9513, 2300 RA Leiden, The Netherlands
         \and
         MTA-E\"otv\"os University Lend\"ulet Hot Universe Research Group, P\'azm\'any P\'eter s\'et\'any 1/A, Budapest, 1117, Hungary
         \and
      Institute of Physics, E\"otv\"os University, P\'azm\'any P\'eter s\'et\'any 1/A, Budapest, 1117, Hungary
       \and Kavli Institute for the Physics and Mathematics of the Universe (WPI), University of Tokyo, Kashiwa 277-8583, Japan
        \and Department of Physics, Bogazici University, 34342 Istanbul, Turkey
       \\}


 
  \abstract
   {We present \textit{XMM-Newton}/EPIC observations of six merging galaxy clusters and study the  distributions of their temperature, iron (Fe) abundance and pseudo-entropy along the merging axis. For the first time, we focus simultaneously, and in a comprehensive way, on the chemical and thermodynamic properties of the freshly collided intracluster medium (ICM). The Fe distribution of these clusters along the merging axis is found to be in good agreement with the azimuthally-averaged Fe abundance profile in typical non-cool-core clusters out to $r_{500}$. In addition to showing a moderate central abundance peak, though less pronounced than in relaxed systems, the Fe abundance flattens at large radii towards $\sim$0.2--0.3 $Z_\odot$. Although this shallow metal distribution is in line with the idea that disturbed, non-cool-core clusters originate from the merging of relaxed, cool-core clusters, we find that in some cases, remnants of metal-rich and low entropy cool cores can persist after major mergers.   While we obtain a mild anti-correlation between the Fe abundance and the pseudo-entropy in the (lower entropy, $K$~=~200-500 keV cm$^2$) inner regions, no clear correlation is found at (higher entropy, $K$~=~500-2300 keV cm$^2$) outer radii. The apparent spatial abundance uniformity that we find at large radii is difficult to explain through an efficient mixing of freshly injected metals, particularly in systems for which the time since the merger is short. Instead, our results provide important additional evidence in favour of the early enrichment scenario -- in which the bulk of the metals are released outside galaxies at $z$~>~2--3 -- and extend it from cool-core and (moderate) non-cool-core clusters to a few of the most disturbed merging clusters as well. These results constitute a first step towards a deeper understanding of the chemical history of merging clusters.}

    
 

   \maketitle
%

\section{Introduction}

In order to become the largest gravitationally bound structures seen in today's Universe, galaxy clusters grow in a hierarchical way, via not only accretion but also merging of surrounding (sub) haloes. The latter case is particularly interesting as  major cluster mergers are the most energetic events since the Big Bang.  During the merging process, the hot intra-cluster medium (ICM) in galaxy clusters is violently compressed and heated, and large amounts of thermal and non-thermal energy are released, giving rise to shock fronts and turbulence \citep{Markevitch2007}.  Shocks can propagate into the ICM (re)accelerating electrons \citep{Bell1987,Blandford1987}, which may produce elongated and polarized  radio structures known as radio relics via synchrotron radio emission \citep[for a review, see][]{Ferrari2008,Brunetti2014}. On the other hand,  particle (re)acceleration by means of the turbulence in the ICM could generate unpolarized cluster-wide sources known as radio haloes \citep[for a review, see][]{Brunetti2001,Feretti2012,VanWeeren2019}.

 The ICM is also known to be rich in metals, which originate mainly from core-collapse (SNcc) and type Ia (SNIa) supernovae, which continuously release their nucleosynthesis products since the epoch of major star formation, about 10 billion years ago ($z$~$\sim$~2--3) \citep{Hopkins2006,Madau2014}. SNcc contribute primarily to the synthesis of light metals (O, Ne, Mg, Si and S). On the other hand, heavier metals (such as Ar, Ca, Mn, Fe and Ni) together with smaller relative contribution of Si and S are produced by SNIa. Finally, C and N are mostly released by low-mass stars on the asymptotic giant branch (AGB). Once the metals are ejected to the ICM by SN or AGB stars, they are later transported, mixed and redistributed via other processes, such as galactic winds \citep{Kapferer2006,Kapferer2007,Baumgartner2009}; ram-pressure stripping \citep{Schindler2005,Kapferer2007}, active galactic nucleus (AGN) outflows \citep{Simionescu2008,Simionescu2009,Kapferer2009} or gas sloshing \citep{Simionescu2010,Ghizzardi2014}, among other processes. Detailed reviews on the observation of metals in the ICM include, e.g. \cite{Werner2008,Bohringer2010,DePlaa2013} and \cite{Mernier2018b}.
 
 \begin{table*}[!htb]
 \begin{center}
  \caption{Observations and exposure times.}
 \label{tab:tab1}
 \begin{threeparttable}
\begin{tabular}{cccccccc}
\hline
\hline
 \noalign{\smallskip}
 Name & Label & Sequence ID &Position (J2000)&Observation& MOS1 &MOS2& pn \\
&&&(RA, Dec.)&starting date&Net (ks)&Net (ks)&Net (ks)\\
 \noalign{\smallskip}
\hline
\noalign{\smallskip}
CIZA J2242.8+5301&CIZA2242&0654030101&(22:43:01.99, 53:07:30.0)&2010-12-13&71.5&71.9&68.4 \\
\noalign{\smallskip}
 1RXS J0603.3+4214&1RXSJ0603&0675060101&(06:03:13.39, 42:12:31.0)&2011-10-03&75.4&75.5&71.4 \\
 \noalign{\smallskip}
 Abell 3376 East&A3376&0151900101&(06:02:08.59, -39:57:18.0)&2003-04-01&27.9&27.8&25.4 \\

 Abell 3376 West&&0504140101&(06:00:51.14, -40:00:31.6)&2007-08-24&40.8&42.7&36.4 \\
 \noalign{\smallskip}
  Abell 3667&A3667&0206850101&(20:13:04.79, -56:53:60.0)&2004-05-03&59.7&60.7&57.3 \\
   Abell 3667 offset&&0653050201&(20:12:31.46, -56:21:39.3)&2010-09-21&28.9&28.6&24.7 \\
 \noalign{\smallskip}
   Abell 665&A665&0109890501&(08:30:57.99, 65:50:20.0)&2001-09-23&80.3&81.4&-- \\
 
 \noalign{\smallskip}
    Abell 2256&A2256&0401610101&(17:03:02.55, 78:45:00.0)&2006-08-04&39.7&42.6&6.6 \\
      &&0141380101&(17:03:02.55, 78:45:00.0)&2003-04-27&9.8&10.1&9.5 \\
 \noalign{\smallskip}
\hline
\end{tabular}
 
 \end{threeparttable}
 \end{center}
\end{table*}

 In the last years, the iron (Fe) abundance distribution in galaxy clusters has been extensively studied using the different X-ray observatories \citep{DeGrandi2001,DeGrandi2004,Baldi2007,Leccardi2008,Maughan2008,Matsushita2011,Mernier2016a,Mernier2017,Mantz2017,Simionescu2017,Simionescu2018}. The Fe radial distribution is different for cool-core (CC, i.e. dynamically relaxed) clusters and  non-cool-core (NCC, i.e. dynamically disturbed) clusters. In the case of CC clusters, the Fe distribution shows a peak in the core, which decreases up to $\sim$0.3$r_{500}$ and flattens for larger radii. NCC clusters, however, have a flat distribution in the core and follow the same universal distribution as CC for outer radii. These abundance distributions do not present evidence of evolution up to $z$~$\sim$~1 \citep{McDonald2016,Mantz2017,Ettori2017}. Moreover, recent X-ray observations with \textit{Suzaku} \citep{Fujita2008,Werner2013,Urban2017,Ezer2017,Simionescu2017} reveal an uniform radial distribution  around $\sim$0.2--0.3 $Z_\odot$ in the outskirts of the galaxy clusters, which suggests an early enrichment of the ICM (i.e. $z$~$\sim$~2--3). Numerical simulations later showed that active galactic nuclei (AGN) feedback at early times is needed to explain these observations \citep[e.g.][]{Biffi2017,Biffi2018a,Biffi2018b}.

Whereas metals in the ICM have been extensively and systematically 
studied in CC and, to some extent, NCC clusters, the ICM enrichment in the extreme case 
of merging clusters has been poorly explored so far (except in a few 
specific cases; e.g. A3376, \citealt{Bagchi2006};  A3667, 
\citealt{Lovisari2009}; \citealt{Lagana2019}). Knowing in detail how 
metals are distributed across merging clusters, however, is a key 
ingredient in our understanding of the chemical evolution of large scale structures. On one hand metals act as passive tracers and may reveal 
valuable information on the dynamical history of the recently merged 
ICM. On the other hand, merging shocks could in principle trigger star 
formation in neighbouring galaxies \citep{Stroe2015a,Sob15}, hence potentially contributing to the ICM enrichment to some extent \citep{Elkholy2015}. Metallicities in the shocked gas of such disturbed clusters, and their connection with their thermodynamical properties and star formation of their galaxies remain  widely 
unexplored.
    
 In this work, we present for the first time a study specifically devoted to the chemical state of merging clusters. For this purpose, we derive the spatial distribution of temperature, abundance (Fe), and pseudo-entropy from specifically selected regions of six merging galaxy clusters:  CIZA J2242.8+5301, 1RXS J0603.3+4214, Abell 3376, Abell 3667, Abell 665 and  Abell 2256. The principal characteristics of these clusters are summarized in Table  \ref{tab:tab2}. The first four clusters of the list are essentially major mergers hosting a double radio relic, while Abell 665 hosts a giant radio halo and Abell 2256 a prominent single radio relic together with a radio halo. The selection criteria of these clusters have been: (i) \textit{XMM-Newton} observation availability, (ii) net exposure time > 25 ks (iii) bright and disturbed central ICM and (iv) probing a variety of diffuse radio features: we include merging clusters with double radio relics, a single radio relic and strong radio halo. This is not intended as a complete sample but rather as an exploratory first step towards future studies on metals in the ICM of merging clusters from larger observational samples. 
 For this study, we report our results with respect to the  protosolar abundances ($Z_\odot$) reported by \cite{Lodders2009}.   We assume the cosmological parameters $H_0$ = 70 km/s/Mpc, $\Omega_{\rm{M}}$ = 0.27 and $\Omega_\Lambda$= 0.73. All errors are given at 1$\sigma$ (68$\%$) confidence level unless otherwise stated and  the spectral analysis uses the modified Cash statistics \citep{Cash1979,Kaastra2017a}.

\begin{table*}[!htb]
 \begin{center}
  \caption{Cluster sample description.}
 \label{tab:tab2}
 \begin{threeparttable}
\begin{tabular}{cccccccc}
\hline
\hline
 \noalign{\smallskip}
  Label & $z$ & $N_{\rm{H}}$\tnote{a} & Radio features & Mass ratio & Dyn. stage\tnote{b} & $\mean{kT}$ & Scale f.\tnote{c} \\
  &&(10$^{20}$ cm$^{-2}$)&&&(Gyr)&(keV)&(kpc/$\arcmin$)\\
 \noalign{\smallskip}
\hline
\noalign{\smallskip}
 CIZA2242&0.192&25.80--55.70&Doble relic, faint halo\tnote{[1]}&1--2:1\tnote{[2,3]}&$\sim$0.6\tnote{[4]}&7.9\tnote{[4]} &193.0\\
\noalign{\smallskip}
 1RXSJ0603&0.225&21.50&Triple relic, halo\tnote{[5]}&3:1:0.1:0.1\tnote{[6]}& $\sim$2.0\tnote{[7]}&7.8\tnote{[8]}&218.0   \\
 \noalign{\smallskip}
 A3376&0.046&4.84&Double relic\tnote{[9]}&3--6:1\tnote{[10]}&$\sim$0.5--0.6\tnote{[10]}&4.2\tnote{[11]}&54.0 \\
 
 \noalign{\smallskip}
 A3667&0.055&4.44&Double relic, mini-halo\tnote{[12,13]} & 5:1\tnote{[14]}&$\sim$1.0\tnote{[14]}&6.3\tnote{[15]}&64.5\\
 \noalign{\smallskip}
   A665 &0.182&4.31&Halo\tnote{[16]}&1--2:1\tnote{[17]}&$\sim$1--2\tnote{[17]}&8.3\tnote{[18]}&184.5\\
 
 \noalign{\smallskip}
    A2256&0.058&4.24&Relic, halo\tnote{[19,20]}& 3:1:0.3\tnote{[21]}&$\sim$--0.2\tnote{[22]} &6.4\tnote{[23]}&67.5\\
   
 \noalign{\smallskip}
\hline
\end{tabular}
 \tiny
 \begin{tablenotes}
    \item[a] \cite{Willingale2013} (http://www.swift.ac.uk/analysis/nhtot/)
   \item[b] All dynamical stages are after core passage except A2256 (pre-merger).
     \item[c] 1 arcmin value in kpc assuming the cosmological parameters $H_0$ = 70 km/s/Mpc, $\Omega_{\rm{M}}$ = 0.27 and $\Omega_\Lambda$= 0.73 for each $z$.

      References: [1] \cite{VanWeeren2010}, [2] \cite{VanWeeren2011}, [3] \cite{Jee2015}, [4] \cite{Akamatsu2015}, [5] \cite{VanWeeren2012a}, [6] \cite{Jee2016}, [7] \cite{Bruggen2012a}, [8] \cite{Ogrean2013b}, [9] \cite{Bagchi2006}, [10] \cite{Machado2013}, [11] \cite{Urdampilleta2018}, [12] \cite{Rottgering1997}, [13] \cite{Riseley2015}, [14] \cite{Roettiger1999}, [15] \cite{Sarazin2016}, [16] \cite{Moffet1989},  [17] \cite{Gomez2000}, [18] \cite{Hughes1992}, [19] \cite{Bridle1976}, [20] \cite{Clarke2006}, [21] \cite{Berrington2002}, [22] \cite{Roettiger1995}, [23] \cite{Trasatti2015}
      \end{tablenotes}
 \end{threeparttable}
 \end{center}
\end{table*}

  \section{Observations and data reduction}
 
  Table 1 summarizes the \textit{XMM-Newton} observations, the different pointings and the net exposure time used for the cluster survey analysis. All the data are reduced with the  \textit{XMM-Newton} Science Analysis System (SAS) v17.0.0, with the calibration files dated by June 2018. We use the observations from the MOS and pn detectors of the EPIC instrument and reduce them first with the standard pipeline commands \texttt{emproc} and \texttt{epproc}. Next, we filter the soft-proton (SP) flares by building Good Time Interval (GTI) files following the method detailed in Appendix A.1 of \cite{Mernier2015}. This method consists of extracting light curves at the high energy range of spectrum (10--12 keV for MOS and 12--14 keV for pn) in bins of 100s. Then, the mean count rate, $\mu$, and the standard deviation, $\sigma$, are calculated in order to apply a threshold of $\mu~\pm~2\sigma$ to the generated distribution. Therefore, all the time bins with a number of counts outside the interval $\mu~\pm~2\sigma$ are rejected. Afterwards, we apply the same process at the low energy range 0.3--2 keV with bins of 100s \citep{Lumb2002}. Only events satisfying the 2$\sigma$ threshold in the two energy ranges mentioned above are further considered. The point source identification in the complete field of view (FOV) is done with the SAS task  \texttt{edetect\_chain} applied in four different bands (0.3--2.0 keV, 2.0--4.5 keV, 4.5--7.5 keV, and 7.5--12 keV). After this, we combine the detections in these four bands and exclude the point sources with a circular region of 10$\arcsec$ radius, except some larger sources where an appropriate radius has been applied. The 10$\arcsec$ radius size, as explained in Appendix A.2 of \cite{Mernier2015}, is optimized for avoiding the polluting flux of the point sources and not reducing considerably the emission of the ICM. We keep the single, double, triple and quadruple events in MOS (\texttt{pattern$\leq$}12) and only the single pixel events in pn (\texttt{pattern}==0). We also correct the out-of-time events from the pn detector. \footnote{https://www.cosmos.esa.int/web/xmm-newton/sas-thread-epic-oot}

   \section{Spectral analysis}
    \subsection{Spectral analysis approach}\label{sect:spectral_approach}
\begin{figure}[ht!]
  \centering
        \includegraphics[width=0.5\textwidth]{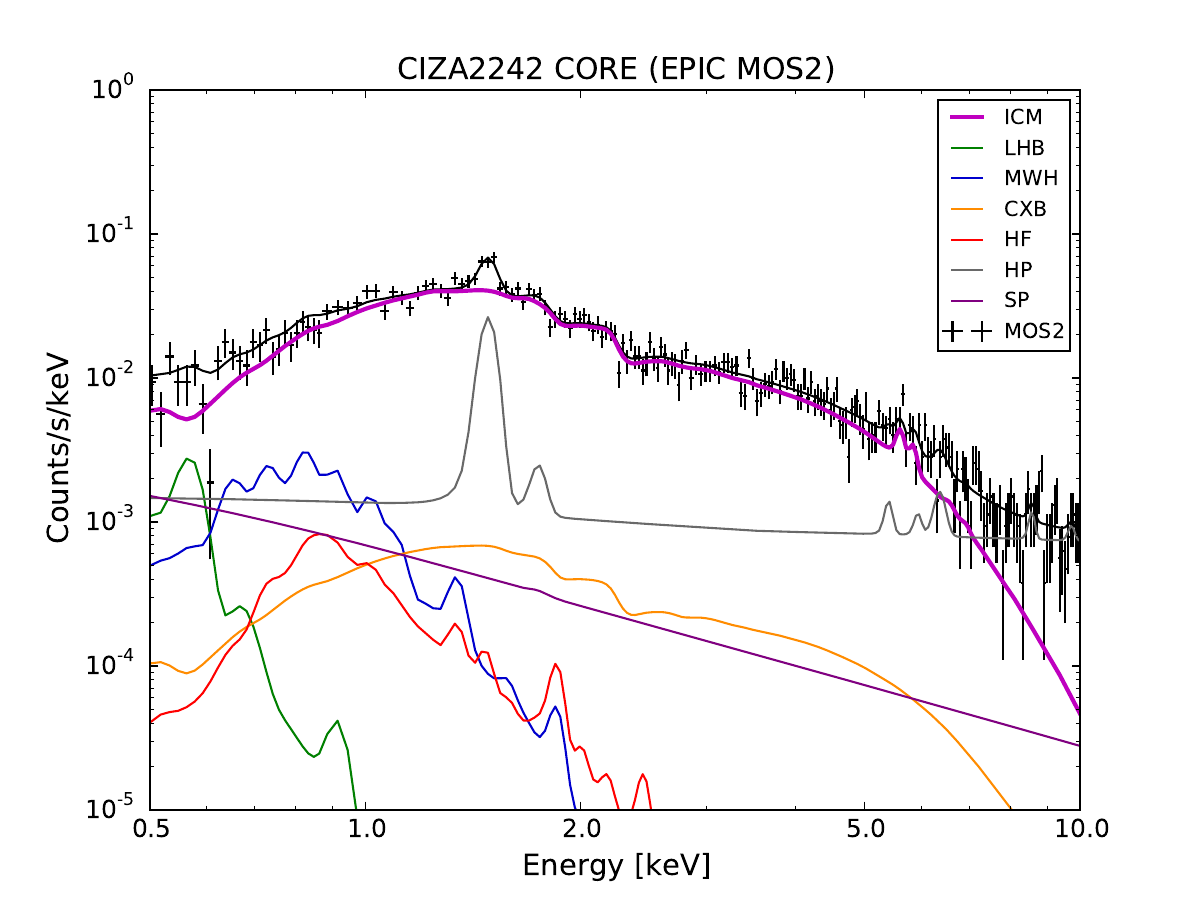}
 
    \caption{EPIC MOS2 spectrum of the CIZA2242 core ($r$~=~1.5$\arcmin$). The solid black line represents the best-fit model. The cluster emission ICM and all background components (LHB, MHW, CXB, HF, HP and SP) are also shown.}
    \label{fig:fig0}
     \end{figure}

  In our spectral analysis of the clusters, we assume that the observed spectra include the following components (see EPIC MOS2 spectrum in Fig. \ref{fig:fig0}): an optically thin thermal plasma emission from the ICM in collisional ionization equilibrium (CIE),  the local hot bubble (LHB), the Milky Way halo (MWH), the cosmic X-ray background (CXB) and the non X-ray background (NXB), consisting of hard particle (HP) background and residual soft-proton (SP) component (see Sect.~\ref{sect:background}). An additional hot foreground (HF) component is included for CIZA J2242.8+5301 as suggested by \cite{Ogrean2013a} and \cite{Akamatsu2015}. Further details about these components (and how we include them in the fits) are given in Sect. 3.2. The emission models are corrected for the cosmological redshift and absorbed by the galactic interstellar medium. The hydrogen column density value adopted for all the clusters is the weighted neutral hydrogen column density, $N_{\rm{HI}}$, estimated using the method of \cite{Willingale2013}\footnote{http://www.swift.ac.uk/analysis/nhtot/}. Previous work (e.g. \citealt{Mernier2016a,dePlaa2017}) noted that the total $N_{\rm{H}}$ value (neutral + molecular hydrogen) may sometimes provide inaccurate fits whereas the weighted neutral hydrogen $N_{\rm{HI}}$ values are generally closer to the best-fit $N_{\rm{H}}$ values. Specifically, we have checked that leaving free the  $N_{\rm{H}}$ in our fits provides values that do not deviate more than  $\sim$10$\%$ from the above estimates. The only exception is CIZA J2242.8+530, for which we need to leave the value of $N_{\rm{H}}$  free within the range 0.8~x~$N_{\rm{HI}}~\leq~N_{\rm{H}}~\leq$~1.2~x~$N_{\rm{H,tot}}$, where $N_{\rm{H,tot}}$ is the weighted total (neutral and molecular) hydrogen column density \citep{Willingale2013}.
 
 In our spectral analysis, we use SPEX\footnote{https://www.sron.nl/astrophysics-spex}  \citep{Kaastra1996,Kaastra2017b} version 3.04.00 with SPEXACT (SPEX Atomic Code and Tables) version 3.04.00. We carry out the spectral fitting in different regions as detailed in the sections below. The redistribution matrix file (RMF)  and the ancillary response file (ARF) are processed using the SAS tasks \texttt{rmfgen} and \texttt{arfgen}, respectively. The spectra of the MOS (energy range from 0.5 to 10 keV) and pn (0.6--10 keV) detectors are fitted simultaneously and binned using the method of optimal binning described in \citet{Kaastra2016}. We assume a single temperature structure in the Collisional Ionization Equilibrium (\textit{cie}) component for all the clusters (this assumption is justified in Sect. \ref{sect:systematics}). Because the high temperature of merging clusters does not allow to measure other metals with reasonable accuracy, their abundances are coupled to Fe. The free parameters considered in this study are the temperature \textit{kT}, the metal abundance \textit{Z} and the normalization \textit{Norm}. In the case of CIZA J2242.8+5301, the column density $N_{\rm{H}}$ is also free to vary.

    \subsection{Estimation of background spectra}\label{sect:background}
    
    A proper estimation of the background components is important especially in the regions where the emission of the cluster is low. This is particularly relevant outside of the core, where the spectrum can be background dominated. In this work, we model the background components directly from our spectra (Sect.~\ref{sect:spectral_approach}) using the following list of components: 
    
\begin{itemize}
 
\item The LHB, modelled by a non-absorbed $cie$ component with proto-solar abundance. The temperature is fixed to 0.08 keV.

\item The MWH, modelled by an absorbed $cie$ component with proto-solar abundance. The temperature is free to vary except for CIZA J2242.8+5301 and Abell 3376, where it is fixed to 0.27 keV, following the values of \cite{Akamatsu2015} and \cite{Urdampilleta2018}, respectively.

\item The CXB, modelled by an absorbed power law with a fixed $\Gamma$~=~1.41 \citep{DeLuca2004}. The normalization is a fixed value for each cluster assuming the CXB flux of 8.07~x~10$^{-15}$ W m$^{-2}$ deg$^{-2}$, taking into account a detection limit of S$_c$~=~3.83~x~10$^{-15}$ W m$^{-2}$ and calculated with the CXBTools \citep{dePlaa2017}. See details in Appendix B.2 of \cite{Mernier2015}.\footnote{Note that the flux of the faintest resolved and excluded point source, and hence the remaining unresolved CXB flux, depend on the exposure time. In our case, most of the observations have similar a exposure except two: 0151900101 and 0653050201. Even for these data sets, the SP power law norm is free to vary and will compensate for any residuals in the CXB subtraction.}

\item  The HF, modelled by an absorbed $cie$ component with proto-solar abundance. The temperature is fixed to 0.7 keV.

\item  The HP background, modelled by a broken power law (unfolded by the effective area) together with several instrumental Gaussian profiles (eight and nine for MOS and pn, respectively). The photon indices of the power laws and the centroid energies of the profiles have been taken from Table B.1 and Table B.2 of \cite{Mernier2015}, see their Appendix B.1 for more details. The normalizations are left free to vary.

\item  The SP background, modelled by a power law (unfolded by the effective area). The power law index of the SP component is left free to vary between 0.1 and 1.4.
\end{itemize}

 The estimation of the sky background components (LHB, MHW and HF) is obtained from the offset regions (see green sectors in Fig. \ref{fig:fig1}, \ref{fig:fig3}, \ref{fig:fig7}, \ref{fig:fig9} and \ref{fig:fig11}) where a negligible (yet still modelled) thermal emission of the cluster is expected and the background is dominant. For Abell 3376 the values of \cite{Urdampilleta2018} are adopted. The parameters of the SP components are obtained fitting the total FOV ($r~=~$15$\arcmin$) EPIC spectra, where the \textit{cie} (i.e. ICM) parameters  are also left free. Finally, the normalization of every background component (except the HP background) has been fixed, rescaled on the sky area of each region and corrected for vignetting if necessary. The best-fit parameters for each cluster are listed in Appendix A. 
 
    \subsection{Systematic uncertainties}\label{sect:systematics}
    
 We account for the effect of systematic uncertainties related to the $\pm$10$\%$ variation in the normalization of the sky background and non X-ray background components \citep{Mernier2017}. The contribution of these systematic uncertainties are further  included in our analysis and results.

 \cite{Mernier2017} and the posterior review of \cite{Mernier2018b} list other potential systematic uncertainties, which could affect the abundance measurements of this work. These include MOS-pn discrepancies due to residual EPIC cross-calibration, projection effects, atomic code and thermal structure uncertainties. The former has limited impact on the relative Fe abundance distribution as explained by \cite{Mernier2017}. Although recent work of \cite{Liu2018} suggests that projection effects may have non-negligible observational biases on the observed abundance evolution in the core of CC clusters, the lack of strong emission gradients in merging clusters makes us confident that such a bias is limited in our case. Moreover, deprojecting merging cluster emission is very challenging, as by definition these systems are far from being spherically symmetric.
 
Uncertainties regarding the atomic codes are limited by the fact that we are using the update of the code SPEXACT and SPEX, which incorporate more precise modelling of the atomic processes and extensive compilation of transitions. However, uncertainties due to a not complete atomic database are still present \citep{Hitomi2018} and their effects can be non-negligible \citep{Mernier2018a}.  
 
 For the thermal structure,  we assume a single-temperature (1T) model. This is a simplified approximation; other more complex models reproducing Gaussian (\texttt{gdem}, \citealt{dePlaa2006}) or power-law (\texttt{wdem},  \citealt{Kaastra2004}) temperature distributions could represent better the inhomogeneities of the ICM. To quantify these possible effects, we have fitted the spectra assuming a simplified version of the \texttt{gdem} model using a 3T model with one main temperature free and the other two coupled with a factor of 0.64 and 1.56, respectively. We  have realized the fits in the SPEXACT version 2.07.00 in order to significantly reduce the computing time. This version includes simpler atomic code and tables, which allows to fit the spectra faster, but with a similar accuracy as the most updated version \citep[for more details, see][]{Mernier2018a}. We verify that the best-fit temperature of the main component in our 3T model remains within $\pm$~5$\%$ of the best-fit temperature obtained when assuming a 1T model. The relative emission measure of the other two components are less than 1$\%$ of the best-fit emission measure of the 1T model, and the variation in the abundance is less than 10$\%$. This 1T-like behaviour is not surprising as, unlike CC clusters, merging clusters do not exhibit a strong temperature gradient, and at these high temperatures the Fe abundance is almost entirely determined from the Fe-K line (which is unaffected by the Fe-bias or the inverse Fe-bias, e.g. \citealt{Buote2000}, \citealt{Simionescu2009}). For this reason and for simplicity we decide to maintain the 1T model and we will add these uncertainties to the other background systematic uncertainties as the thermal structure contribution.

    \section{Spectral analysis: individual clusters}

        \begin{figure*}[ht!]
  \centering
        \begin{minipage}[h]{0.40\textwidth}
    \centering
    \includegraphics[width=1\textwidth]{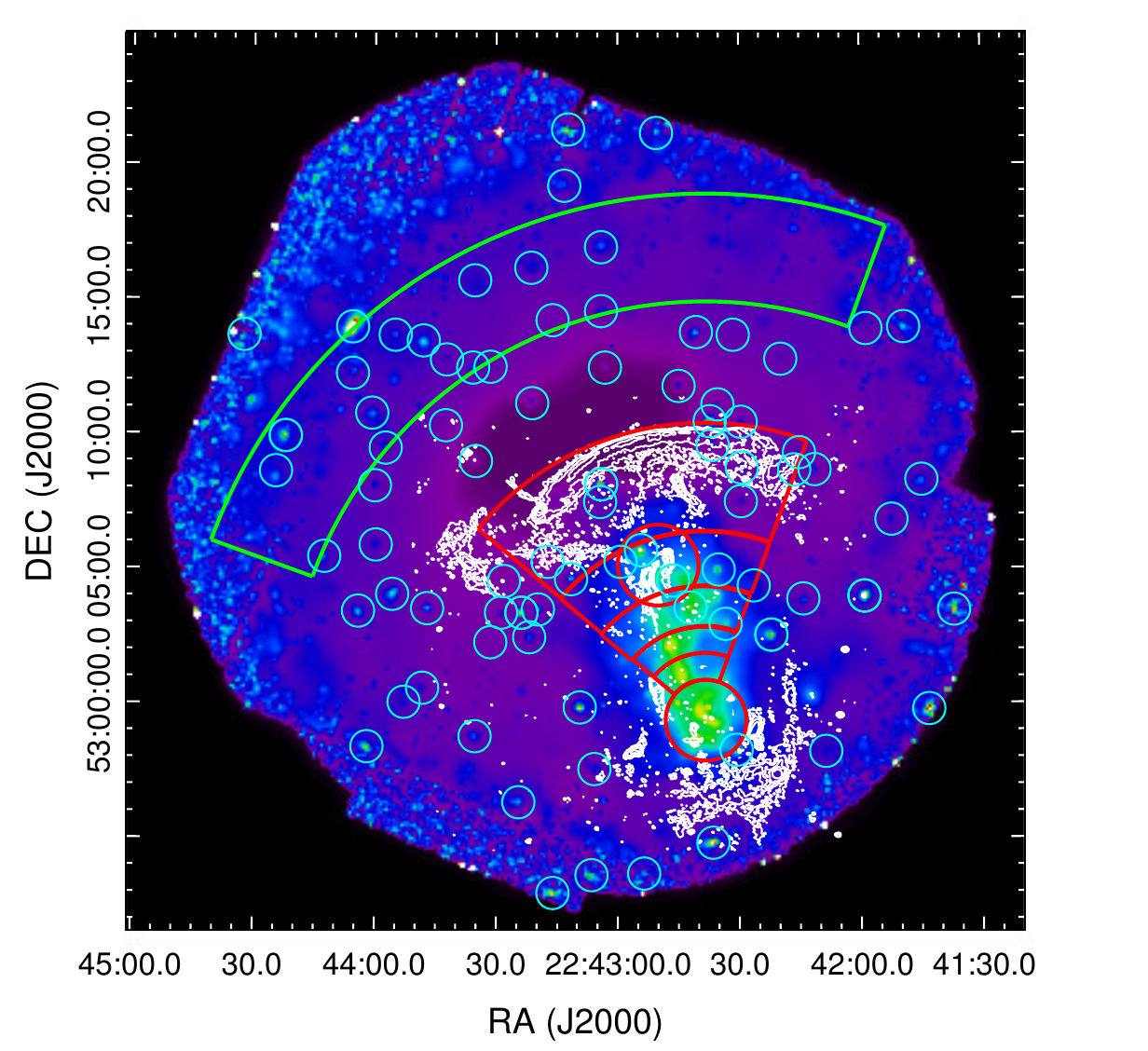}
  \end{minipage}
 \begin{minipage}[h]{0.39\textwidth}
    \centering
    \includegraphics[width=1\textwidth]{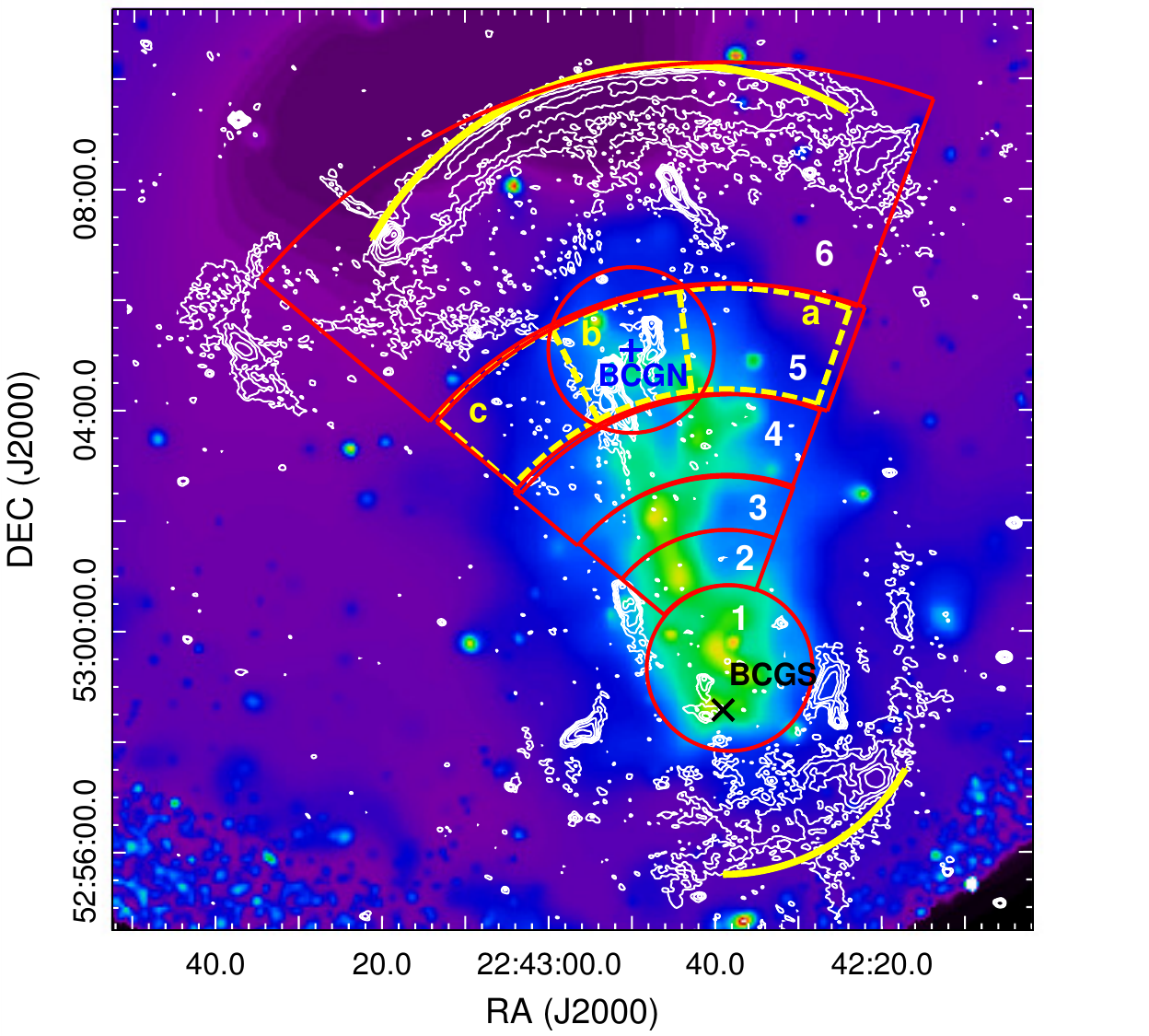}
 \end{minipage}
 
    \caption{\textit{Left panel}: \textit{XMM-Newton} smoothed image in the 0.3--10 keV band of CIZA2242.The red sectors represent the regions used for the radial temperature and abundance profiles. The green sector is the offset region used for the sky background estimation. Cyan circles are the point sources removed (with enlarged radius for clarity purpose). White contours are LOFAR radio contours. \textit{Right panel}: Enlarged image of CIZA2242. The BCGN and BCGS are marked with blue and black crosses, respectively. The solid yellow lines show the X-ray shocks position.}
    \label{fig:fig1}
  \centering
  \begin{minipage}[h]{0.45\textwidth}
    \centering
    \includegraphics[width=1\textwidth]{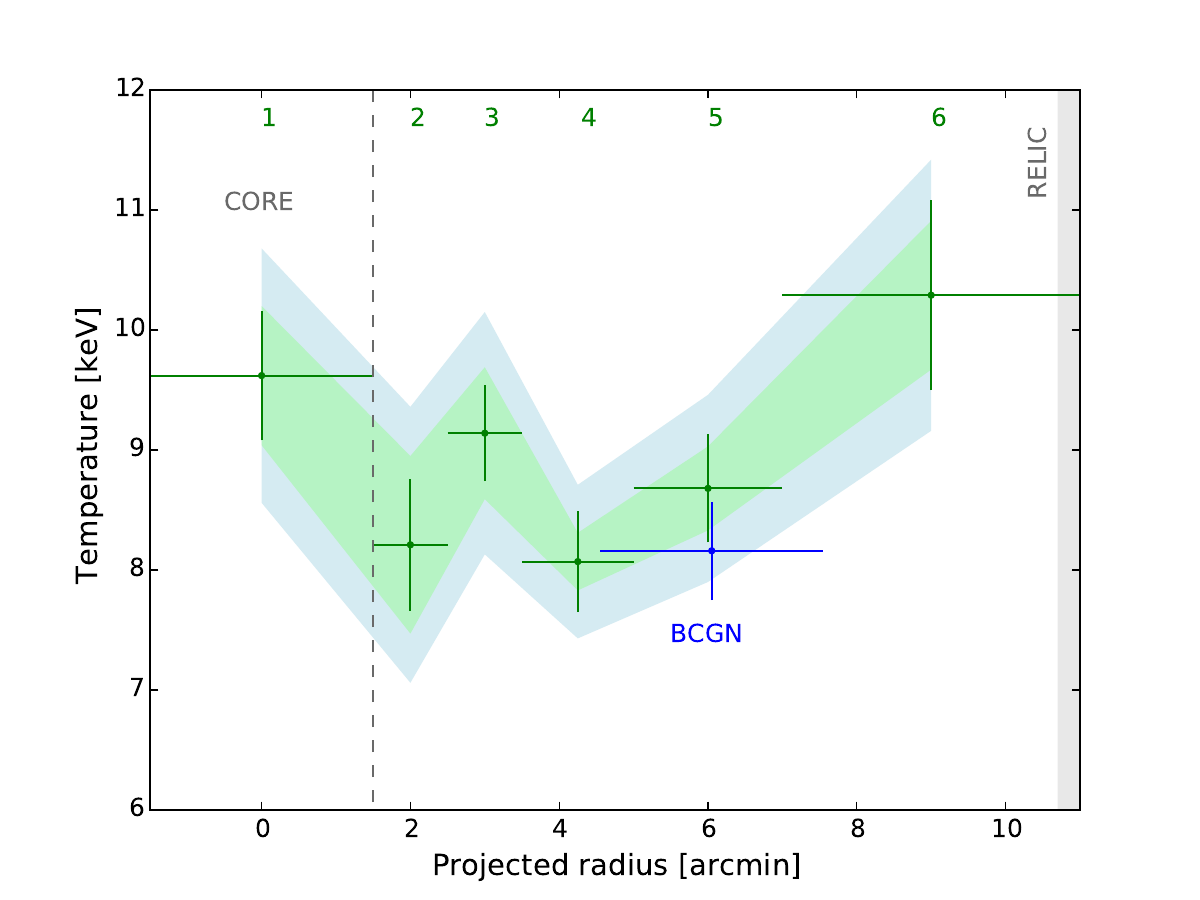}
  \end{minipage}
 \begin{minipage}[h]{0.45\textwidth}
    \centering
    \includegraphics[width=1\textwidth]{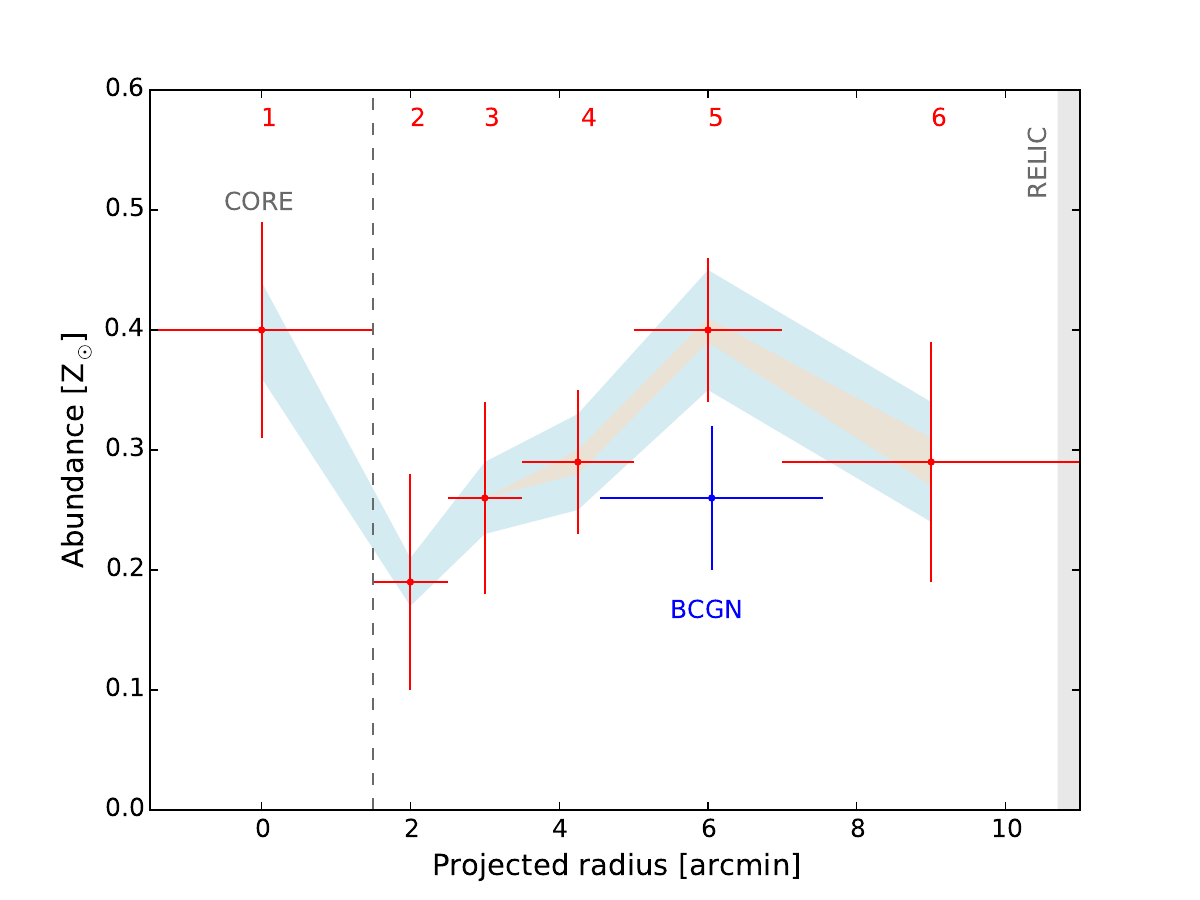}
 \end{minipage}
\caption{\textit{Left panel}:  Temperature distribution along the merging axis of CIZA2242. \textit{Right panel}: Abundance distribution along the merging axis of CIZA2242.  The shaded areas represent the systematic uncertainties: green and orange shadows correspond to the background normalization variation, and the blue one to the temperature structure. The shaded grey area shows the radio relic position. The dashed line shows the outer edge of the core spectral extraction region. The numbers and the BCGN blue point correspond to the regions and the red circle centered on BCGN, respectively, shown in Fig. \ref{fig:fig1}.}
\label{fig:fig2}
\end{figure*}

      \subsection{CIZA J2242.8+5301}

        CIZA J2242.8+5301 (hereafter CIZA2242, also known as the "Sausage" cluster) is a massive ($\sim$10$^{15}$M$_\odot$) and two-body \citep{Jee2015} merging galaxy cluster at a redshift of $z$~=~0.192. It was discovered in the second CIZA sample from the \textit{ROSAT} All-Sky Survey and identified as a major-merger by \cite{Kocevski2007}. \cite{VanWeeren2010} studied CIZA2242 for the first time using WSRT, GMRT and VLA. They observed a double radio relic feature located at $\sim$1.5 Mpc from the cluster center.   \textit{XMM-Newton} and  \textit{Chandra} observations \citep{Ogrean2013a,Ogrean2014} reveal a strongly disturbed and complex merger. In fact, the ICM presents an elongated X-ray morphology with a bullet shape bright edge in the south (S) and more irregular edge in the north (N), suggesting that the infalling direction is N to S. \cite{Akamatsu2015} estimate the time after the core passage to be $\sim$0.6 Gyr, in good agreement with the Sunyaev–-Zel'dovich (SZ) analysis of \cite{Rumsey2017}.

    Finally, \citet{Stroe2014a, Stroe2015b, Sob15, Stroe2017} confirm an overdensity of H$_\alpha$ emitters, mainly cluster star-forming galaxies and AGNs, concentrated close to the subcores and northern post-shock region. The star-forming galaxies are found to be highly metal rich and to have strong outflows. The aboved mentionned authors propose that such an enhancement of the star formation ratio could be a shock-induced effect.

        \subsubsection{Results}

       \begin{figure*}[ht!]
  \centering
  \begin{minipage}[h]{0.40\textwidth}
    \centering
        \includegraphics[width=1\textwidth]{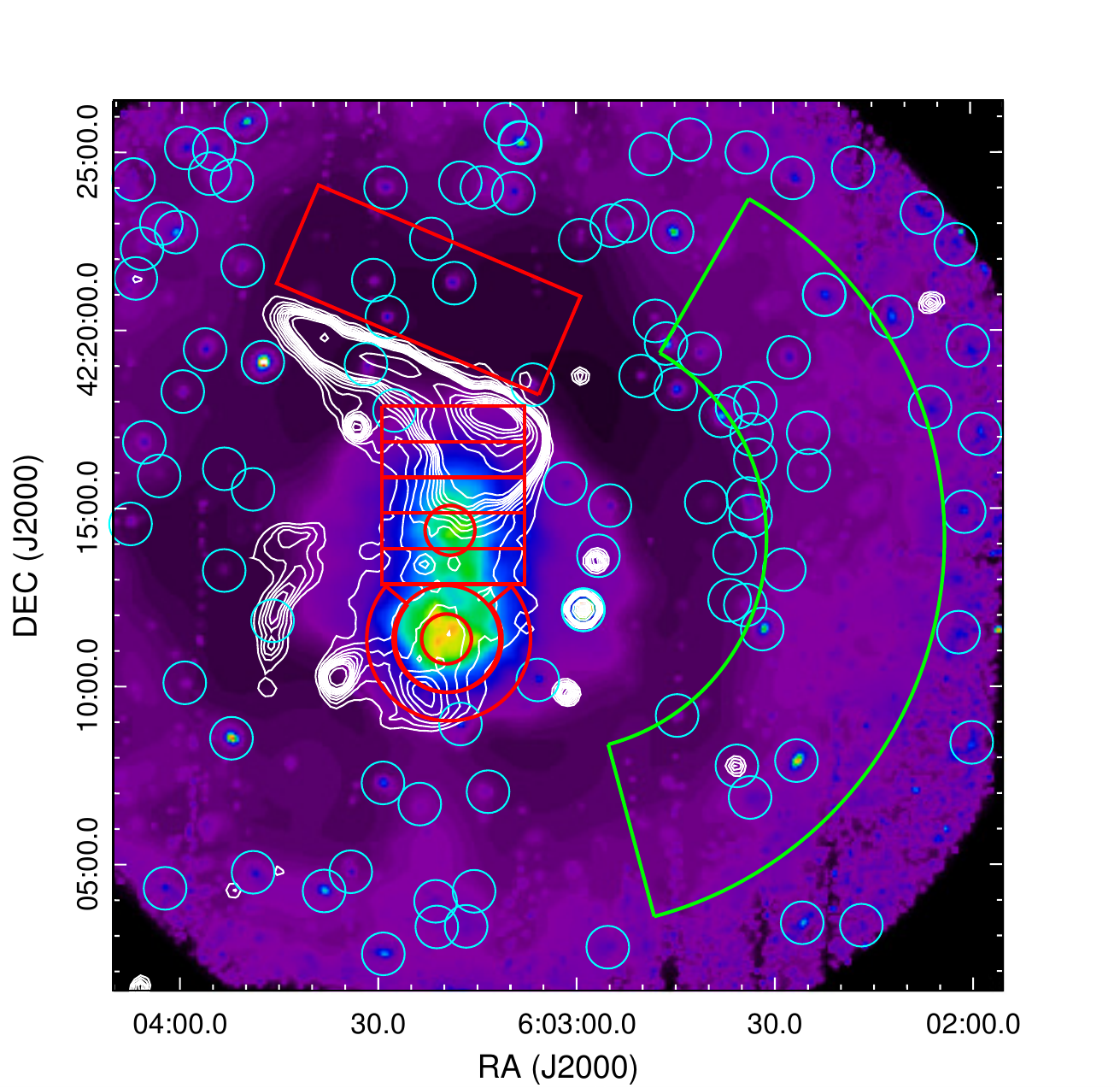}
 \end{minipage}
  \begin{minipage}[h]{0.42\textwidth}
    \centering
         \includegraphics[width=1\textwidth]{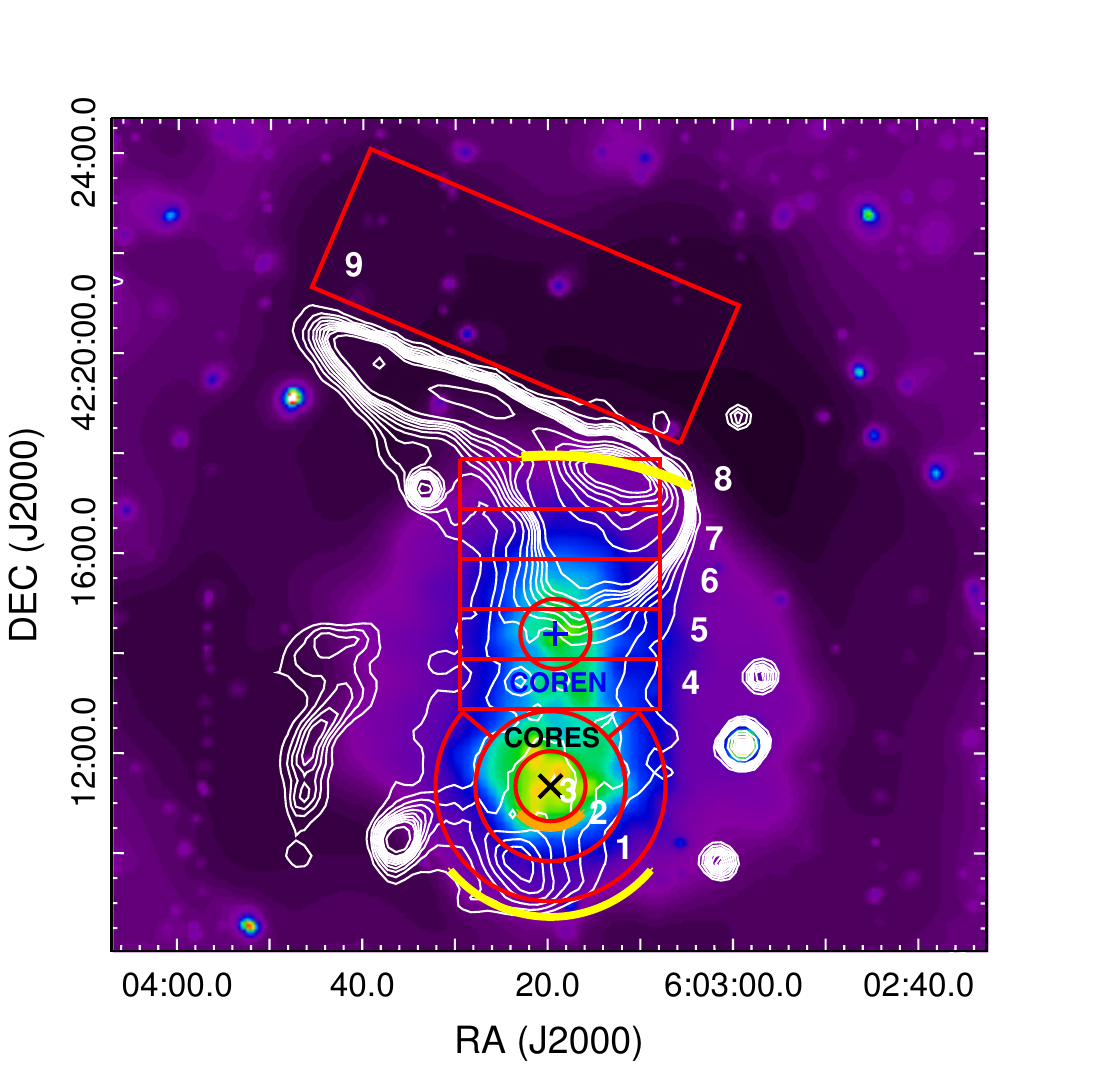}
 \end{minipage}
    \caption{\textit{Left panel}: \textit{XMM-Newton} smoothed image in the 0.3--10 keV band of 1RXSJ0603. The red sectors represent the regions used for the temperature and abundance  distributions. The green sector is the offset region used for the sky background estimation. Cyan circles are the point sources removed, with enlarged radius for clarity purpose. White contours are LOFAR HBA radio contours. \textit{Right panel}: Enlarged image of 1RXSJ0603. The COREN and CORES are marked with blue and black crosses, respectively. The solid yellow and orange lines show the X-ray shock and cold front position.  The numbers label the regions used in the spectral fitting.}
    \label{fig:fig3}
  \centering
  \begin{minipage}[h]{0.45\textwidth}
    \centering
    \includegraphics[width=1\textwidth]{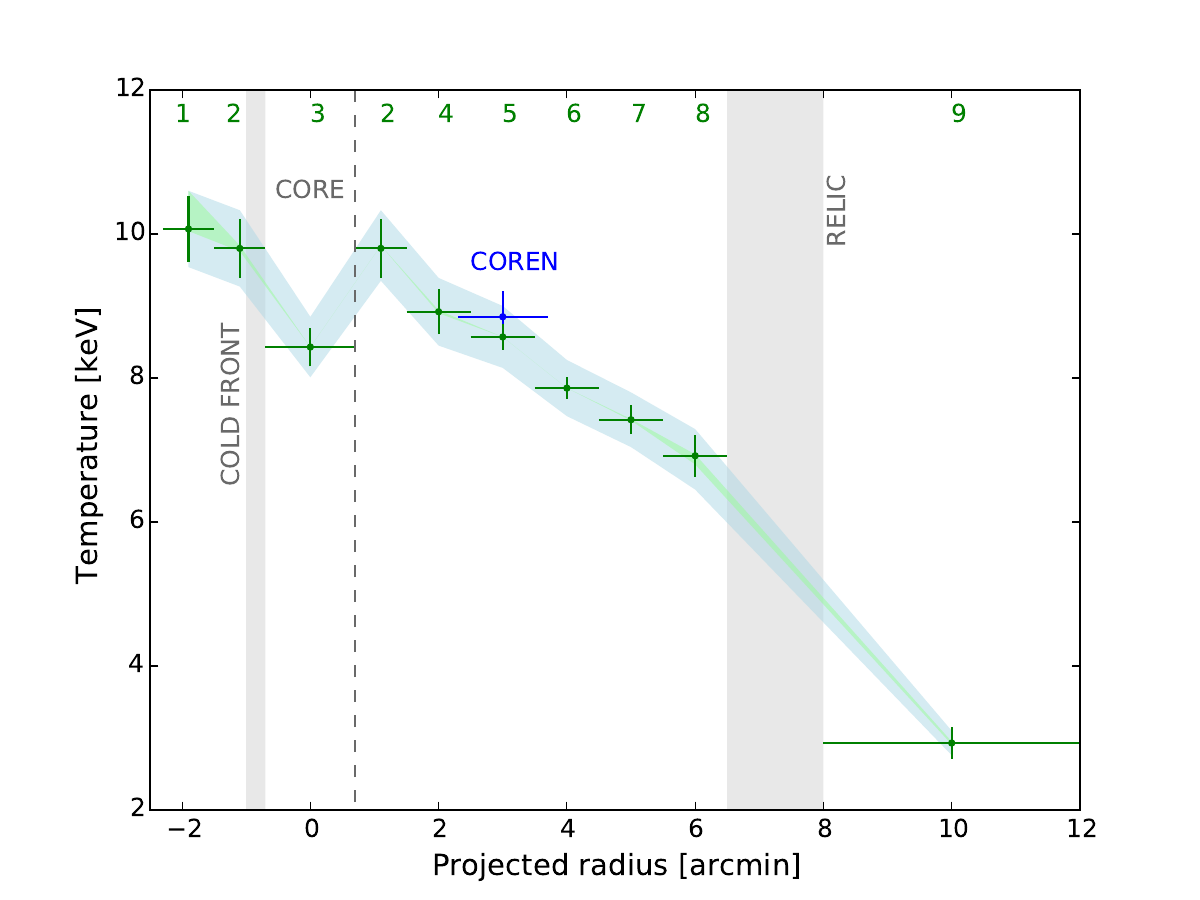}
  \end{minipage}
 \begin{minipage}[h]{0.45\textwidth}
    \centering
    \includegraphics[width=1\textwidth]{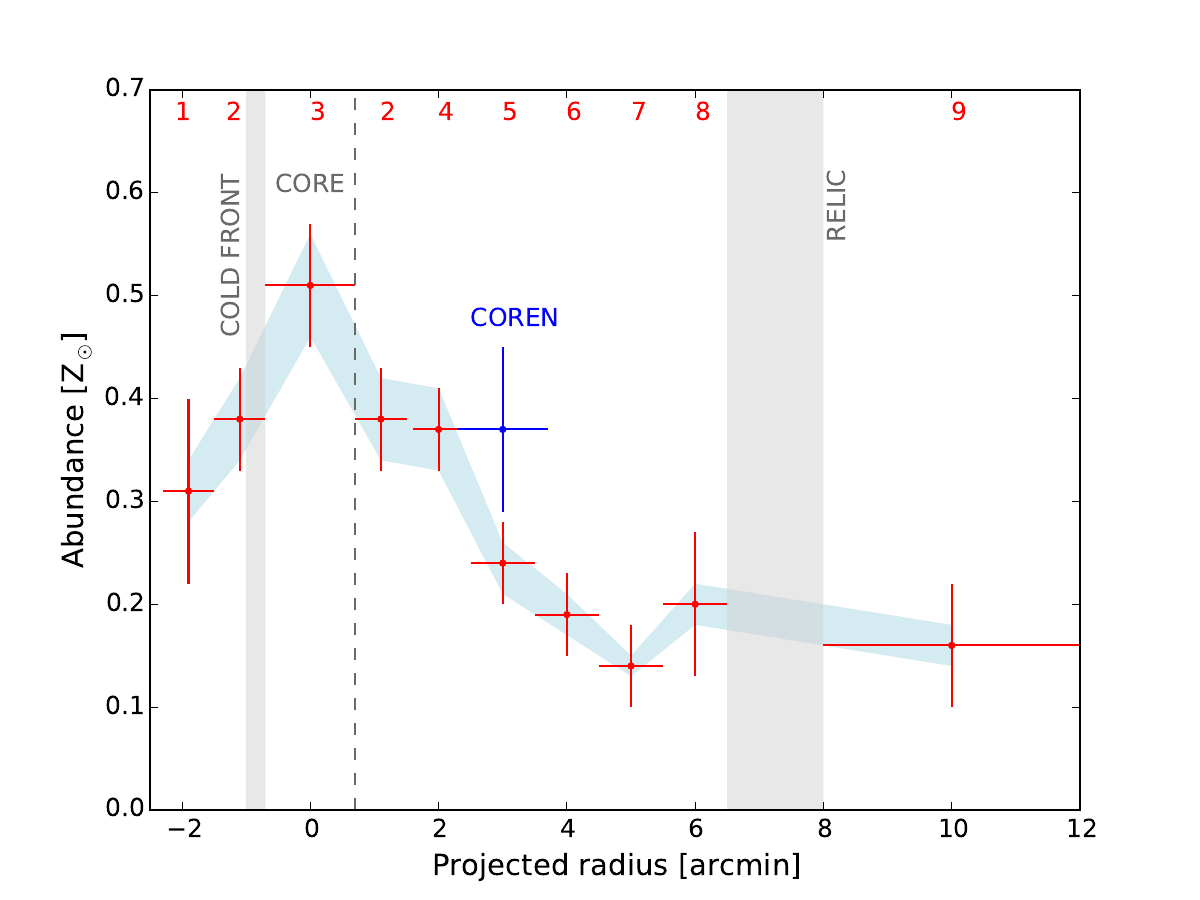}
 \end{minipage}
\caption{\textit{Left panel}:  Temperature distribution along the merging axis of 1RXSJ0603. \textit{Right panel}:  Abundance distribution along the merging axis of 1RXSJ0603. The shaded areas represent the systematic uncertainties: the green shadow correspond to the background normalization variation, and the blue one to the temperature structure.  The shaded grey areas show the cold front and radio relic position. The dashed line shows the outer edge of the core spectral extraction region. The numbers and COREN blue point correspond to the regions and the red circle centered on COREN, respectively shown in Fig. \ref{fig:fig3}.}
\label{fig:fig4}
\end{figure*}

      We fit the cluster emission and the background spectra (see Table \ref{tab:taba1}) in the sectors shown in Fig. \ref{fig:fig3}, having excluded the point sources. We leave $N_{\rm{H}}$ free to vary between 0.8~x~$N_{\rm{HI}}$~$\leq$~ $N_{\rm{H}}$~$\leq$~1.2~x~$N_{\rm{HTotal}}$, where $N_{\rm{HI}}$ = 3.22~$\times$~10$^{21}$ cm$^{-2}$ and  $N_{\rm{HTotal}}$ = 4.64~$\times$~10$^{21}$ cm$^{-2}$ \citep{Willingale2013}, as explained in Section 3.1. The two circular regions are centred close to the two brightest cluster galaxies (BCGs) or subcores of the bimodal merger system described by \cite{Dawson2015} (BCGN: $22^{\rm{h}} 42^{\rm{m}} 50\fs00$, $53\degr 05\arcmin 06\farcs 00$ ; BCGS: $22^{\rm{h}} 42^{\rm{m}} 39\fs00$, $52\degr 58\arcmin 35\farcs 00$). We use as centroid the southern "core", because it contains the X-ray peak emission and is close to the southern BCG.

      The best-fit parameters are shown in  Table \ref{tab:tabb1}, while the  temperature and abundance  distributions along the merging axis are plotted in Fig. \ref{fig:fig2}.  By convention, and for all the further relevant figures in this paper, the regions lying behind the projected main BCG trajectory are plotted with positive values on the x-axis. The temperature distribution presents a  variation in the range between $\sim$8--10 keV within the inner regions of the X-ray emission. This temperature is higher in  the last annular region, coincident with the post-shock region as described in \cite{Ogrean2014} and \cite{Akamatsu2015}. These results are in good agreement within the uncertainties with the temperature map obtained by \cite{Ogrean2013a} and the radial profile of \cite{Akamatsu2015}.

       Between the two BCGs (sectors 2 to 4) the Fe abundance is measured at its lowest value, between $\sim$0.2--0.3 $Z_\odot$. Moreover, the abundance profile shows an apparent enhancement with a maximum value of $\sim$0.4 $Z_\odot$ in sectors 1 and 5, where the two BGCs are located.  The northern BCG region evinces a lower abundance than the entire sector 5. If we divide sector 5 in three regions, we obtain a gradient in the abundance and slight increase in the temperature towards the east, see Table \ref{tab:tabb1}. BCGN coincides mostly with region 5b and the enhancement in abundance in sector 5 is mainly driven by region 5a.

           \subsection{1RXS J0603.3+4214}

        1RXS J0603.3+4214 (hereafter 1RXSJ0603), also known as the "Toothbrush" cluster, is a bright, rich and massive cluster discovered by \cite{VanWeeren2012a} and located at $z$~=~0.225. It hosts three radio relics and a giant ($\sim$2 Mpc) radio halo elongated along the main merger axis S--N, which follows the X-ray emission morphology excluding the southernmost part \citep{Ogrean2013b,Rajpurohit2018}. \cite{Stroe2014b} found no enhancement of H$_\alpha$ emitters near the northern radio relic, with a star formation rate (SFR) consistent with that of blank field galaxies at $z$~=~0.2. They concluded that the lack of H$_\alpha$ emitters could be possible because 1RXSJ0603 is a well evolved system, approximately $\sim$2 Gyr after the core passage.

           \begin{figure*}[ht!]
               \centering
                \begin{minipage}[h]{0.40\textwidth}
    \centering
        \includegraphics[width=1\textwidth]{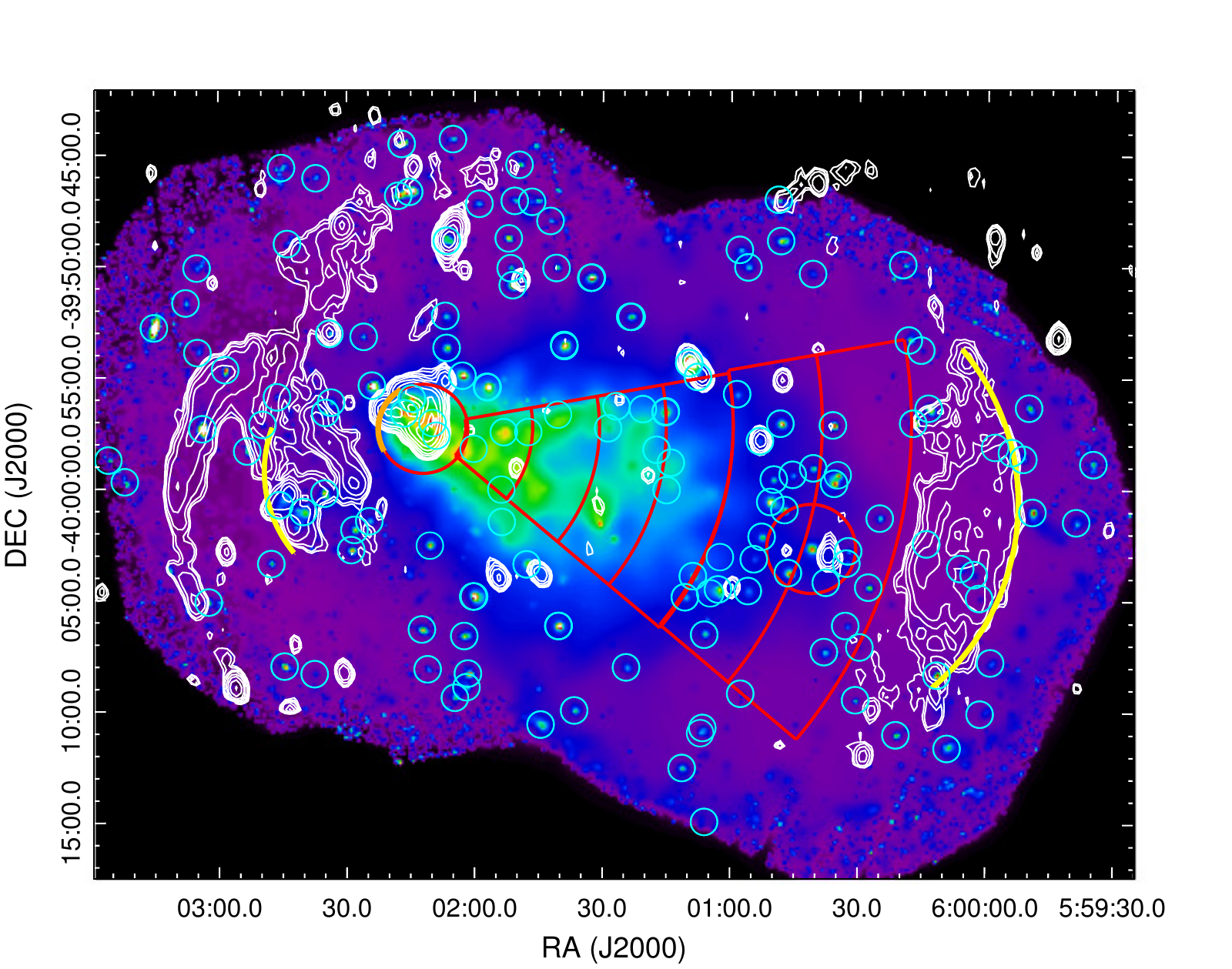}
 \end{minipage}
  \begin{minipage}[h]{0.39\textwidth}
    \centering
         \includegraphics[width=1\textwidth]{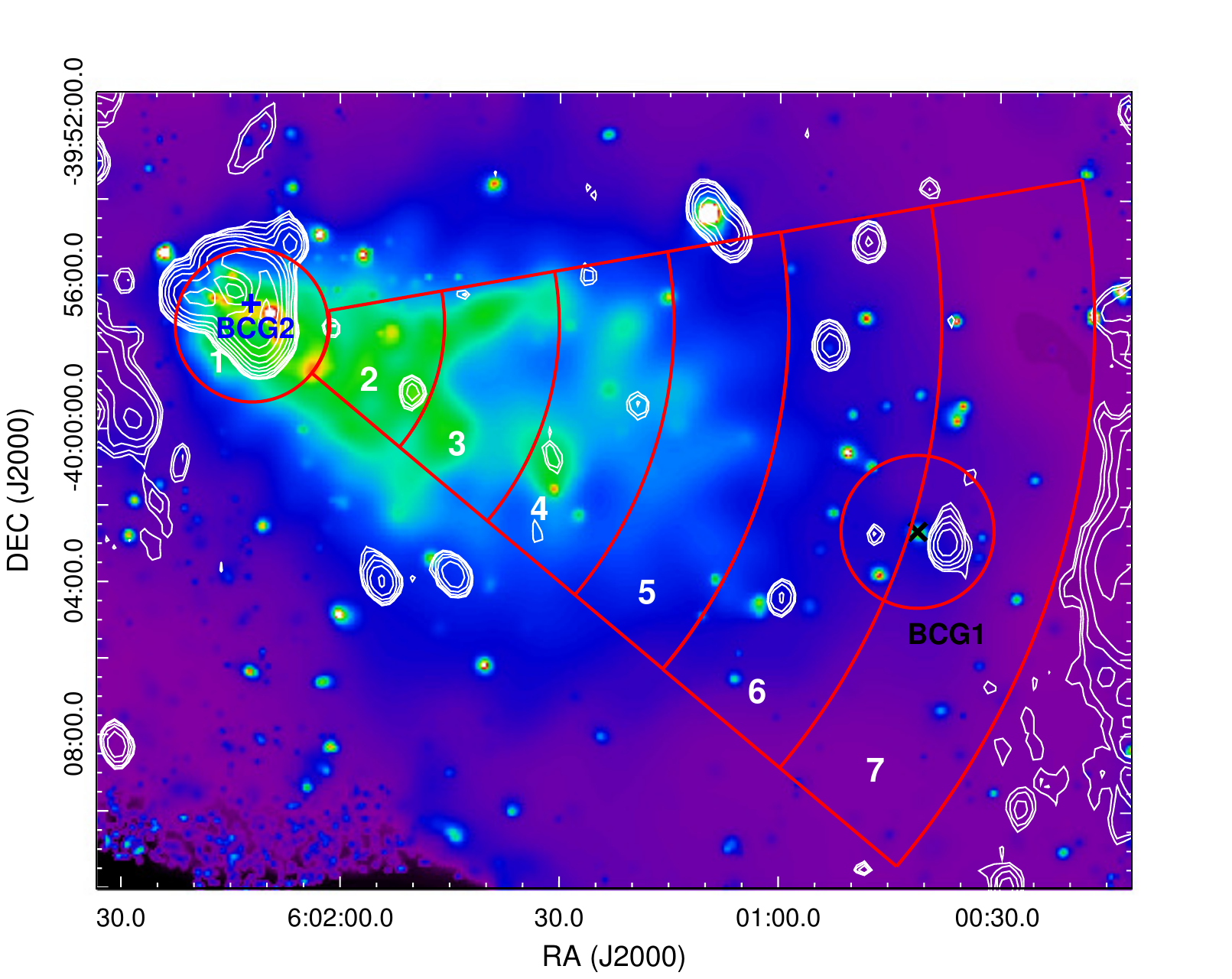}
 \end{minipage}
    \caption{\textit{Left panel}: \textit{XMM-Newton} smoothed image in the 0.3--10 keV band of A3376. The red sectors represent the regions used for the temperature and abundance  distributions. Cyan circles are the point sources removed, with enlarged radius for clarity purpose. White contours are VLA 1.4 GHz radio contours. The solid yellow and orange lines show the X-ray shocks and cold front position, respectively.  \textit{Right panel}: The BCG2 and BCG1 are marked with blue and black crosses. The numbers label the regions used in the spectral fitting.}
    \label{fig:fig5}
  \centering
  \begin{minipage}[h]{0.45\textwidth}
    \centering
    \includegraphics[width=1\textwidth]{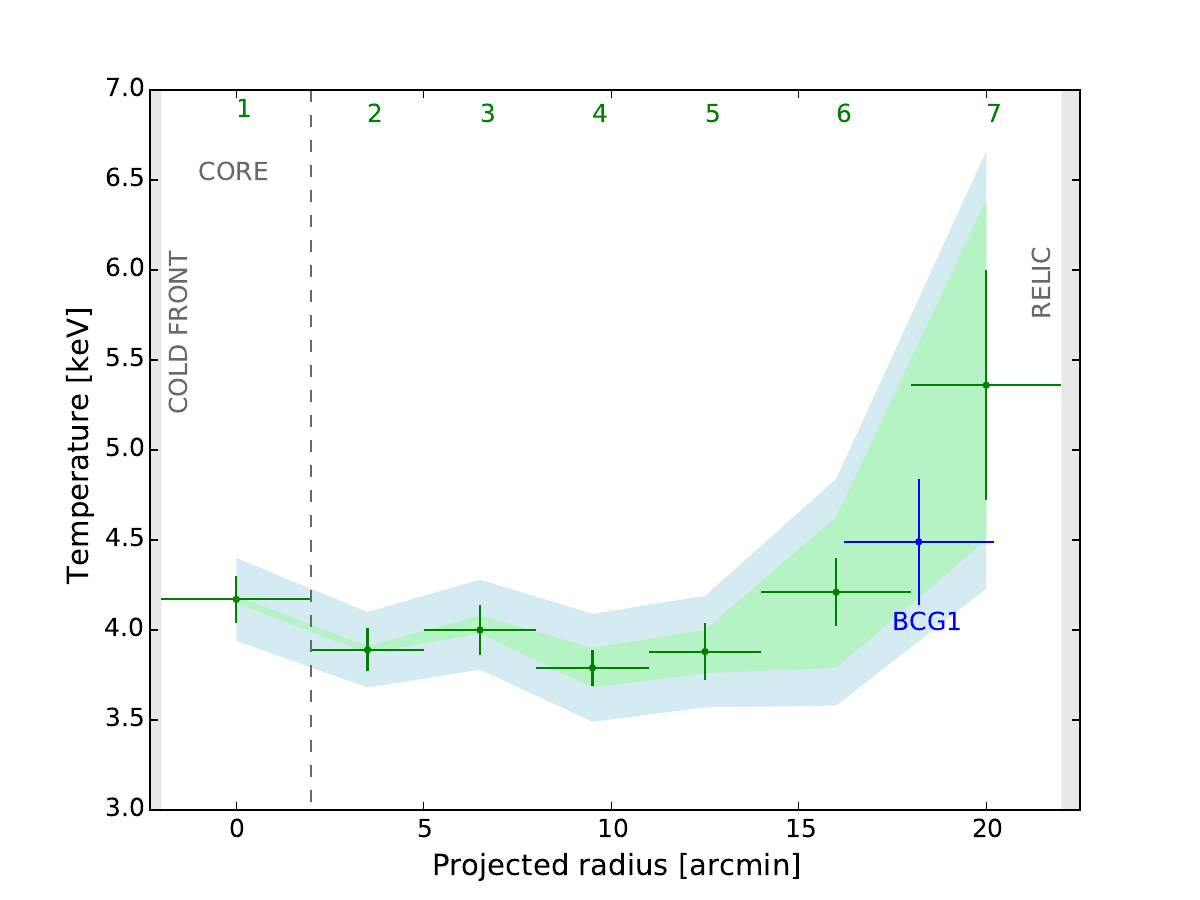}
  \end{minipage}
 \begin{minipage}[h]{0.45\textwidth}
    \centering
    \includegraphics[width=1\textwidth]{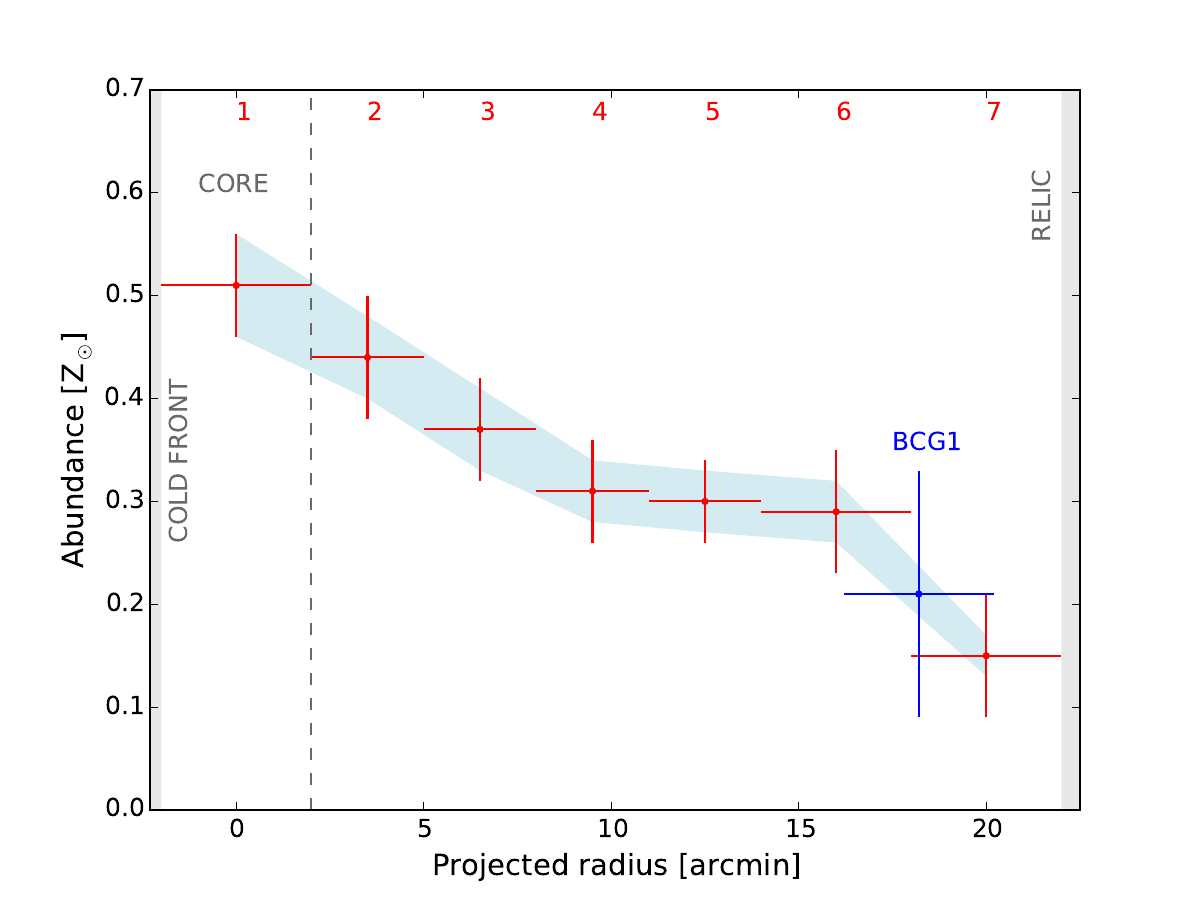}
 \end{minipage}
\caption{\textit{Left panel}:  Temperature distribution along the merging axis of A3376. \textit{Right panel}:  Abundance distribution along the merging axis of A3376. The shadow area represent the systematic uncertainties. The shaded areas represent the systematic uncertainties: the  green shadow correspond to the background normalization variation, and the blue one to the temperature structure.  The shaded grey areas show the cold front and radio relic position. The dashed shows the outer edge of the core spectral extraction region. The numbers and BCG1 blue point correspond to the regions and the red circle centered on BCG1, respectively shown in Fig. \ref{fig:fig5}.}
\label{fig:fig6}
\end{figure*}

\subsubsection{Results}
        
          As shown in Fig. \ref{fig:fig3},  we use various sectors to fit the cluster emission and the background spectra (see Table \ref{tab:taba2}), excluding the point sources and using $N_{\rm{HI}}$ = 2.15~$\times$~10$^{21}$ cm$^{-2}$ \citep{Willingale2013}. We follow the X-ray emission S--N shape, coincident with the radio halo elongation with boxes  4--8, as well as the pre-shock region 9. We dedicate two circular sectors to the subcores of the main two subclusters, locating the brightest one in X-rays at the south (CORES: $6^{\rm{h}} 03^{\rm{m}} 13\fs39$, $42\degr 12\arcmin 31\farcs 00$) and the second at the north (COREN: $6^{\rm{h}} 03^{\rm{m}} 19\fs17$, $42\degr 14\arcmin 24\farcs 00$). We use CORES as centroid for our temperature and abundance  distributions along the merging axis (see Fig. \ref{fig:fig4}).

         The best-fit parameters are presented in  Table \ref{tab:tabb2} and are in good agreement with the previous X-ray studies by \cite{Ogrean2013b,Itahana2015}; and \cite{VanWeeren2016}. The  subcore temperatures ($kT$~=~8.43~$\pm$~0.27 keV for CORES and $kT$~=~8.85~$\pm$~0.36 keV for COREN) are consistent with \cite{VanWeeren2016}. Interestingly, we find a peak in Fe abundance    located in the inner core, just behind the cold front (CF) (see Fig. \ref{fig:fig4}; its location is approximated based on \citealt{VanWeeren2016}). In the south, we see a slight temperature gradient towards the southern shock front, anticorrelated with a decrease in Fe abundance. The COREN shows the same abundance as the adjacent inner region, but a higher value than in the outer sectors (6--9). It is possible that the gas belonging to this subcluster has been stripped during the merger and is mixing in that region. In the north, a gradual smooth decrease of temperature  can be observed up to the post-shock region, followed by a significant drop after the radio relic. Meanwhile, the Fe abundance decreases in sector 5 and flattens later around $\sim$0.2 $Z_{\odot}$ for the outer sectors.

              \begin{figure*}[ht!]
             \centering
             
               \begin{minipage}[h]{0.40\textwidth}
    \centering
    \includegraphics[width=1\textwidth]{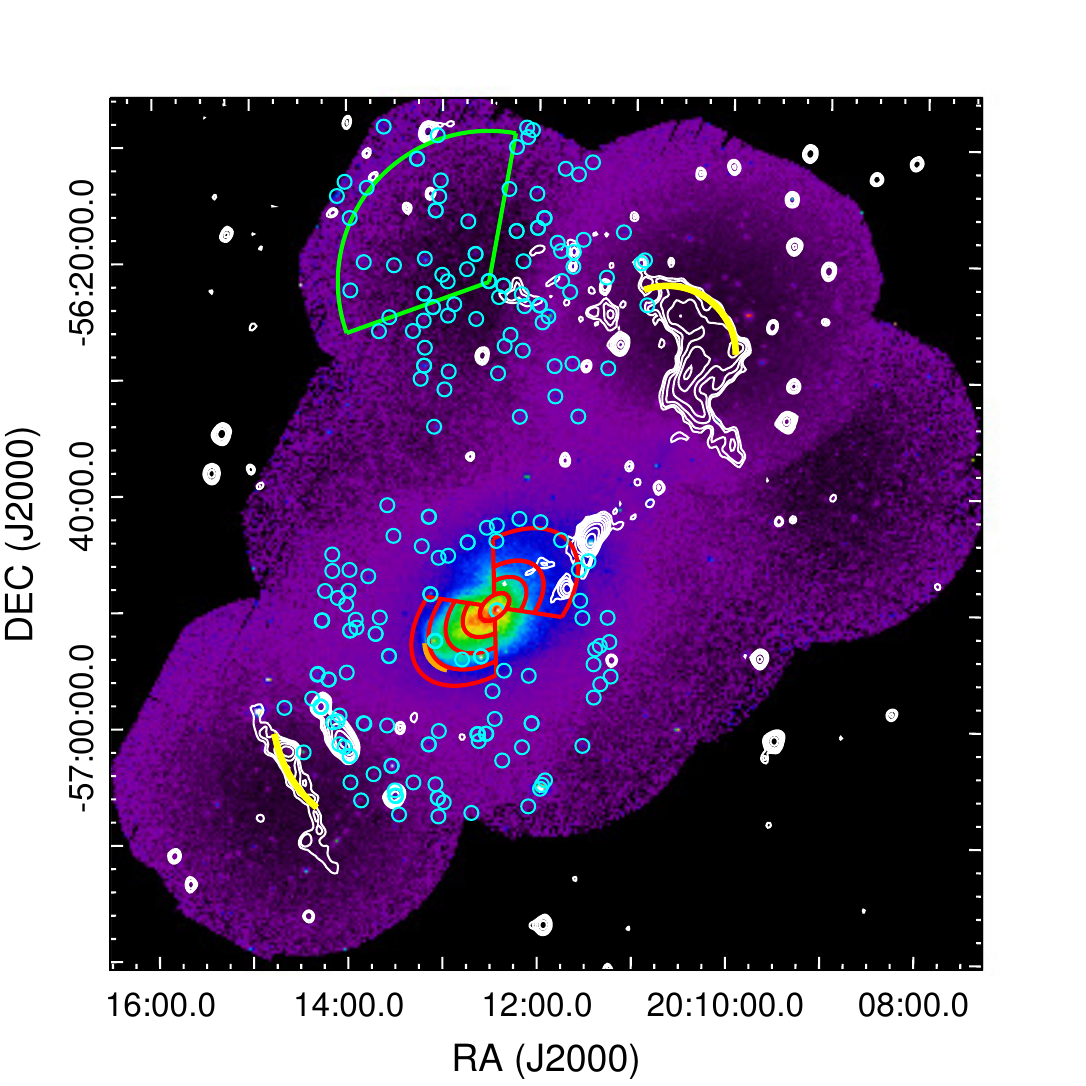} 
  \end{minipage}
 \begin{minipage}[h]{0.40\textwidth}
    \centering
    \includegraphics[width=1\textwidth] {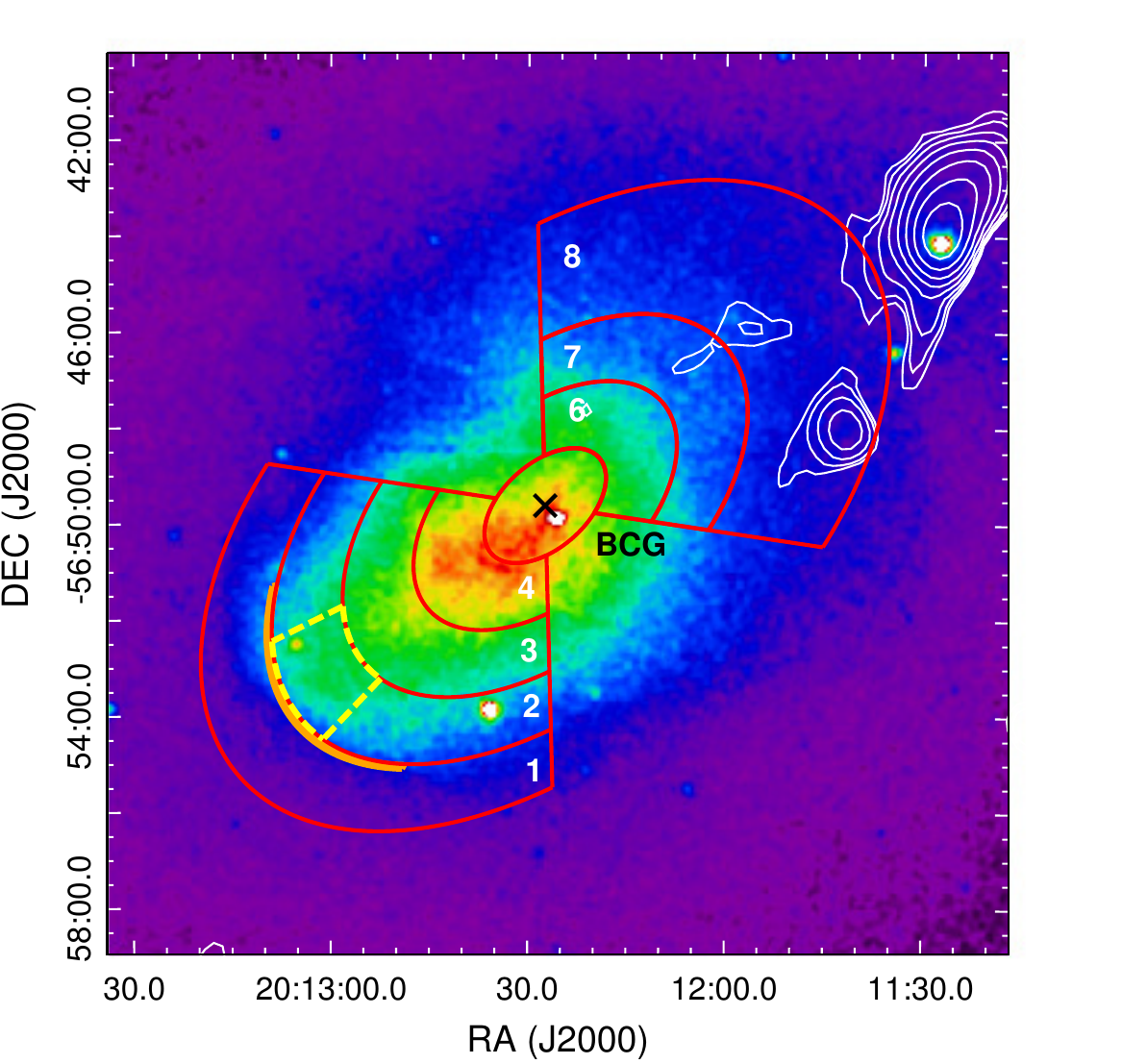}
 \end{minipage}

    \caption{\textit{Left panel}: \textit{XMM-Newton} smoothed image in the 0.3--10 keV band of A3667. The red elliptical sectors represent the regions used for the temperature and abundance distributions. The green sector is the offset region used for the sky background estimation. Cyan circles are the point sources removed, with enlarged radius for clarity purpose. White contours are SUMSS radio contours. The solid yellow and orange lines show the X-ray shocks and cold front position, respectively. \textit{Right panel}: Enlarged image of A3667. The BCG is marked with black cross. The solid  orange line shows the cold front position.  In addition to sector 2, the dashed yellow sector corresponds to a region considered also separately (see text). The numbers label the regions used in the spectral fitting.}
    \label{fig:fig7}
  \centering
  \begin{minipage}[h]{0.45\textwidth}
    \centering
    \includegraphics[width=1\textwidth]{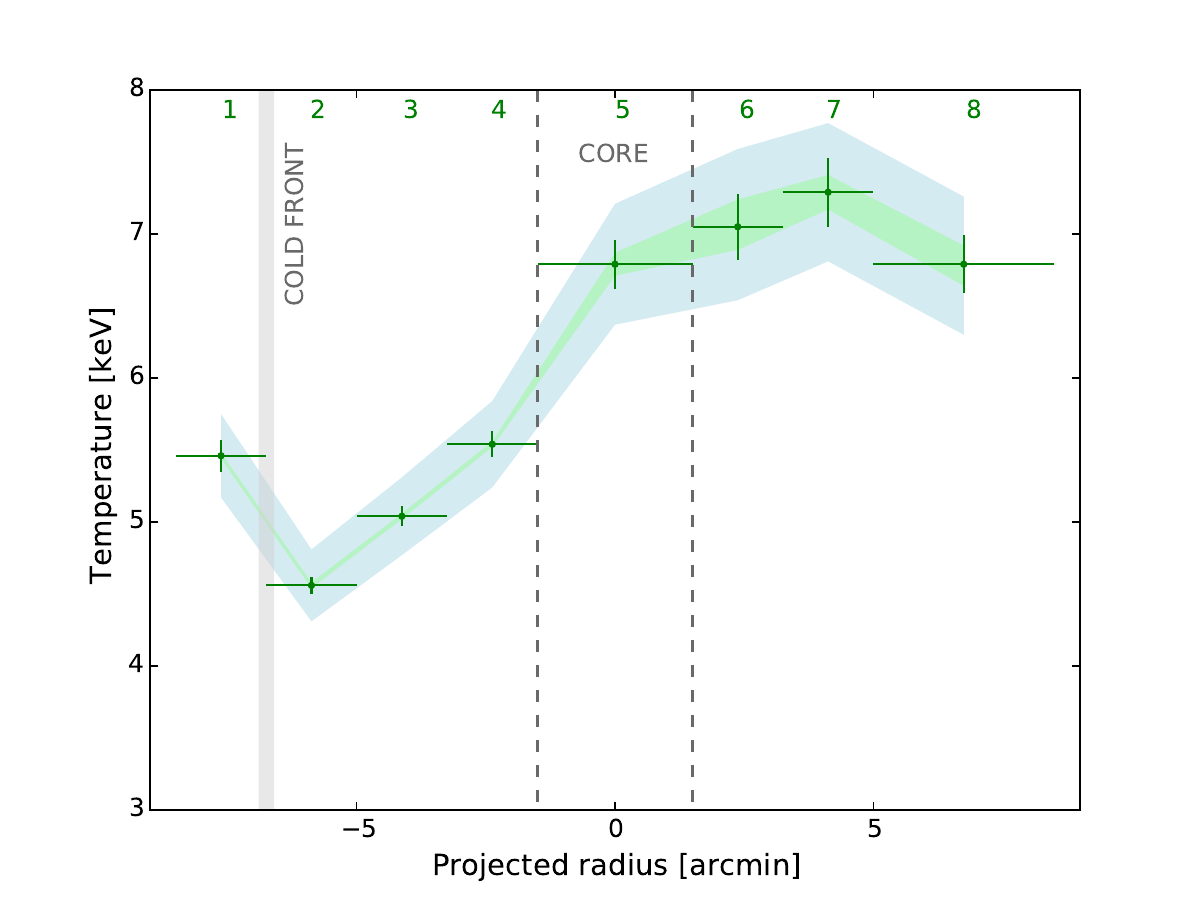}
  \end{minipage}
 \begin{minipage}[h]{0.45\textwidth}
    \centering
    \includegraphics[width=1\textwidth]{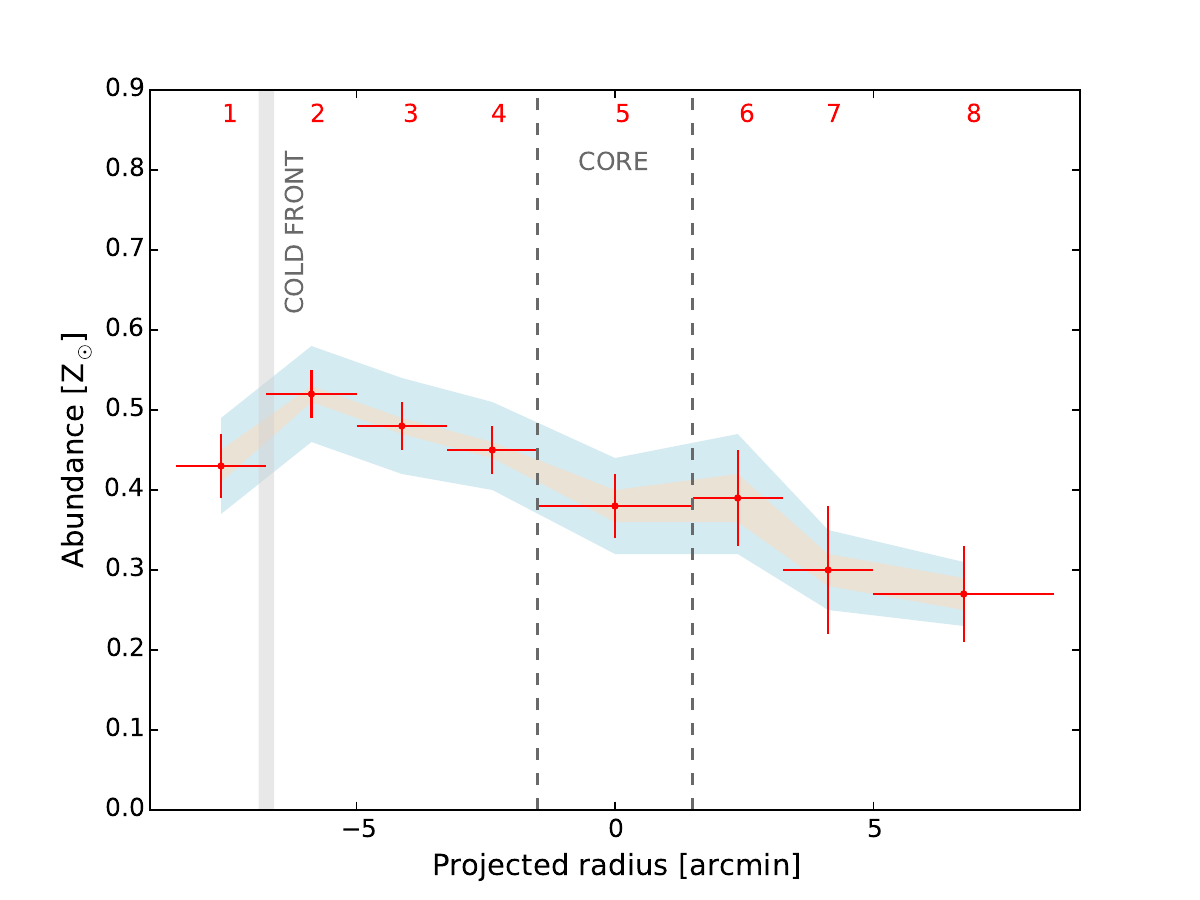}
 \end{minipage}
\caption{\textit{Left panel}:  Temperature distribution along the merging axis of A3667. \textit{Right panel}: Abundance distribution along the merging axis of A3667. The shaded areas represent the systematic uncertainties: green and orange shadows correspond to the background normalization variation, and the blue one to the temperature structure. The shaded grey area shows the cold front position. The dashed lines show the outer edges of the core spectral extraction region. The numbers correspond to the regions shown in Fig. \ref{fig:fig7}.}
\label{fig:fig8}
\end{figure*}

        \begin{figure*}[ht!]
               \centering
   \begin{minipage}[h]{0.40\textwidth}
    \centering
    \includegraphics[width=1\textwidth]{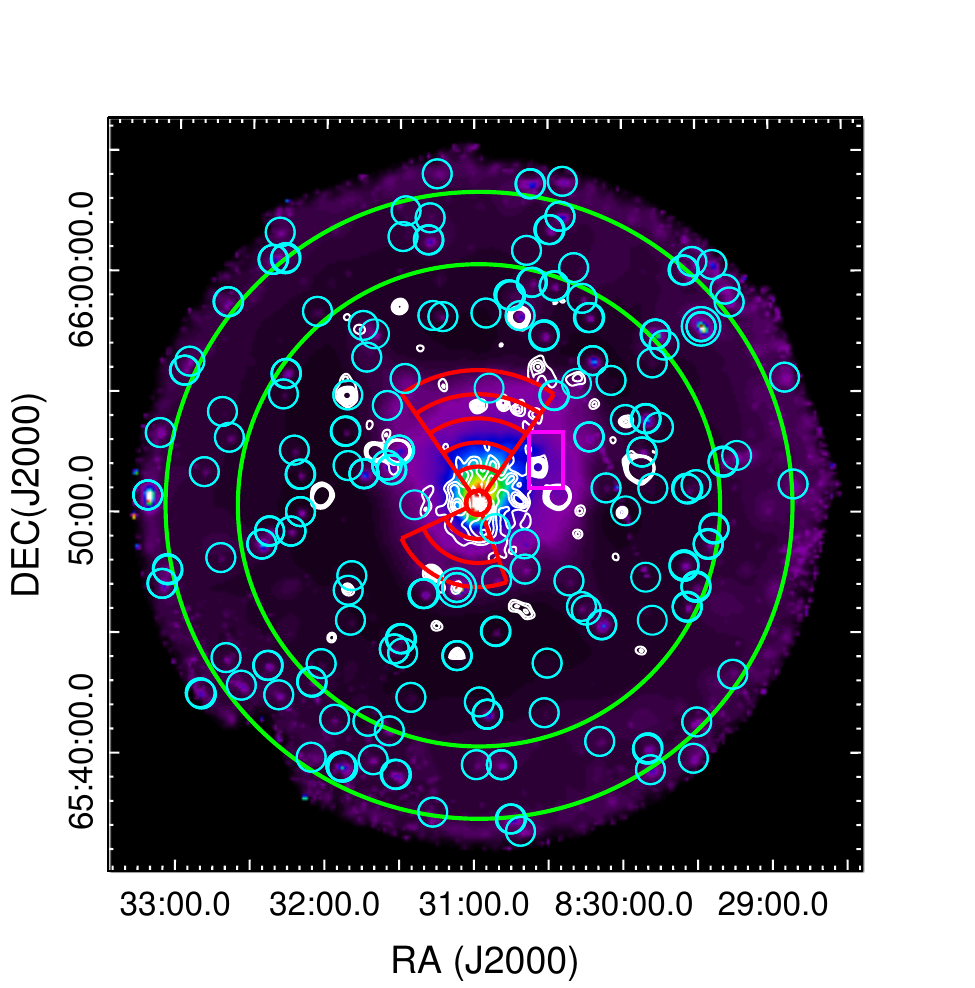} 
  \end{minipage}
 \begin{minipage}[h]{0.41\textwidth}
    \centering
    \includegraphics[width=1\textwidth] {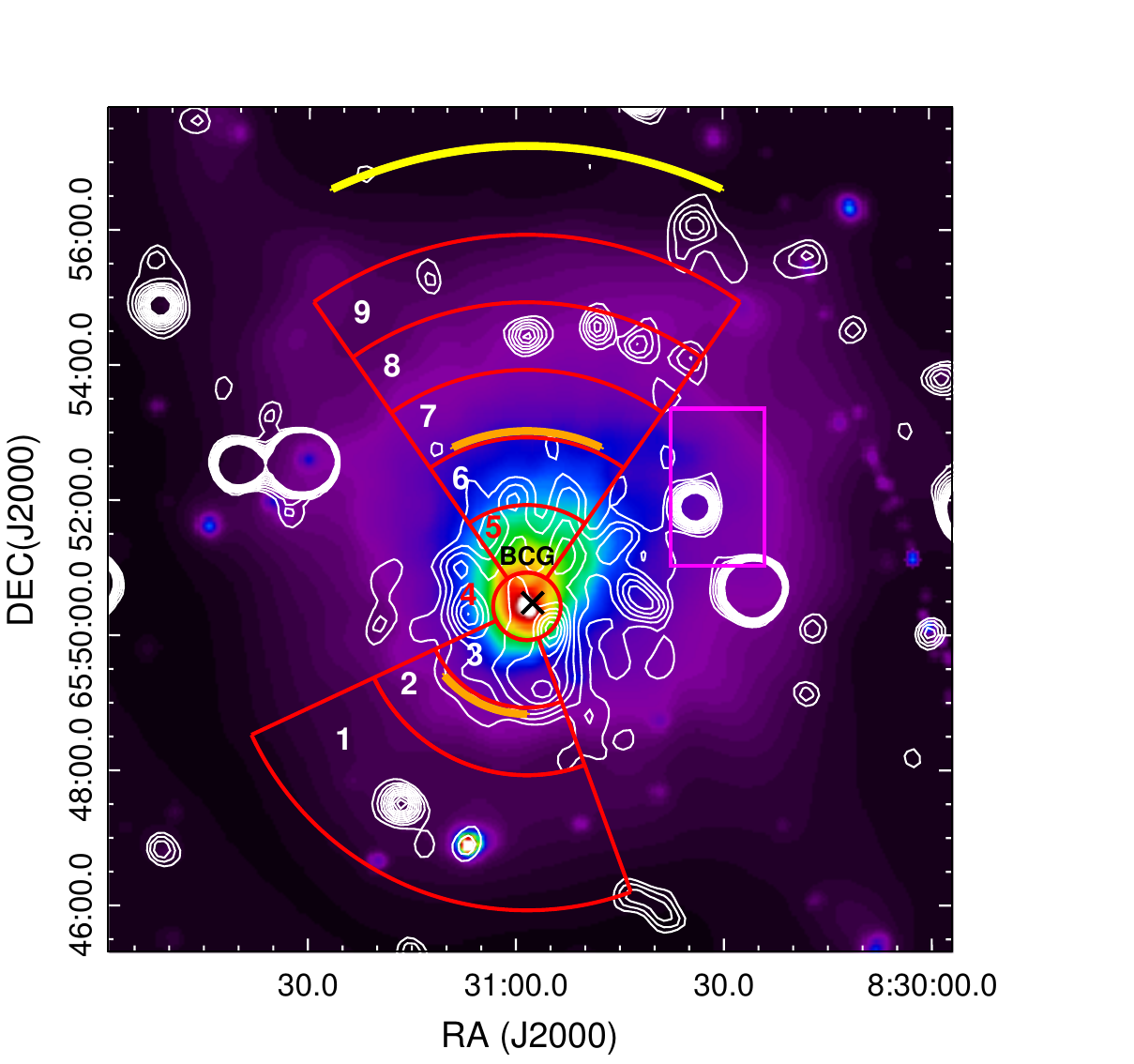}
 \end{minipage}
    \caption{\textit{Left panel}: \textit{XMM-Newton} smoothed image in the 0.3--10 keV band of A665. The red sectors represent the regions used for the temperature and abundance  distributions. The green annulus is the offset region used for the sky background estimation. Cyan circles are the point sources removed, with enlarged radius for clarity purpose. White contours are VLA 1.4 GHz radio contours.   \textit{Right panel}: Enlarged image of A665. The BCG is marked with a black cross. The magenta box includes the cooler gas region observed by \cite{Markevitch2001}. The solid yellow and orange lines show the X-ray shock and cold fronts position, respectively. The numbers label the regions used in the spectral fitting.} 
    \label{fig:fig9}
  \centering
  \begin{minipage}[h]{0.45\textwidth}
    \centering
    \includegraphics[width=1\textwidth]{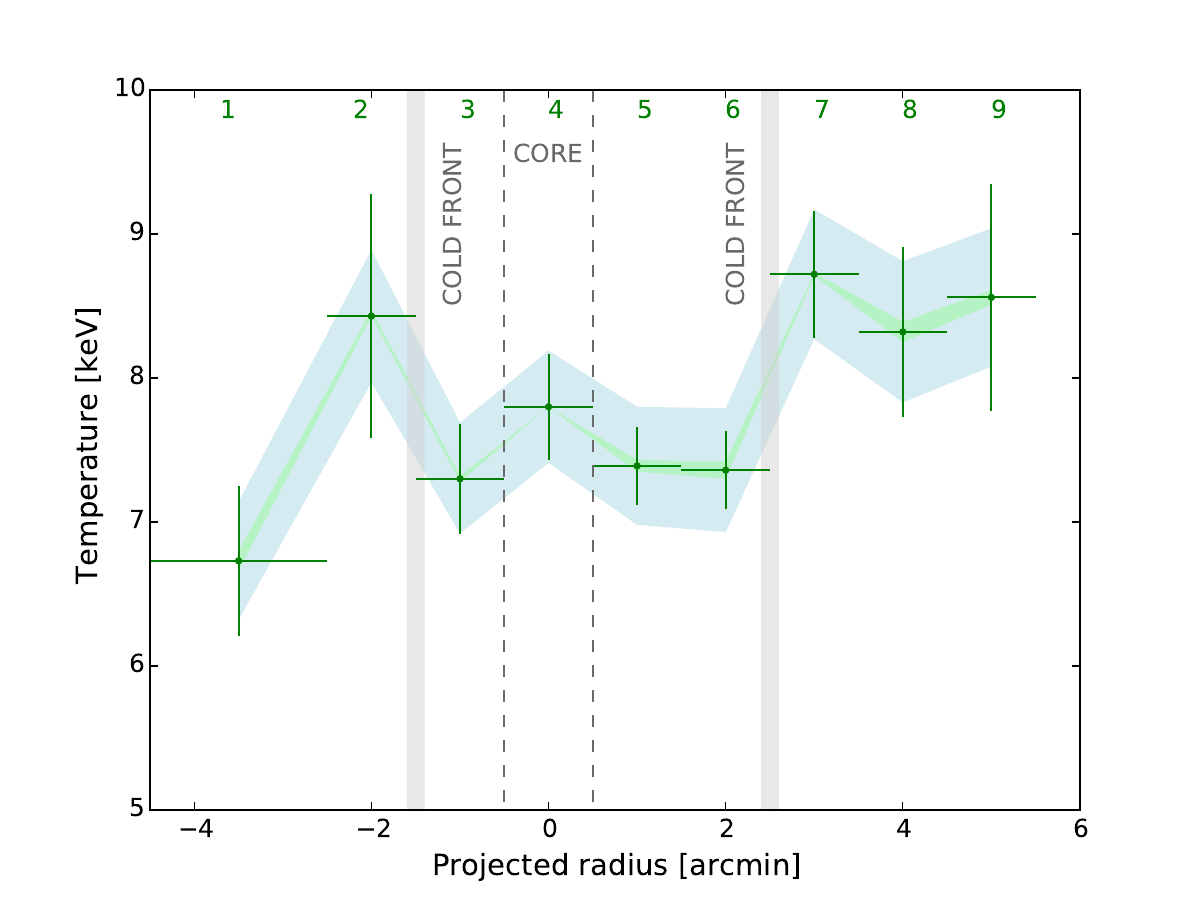}
  \end{minipage}
 \begin{minipage}[h]{0.45\textwidth}
    \centering
    \includegraphics[width=1\textwidth]{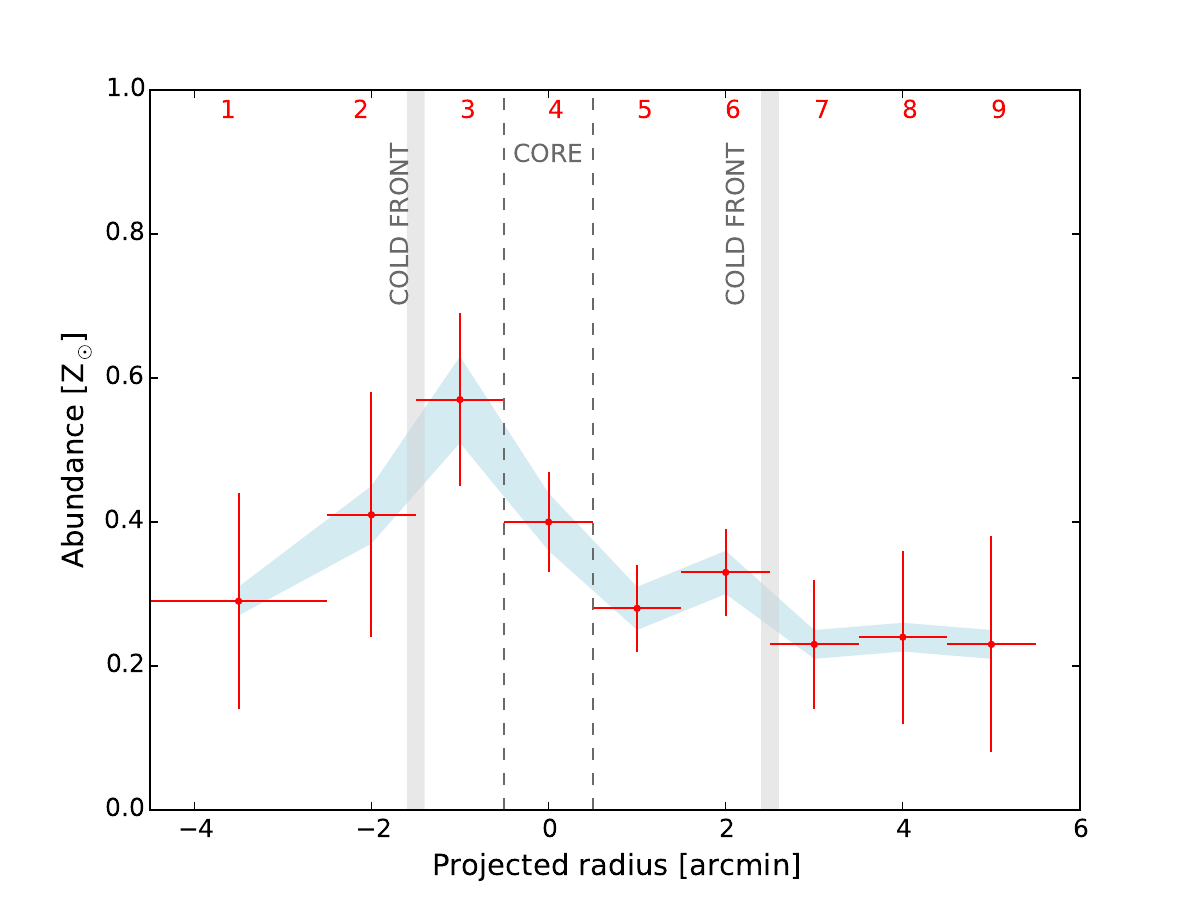}
 \end{minipage}
\caption{\textit{Left panel}:  Temperature distribution along the merging axis of A665. \textit{Right panel}: Abundance distribution along the merging axis of A665. The shaded area represent the systematic uncertainties: green shadow correspond to the background normalization variation, and the blue one to the temperature structure. The shaded grey areas show the cold fronts position. The dashed lines show the outer edges of the core spectral extraction region. The numbers correspond to the regions shown in Fig. \ref{fig:fig9}. }
\label{fig:fig10}
\end{figure*}

  \subsection{A3376}

        Abell 3376 (hereafter A3376) is a bright and nearby ($z$~=~0.046) merging galaxy cluster located in the southern hemisphere. The N-body hydrodynamical simulations of \cite{Machado2013} suggest a two body merger scenario with a mass ratio of 3--6:1. The more massive and diffuse western subcluster core (BCG1: $6^{\rm{h}} 00^{\rm{m}} 41\fs10$, $-40\degr 02\arcmin 40\farcs 00$) has been disrupted by a dense and compact eastern subcluster (BCG2: $6^{\rm{h}} 02^{\rm{m}} 09\fs70$, $-39\degr 57\arcmin 05\farcs 00$).  Recent studies suggest that the core-passing took place $\sim$0.5--0.6 Gyr ago in the W--E direction \citep{Durret2013,Monteiro-Oliveira2017,Urdampilleta2018}. Due to the merger, the inner gas of A3376 follows a cometary tail X-ray morphology as shown in Fig. \ref{fig:fig5}.  A3376 hosts two giant ($\sim$Mpc) arc-shaped radio relics \citep{Bagchi2006} in the periphery of the cluster.  X-ray analyses with the \textit{Suzaku} satellite \citep{Akamatsu2012b,Urdampilleta2018} confirm the presence of a X-ray shock front located at the western radio relic, another shock probably associated with the eastern "notch" \citep{Paul2011,Kale2012} and a cold front at 3$\arcmin$ from the center.

       \subsubsection{Results}
           In the spectral analysis of A3376 we use two different observations as listed in Table 1. Specifically, ObsI:0151900101 is used for regions 1, 2 and 3, while ObsID:0504140101 is used for regions 6 and 7. In addition, both these observations are used for the overlapping regions 4 and 5. The background modelling parameters for each of the observations are shown in Table \ref{tab:taba3a} and Table \ref{tab:taba3b}. We fit the spectra of the annular sectors centred on BCG2, located close to the X-ray peak emission at the eastern subcluster. We assume the column density as $N_{\rm{HI}}$ = 4.84~$\times$~10$^{20}$ cm$^{-2}$ \citep{Willingale2013}. The best-fit parameters (see Table  \ref{tab:tabb3}) are in good agreement with previous studies with the \textit{Suzaku} satellite \citep{Akamatsu2012b,Urdampilleta2018}.
           
The temperature distribution of A3376 (Fig. \ref{fig:fig6}) shows a central region with an average temperature $\sim$4 keV out to the outer radii where an increase of the temperature is seen. It is associated to the post merger region behind the western X-ray shock as described by \cite{Urdampilleta2018}.  The Fe abundance profile shows a peak of  $\sim$0.5 $Z_\odot$ in the core, coincident with the BCG2. At larger distances, the abundance decreases smoothly to reach $\sim$0.3 $Z_\odot$ in regions 4, 5, 6, including BCG1 (dashed annular sector in Fig.  \ref{fig:fig5}) and in the surrounding diffuse gas. The lower abundance value appears in the outermost region of the cluster before the X-ray shock.  
   
         \subsection{A3667}
         
          Abell 3667 (hereafter A3667) is a widely studied bright, low redshift ($z$~=~0.0553) and bimodal merging galaxy cluster \citep{Markevitch1999}. As a consequence of a violent merger along the northwest (NW) direction, A3667 hosts two curved radio relics to the NW and SE  \citep{Rottgering1997,Johnston-Hollitt2003,Johsnton-Hollitt2017,Carretti2013,Hindson2014,Riseley2015}. The inner central region shows an elongated morphology of high-abundance gas in SE-NW direction \citep{Mazzotta2002, Briel2004,Lovisari2009} according to the subcluster infall direction. A3667 also includes a prominent mushroom-like cold front close to the center at SE   \citep{Vikhlinin2001,Briel2004,Owers2009,Datta2014,Ichinohe2017}.

          \subsubsection{Results}
        We fit the cluster emission and the background spectra (see Table \ref{tab:taba4}) in the elliptical regions as shown in Fig. \ref{fig:fig7}, excluding the point sources and using $N_{\rm{HI}}$ = 4.44~$\times$~10$^{20}$ cm$^{-2}$ \citep{Willingale2013}.   We use as centroid  the main BCG ($20^{\rm{h}} 12^{\rm{m}} 27\fs43$, $-56\degr 49\arcmin 35\farcs 85$), which is close to the X-ray emission peak ($20^{\rm{h}} 12^{\rm{m}} 27\fs$,  $-56\degr 50\arcmin 11\farcs 00$). We distribute the elliptical sectors along the merging axis assuming the BCG as the center and following the X-ray morphology described by \cite{Briel2004}.

        \begin{figure*}[ht!]
               \centering
   \begin{minipage}[h]{0.40\textwidth}
    \centering
    \includegraphics[width=1\textwidth]{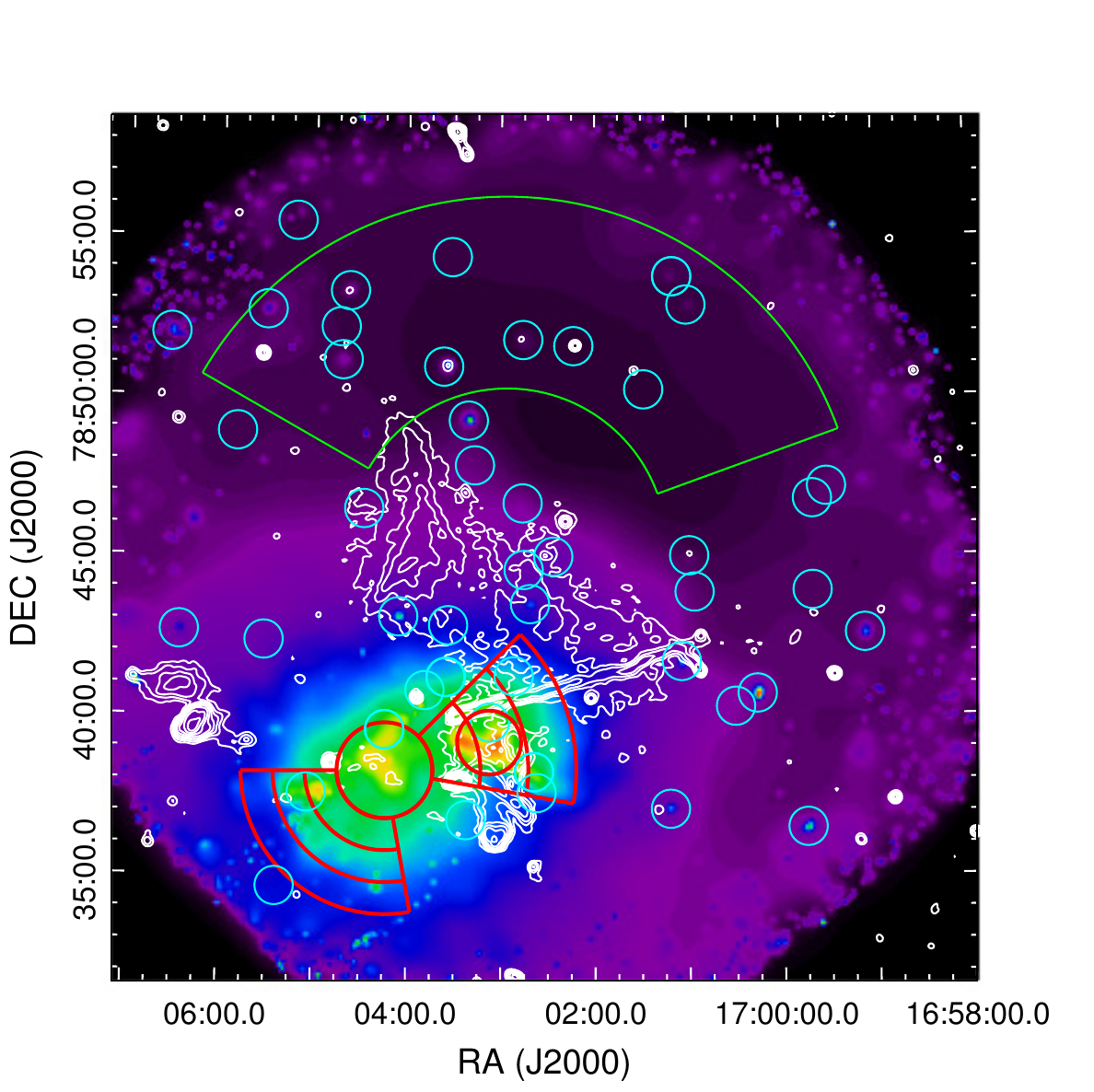} 
  \end{minipage}
 \begin{minipage}[h]{0.40\textwidth}
    \centering
    \includegraphics[width=1\textwidth] {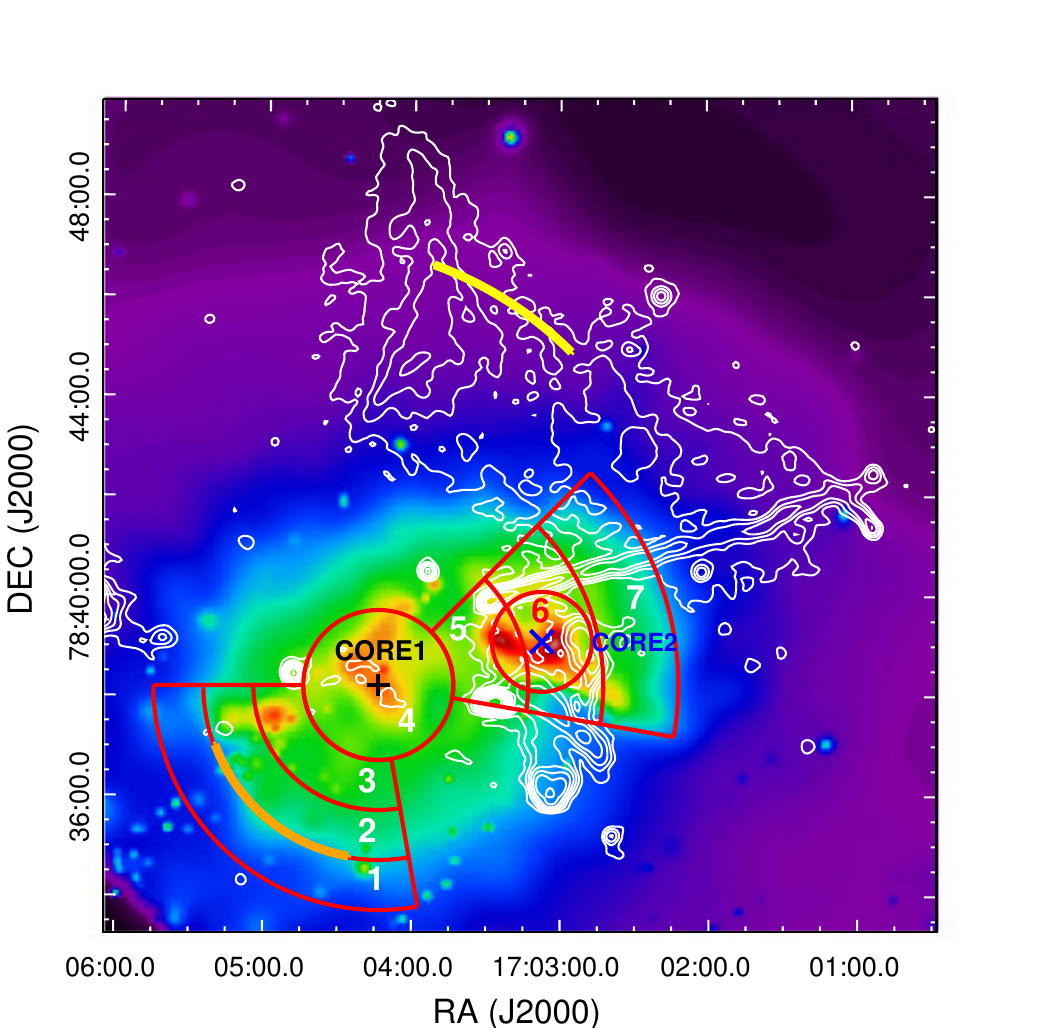}
 \end{minipage}
    \caption{\textit{Left panel}: \textit{XMM-Newton} smoothed image in the 0.3--10 keV band of A2256. The red sectors represent the regions used for the temperature and abundance  distributions. The green sector is the offset region used for the sky background estimation. Cyan circles are the point sources removed, with enlarged radius for clarity purpose. White contours are WRST radio contours. \textit{Right panel}: Enlarged image of A2256. The CORE1 and CORE2 are marked with a black and blue crosses, respectively. The solid yellow and orange lines show the X-ray shocks and cold front position. The numbers label the regions used in the spectral fitting.}
    \label{fig:fig11}
  \centering
  \begin{minipage}[h]{0.45\textwidth}
    \centering
    \includegraphics[width=1\textwidth]{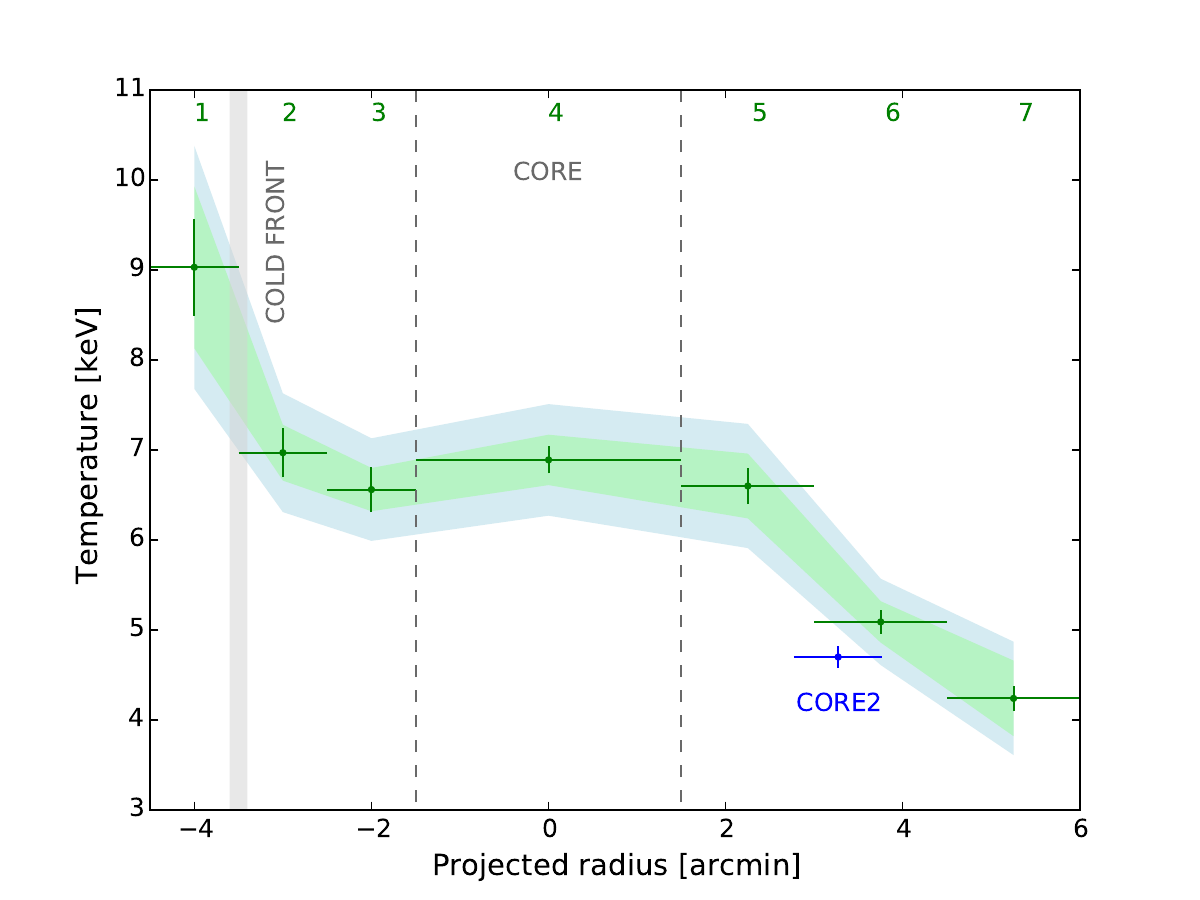}
  \end{minipage}
 \begin{minipage}[h]{0.45\textwidth}
    \centering
    \includegraphics[width=1\textwidth]{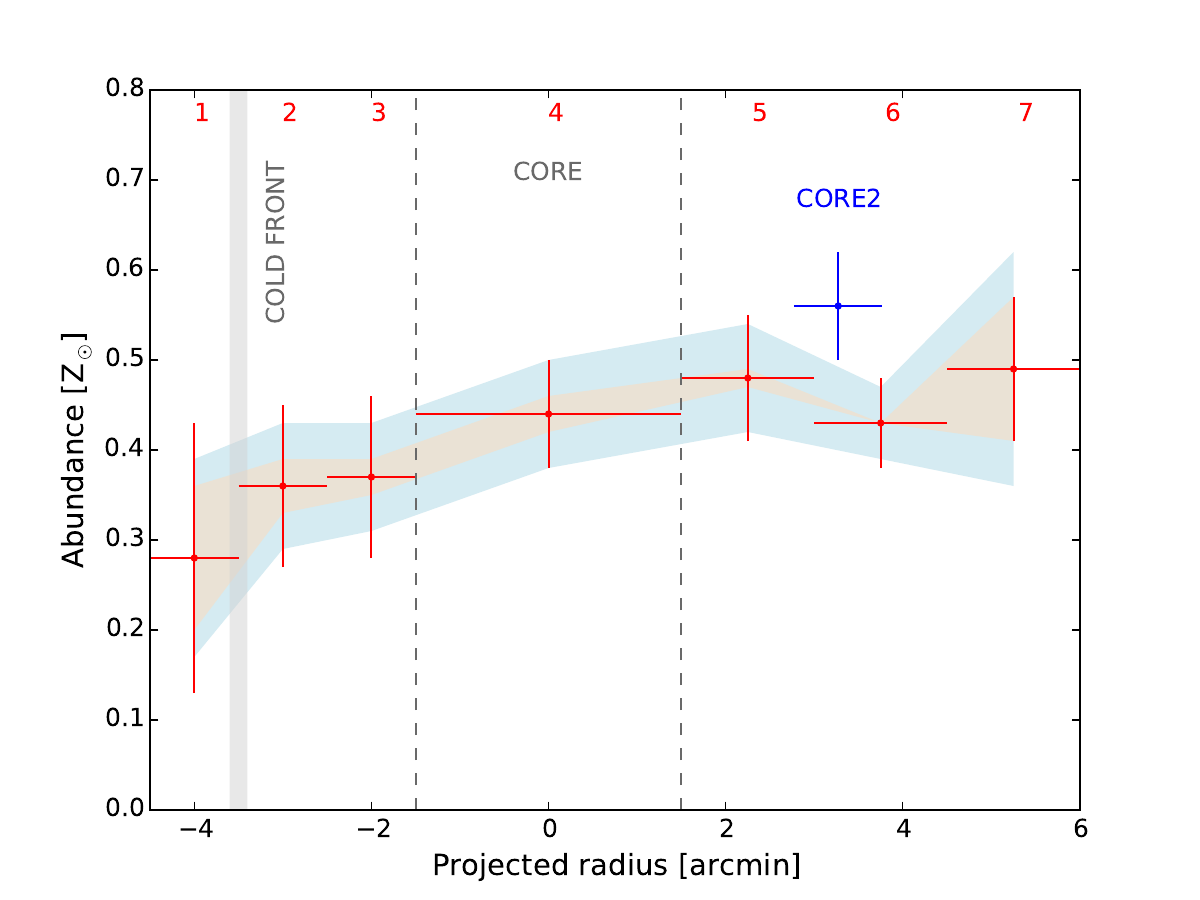}
 \end{minipage}
\caption{\textit{Left panel}: Temperature distribution along the merging axis of A2256. \textit{Right panel}: Abundance distribution along the merging axis of A2256. The shaded area represent the systematic uncertainties: green and orange shadows correspond to the background normalization variation, and the blue one to the temperature structure. The shaded grey area shows the cold front position. The dashed lines show the outer edges of the core spectral extraction region. The numbers and CORE2 blue point correspond to the regions and the red circle centered on CORE2, respectively shown in Fig. \ref{fig:fig11}.}
\label{fig:fig12}
\end{figure*}

The best-fit parameters are summarized in Table \ref{tab:tabb4} and represented in the distributions along the merging axis in Fig \ref{fig:fig8}. The results are in good agreement with previous studies \citep{Briel2004,Lovisari2009,Akamatsu2013a,Datta2014,Ichinohe2017}. The distributions show a clear temperature decrease towards the cold front at SE. It has an opposite trend with the Fe abundance, which reaches a peak just behind the cold front. The value presented here is lower than the maximum found by \cite{Briel2004} and \cite{Lovisari2009} of $\sim$0.7 $Z_\odot$, probably because our value includes larger azimuthal angles with lower abundance. However, if we select a smaller region in sector 2 (yellow dashed lines in Fig. \ref{fig:fig7}), the temperature drops to 3.77~$\pm$~0.09 keV and the abundance increases up to 0.64~$\pm$~0.06 $Z_\odot$. As pointed out by \cite{Lovisari2009}, this behaviour agrees with the simulations of \cite{Heinz2003}, which suggest a possible displacement of the central rich and cold gas of the subcluster center well to the front during the infall, due to the internal dynamics of the cold front. The sectors in the NW direction (6 to 9) show a temperature distribution close to $\sim$7 keV, i.e. the average temperature of the cluster \citep{Markevitch1999}, and consistent with the hot tail described by \cite{Briel2004}. The abundance in the central regions  (2 to 4) is almost uniform, around $\sim$0.4 $Z_\odot$, but higher than the main cluster value ($\sim$0.2--0.3 $Z_\odot$). Finally, the abundance is again flat in the outermost regions (7 to 8) close to  $\sim$0.3 $Z_\odot$.
      
         \subsection{A665}
     
       The two-body merging cluster Abell 665 (hereafter A665) was discovered as a non-relaxed cluster in an optical observation of \cite{Geller1982}. The authors found the first evidence of an elongated morphology and non-relaxed galaxy cluster. A665 is a hot cluster with an average temperature of $\sim$8.3 keV \citep{Hughes1992} and located at $z$~=~0.182 \citep{Gomez2000}. The N-body hydrodynamical simulations and dynamic studies of \cite{Gomez2000} suggest a subcluster merger in the NW--SE direction with two similar masses. The bright core of the infalling subcluster  is moving south, crossing a more diffuse halo, before being stripped by ram pressure. A665 hosts a giant radio halo \citep{Jones1996,Feretti2004,Vacca2010}, which follows the X-ray emission elongation. The X-ray observations of A665 \citep{Markevitch2001,Govoni2004,Dasadia2016a} have revealed only one X-ray peak ($8^{\rm{h}} 30^{\rm{m}} 59\fs8$, $65\degr 50\arcmin 31\farcs 3$), close to the BCG, which indicates that only the infalling cluster core has survived  the merger. The remnant core presents a nearby cold front in the SE, followed by a hot gas ($\sim$8--15 keV) thought to be a post-merger region of a shock \citep{Markevitch2001,Govoni2004}.

        \subsubsection{Results}
        
     We use spectra from MOS, the only detector available for these observations, to fit the cluster emission together with the modelled background (Table \ref{tab:taba5}). We analyse the circular sectors centred on the BCG ($8^{\rm{h}} 30^{\rm{m}} 57\fs.6$, $65\degr 50\arcmin 29\farcs 6$) as shown in Fig. \ref{fig:fig9}, using $N_{\rm{HI}}$ = 4.31~$\times$~10$^{20}$ cm$^{-2}$ \citep{Willingale2013}. The best-fit parameters are presented in the Table \ref{tab:tabb5} and plotted in Fig. \ref{fig:fig10}.
        
    The  temperature and Fe abundance distributions along the merging axis show the presence of the two cold fronts as described by previous X-ray studies \citep{Markevitch2001,Govoni2004,Dasadia2016a}. We see a slight temperature decrease across them towards the center. The abundance peak is found just within the south cold front and reflects the same case described by \cite{Heinz2003} for A3667.  The diffuse gas in the north presents a lower abundance and a more uniform distribution ($\sim$0.2--0.3 $Z_\odot$) than the gas in the core and just behind the cold front in the south. This could indicate that the gas belongs to the more diffuse cluster, which had been already disrupted by the core-crossing. The two cold fronts seem to delimit a central region with a lower temperature ($\sim$7 keV) than the surrounding gas. We found as well the cooler gas region ($kT$~=~7.4~$\pm$~0.2, $Z$~=~0.33$\pm$~0.06 $Z_\odot$) in the north observed by \cite{Markevitch2001} and shown in Fig. \ref{fig:fig9} as a magenta box. This cooler region seems to belong to the infalling cluster coming from the NW direction as described by \cite{Markevitch2001} and \cite{Govoni2004}.

           \subsection{A2256}

         Abell 2256 (hereafter A2256) is a nearby ($z$~=~0.058) rich galaxy cluster with bright X-ray emission. Optical analyses \citep{Fabricant1989,Berrington2002,Miller2003} revealed a complex dynamical state, which consists of at least a triple merging with a mass ratio of 3:1:0.3. \cite{Berrington2002} propose that the system includes a main cluster with an infalling subcluster NW at $\sim$0.2 Gyr from the core passage \citep{Roettiger1995} and another infalling "group"  northeast (NE), possibly associated to a previous merger. \cite{Briel1991} detected two separate X-ray peaks corresponding to the main cluster and the infalling subcluster, lately confirmed by \cite{Sun2002} and \cite{Bourdin2008}. They also found bimodality in the temperature structure along the cluster elongation axis, being the low temperature component ($\sim$4.5 keV) associated with the subcluster and the higher temperature component ($\sim$7--8 keV) with the main cluster. In addition, \cite{Tamura2011} found bulk motion of gas of the cooler temperature component.

        \subsubsection{Results}
     We use two observations (see Table \ref{tab:tab1}) for the analysis of A2256. We divide the ICM emission region in circular sectors along the cluster major elongation axis centered in the X-ray peak of the main cluster (CORE1: $17^{\rm{h}} 04^{\rm{m}} 13\fs 82$, $78\degr 38\arcmin 15\farcs 0$, \citealt{Bourdin2008}). We fit the cluster emission and the background spectra (see Table \ref{tab:taba6}) using $N_{\rm{HI}}$ = 4.24~$\times$~10$^{20}$ cm$^{-2}$ \citep{Willingale2013}. The best-fit parameters are presented in the Table \ref{tab:tabb5} and plotted in Fig. \ref{fig:fig12}.  
     
      The temperature profile shows the bimodal temperature structure already described by \cite{Sun2002} and \cite{Bourdin2008}, as well as the hot ($\sim$9 keV) temperature component south of the cluster center.   It shows a temperature discontinuity from the adjacent region at $\sim$7 keV, suggested to be caused by the presence of a cold front \citep{Bourdin2008}. However, and similarly to \cite{Bourdin2008}, we do not find evidence for the presence of the "shoulder" mentioned by \cite{Sun2002}  east of the CORE1. The Fe abundance distribution is in agreement with the values obtained by \cite{Sun2002}. We obtain a quasi uniform abundance distribution (0.3--0.5 $Z_\odot$)  with a slight decrease at the south. The second X-ray peak, (CORE2: $17^{\rm{h}} 03^{\rm{m}} 07\fs 43$, $78\degr 39\arcmin 07\farcs 8$, \citealt{Sun2002}), associated to the infalling subcluster, presents the highest abundance value of $\sim$0.6 $Z_\odot$. 
   
        \section{Discussion}

          \subsection{Averaged abundance distribution}
          \label{ave}

                       \begin{figure*}[ht!]
               \centering
   \begin{minipage}[h]{0.45\textwidth}
    \centering
    \includegraphics[width=1\textwidth]{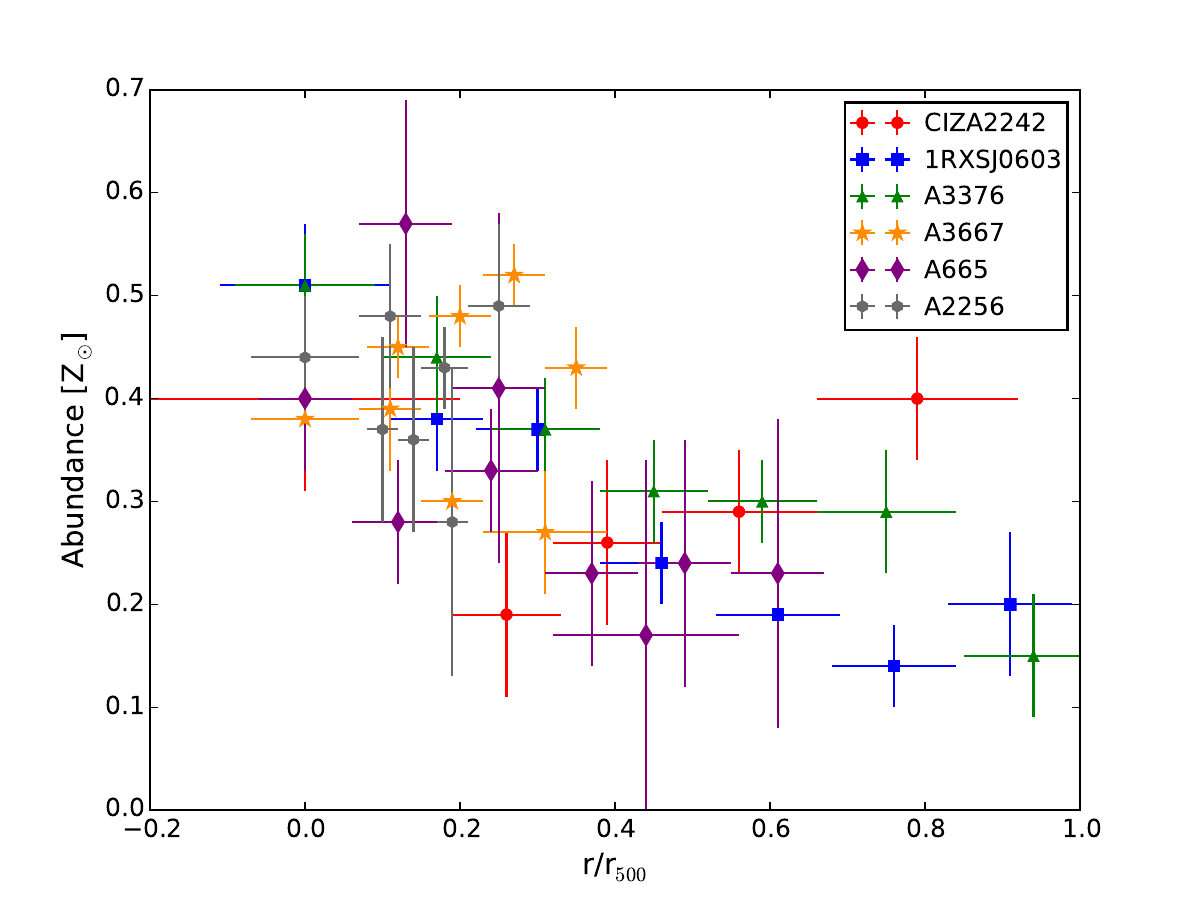}
  \end{minipage}
 \begin{minipage}[h]{0.45\textwidth}
    \centering
    \includegraphics[width=1\textwidth]{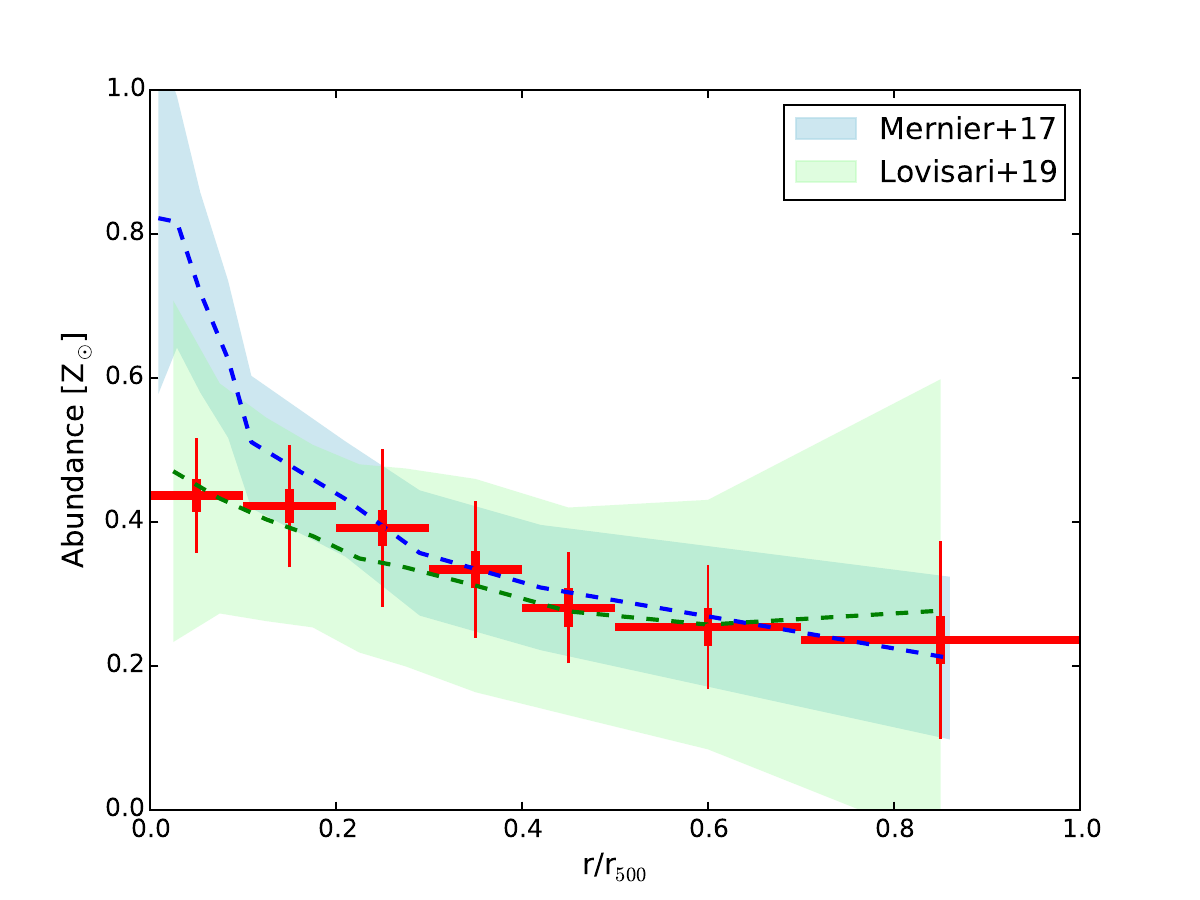}
 \end{minipage}
    \caption{\textit{Left panel}: Abundance distribution along the merging axis scaled by $r_{500}$.  \textit{Right panel}: Averaged abundance distribution scaled by $r_{500}$. Data points in red show the averaged value and the statistical error as a thick errorbar plus the scatter as a thinner errorbar for each bin of Table \ref{tab:tabd2}. The blue shaded area shows the average profile  for clusters (>1.7 keV) , including the statistical error and the scatter, derived by \cite{Mernier2017}. The green shaded area represents the mean values for disturbed systems, together with the scatter, obtained by \cite{Lovisari2019}. The blue and green dashed line follow the average and mean abundance value of \cite{Mernier2017} and \cite{Lovisari2019}, respectively.}
    \label{fig:figd1}

\end{figure*}

       The individual Fe abundance  distributions of the six merging galaxy clusters along their merging axis are compiled in Fig. \ref{fig:figd1} (\textit{left}). We remind the reader that such distributions are not azimuthally averaged. Instead, in this work we have chosen to trace the  central ICM elongation along the merging axis, essentially because the surface brightness allows better constraints on the Fe abundance and because no spectacular metal redistribution is expected to take place perpendicular to the merging axis. In the case of CIZA2242, 1RXSJ0603 and A3376 the interest region clearly extends the core (where the BCG resides) to the outwards direction. For A3667, A665 and A2256, sectors along two opposite directions from the core are used. In Fig. \ref{fig:figd1}, all the clustercentric radii have been rescaled to fractions of  $r_{500}$\footnote{$r_{\Delta}$ is formally defined as the radius within which the density is $\Delta$ times the critical density of the Universe at the redshift of the source.}, adopting $r_{500}$ and $r_{200}$ values found in the literature and assuming the conversion  $r_{500}~\simeq~$0.65$r_{200}$  of \cite{Reiprich2013}. For our cosmology and redshifts we use the approximation of \cite{Henry2009}:
\begin{equation}
\label{eqn:eq0}
r_{200} = 2.77h^{-1}_{70}(\mean{kT}/10~\rm{keV})^{1/2}/\textit{E(z)}~\rm{Mpc},
\end{equation}

where $E(z)$=$(\Omega_{\rm{M}}(1 + z)^3 + 1 - \Omega_{\rm{M}})^{1/2}$. The mean temperature values, $\mean{kT}$, are adopted from references as described in Table \ref{tab:tab2}.  Our averaged Fe abundance  distribution, stacked from all the above measurements, is shown in Fig.  \ref{fig:figd1} (\textit{right}), and the numerical values are reported in Table \ref{tab:tabd2}.

We use the stacking method of \cite{Mernier2017} to determine this  distribution. The averaged abundance  distribution, $Z_\text{ref}(k)$, as a function of the $k$th reference bin is defined as:

\begin{equation}
\label{eqn:eq0}
Z_\text{ref}(k) = \Bigg( \sum_{j=1}^N\sum_{i=1}^Mw_{i,j,k}\frac{Z(i)_j}{\sigma^2_{Z(i)_j}} \Bigg) / \Bigg( \sum_{j=1}^N\sum_{i=1}^Mw_{i,j,k}{\frac{1}{\sigma^2_{Z(i)_j}}}\Bigg),
\end{equation}

where $Z(i)_j$ is the individual abundance of the $j$th cluster at its $i$th region; $\sigma_{Z(i)_j}$ is its statistical error, $N$ is the number of clusters, $M$ is the number of regions analysed in each cluster and $w_{i,j,k}$ is the weighting factor. This factor represents the geometrical overlap of the $i$th region with the $k$th reference bin in cluster $j$. It varies from 0 to 1. The stacked statistical error is calculated as: 

\begin{equation}
\label{eqn:eq1}
\sigma_\text{stat}(k) = {\frac{1}{\sqrt{\sum_{j=1}^N\sum_{i=1}^Mw_{i,j,k}{\frac{1}{\sigma^2_{Z(i)_j}}}}}},
\end{equation}
We also obtained the scatter of the measurements for each $k$th reference bin as:

 \begin{equation}
\label{eqn:eq2}
\sigma_\text{scatter}(k) ={\frac{ \sqrt{\sum_{j=1}^N\sum_{i=1}^Mw_{i,j,k}\left({\frac{Z(i)_j-Z_\text{ref}(k)}{\sigma_{Z(i)_j}}}\right)^2}}{\sqrt{\sum_{j=1}^N\sum_{i=1}^Mw_{i,j,k}{\frac{1}{\sigma^2_{Z(i)_j}}}}}},
\end{equation}

It clearly appears from Fig. \ref{fig:figd1} that abundance distributions of these merging clusters are not flat. Instead, we note a decrease from the core up to $\sim$0.4$r_{500}$, followed by a flattening at larger radii.  Figure \ref{fig:figd1} (\textit{right}) compares our  averaged distribution with the relaxed,  azimuthally-averaged CC systems of \cite{Mernier2017} with $kT$ > 1.7 keV and the disturbed, NCC systems of \cite{Lovisari2019}. In both cases we have included the statistical errors and the scatter. Remarkably, our values are in   good agreement with the  (azimuthally-averaged) "NCC" profile of \cite{Lovisari2019}, which shows lower abundance values in the core than relaxed clusters and a shallower decrease towards the outer radii. The averaged Fe distribution values shown in Table \ref{tab:tabd2} are therefore consistent with \cite{Lovisari2019} results. We have additionally checked the averaged abundance distribution along the BCG displacement direction (assuming only the positive radii). In this case, the agreement with the NCC systems profile of \cite{Lovisari2019} improves. For more details see Appendix A.

     \begin{table}[ht!]
 \begin{center}
  \caption{ Averaged abundance profile.}
   \label{tab:tabd2}
\begin{threeparttable}
 \begin{tabular}{lcccc}
\hline
\hline
 \noalign{\smallskip}
 Radius& $Z$& $\sigma_\text{stat}$& $\sigma_\text{scatter}$\\
 (r/r$_{500}$)& ($Z_\odot$)& &\\
  \noalign{\smallskip}
 \hline
 \noalign{\smallskip}
0.00--0.10&0.437&0.023&0.057\\ 
  \noalign{\smallskip}
 0.10--0.20&0.422&0.024&0.061\\
 \noalign{\smallskip}
  0.20--0.30&0.392&0.025&0.085\\
 \noalign{\smallskip}
   0.30--0.40&0.334&0.026&0.069\\
 \noalign{\smallskip}
   0.40--0.50&0.281&0.027&0.049\\
 \noalign{\smallskip}
   0.50--0.70&0.254&0.026&0.060\\
 \noalign{\smallskip}
   0.70--1.00&0.236&0.033&0.104\\
 \noalign{\smallskip}
  \hline

 \end{tabular}
 \tiny

  \end{threeparttable}
  \end{center}
 \end{table}

\begin{figure*}[ht!]
    \centering
    \includegraphics[width=0.45\textwidth]{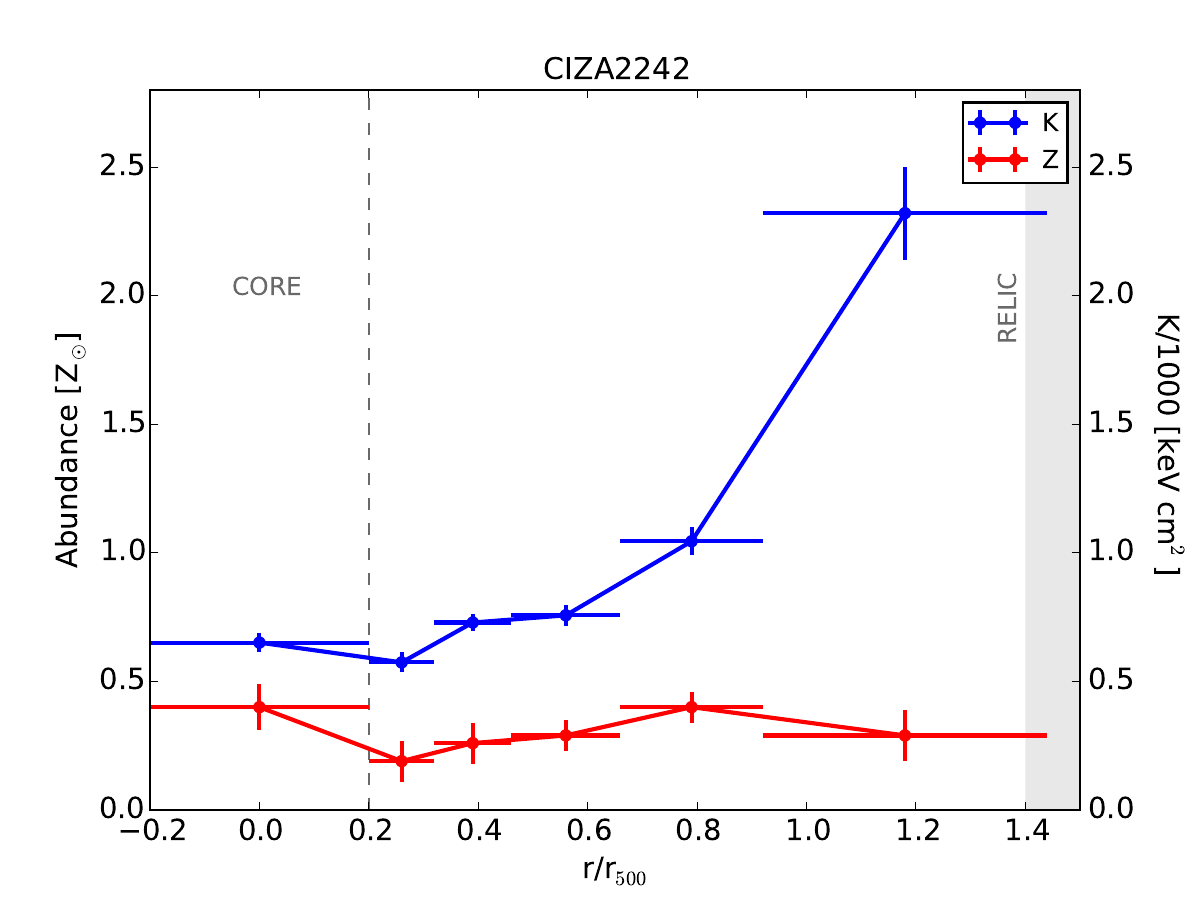} 
    \includegraphics[width=0.45\textwidth] {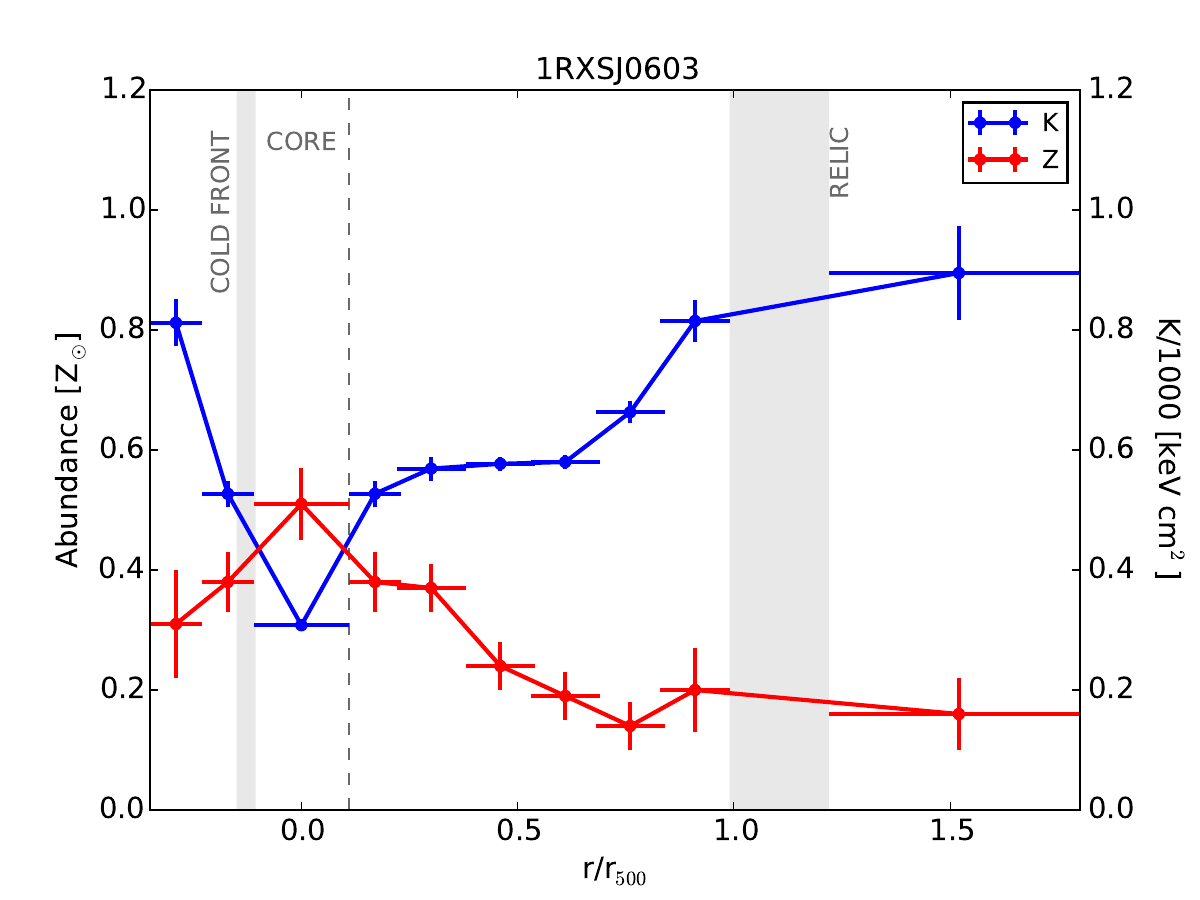} \\
    \includegraphics[width=0.45\textwidth]{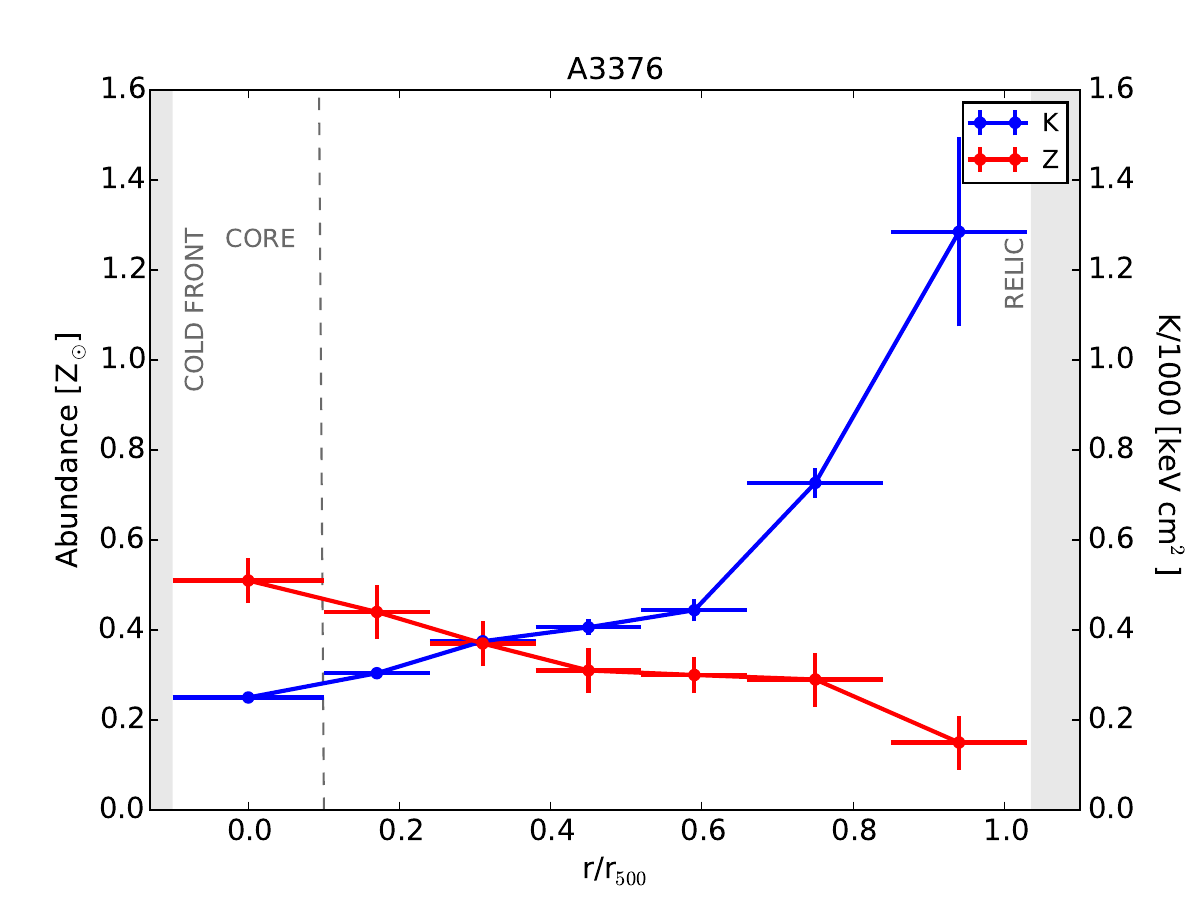} 
    \includegraphics[width=0.45\textwidth]{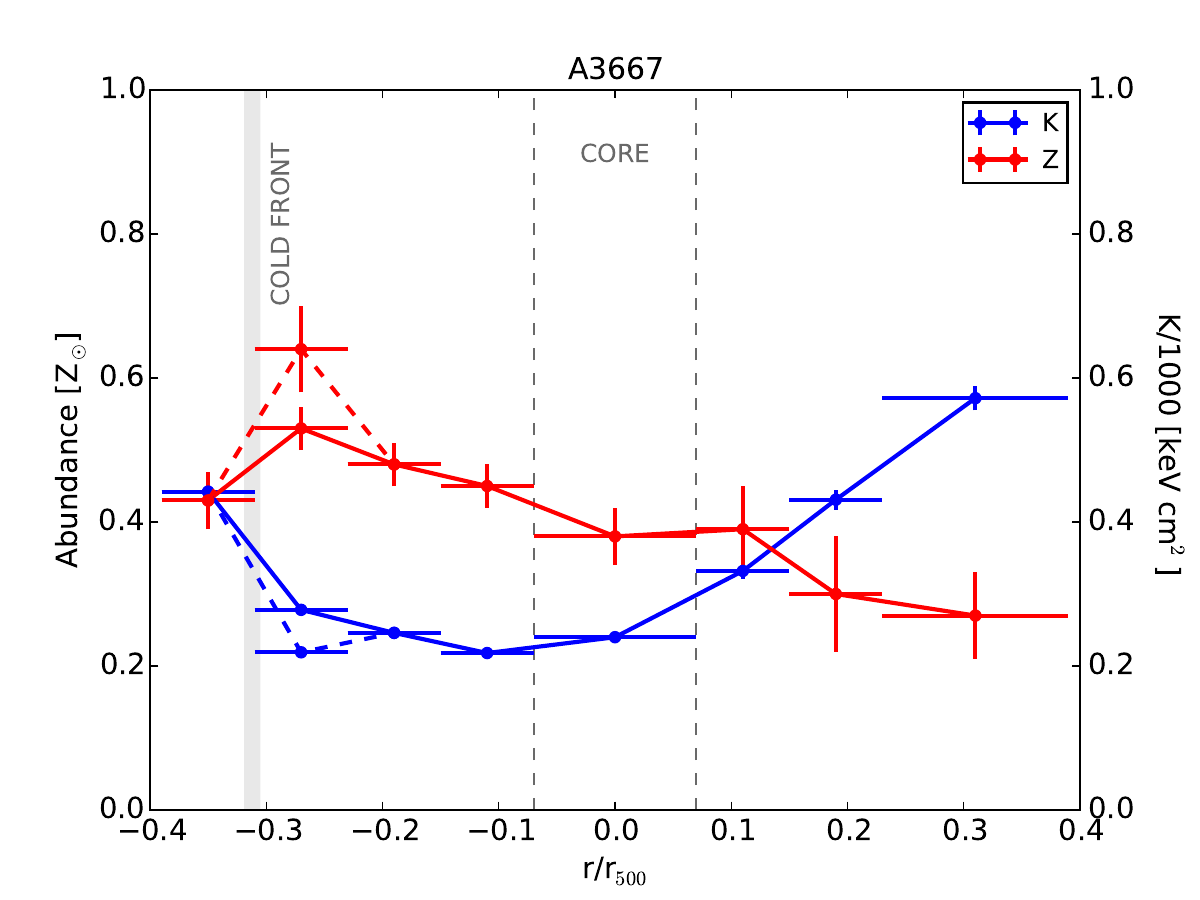} \\
    \includegraphics[width=0.45\textwidth]{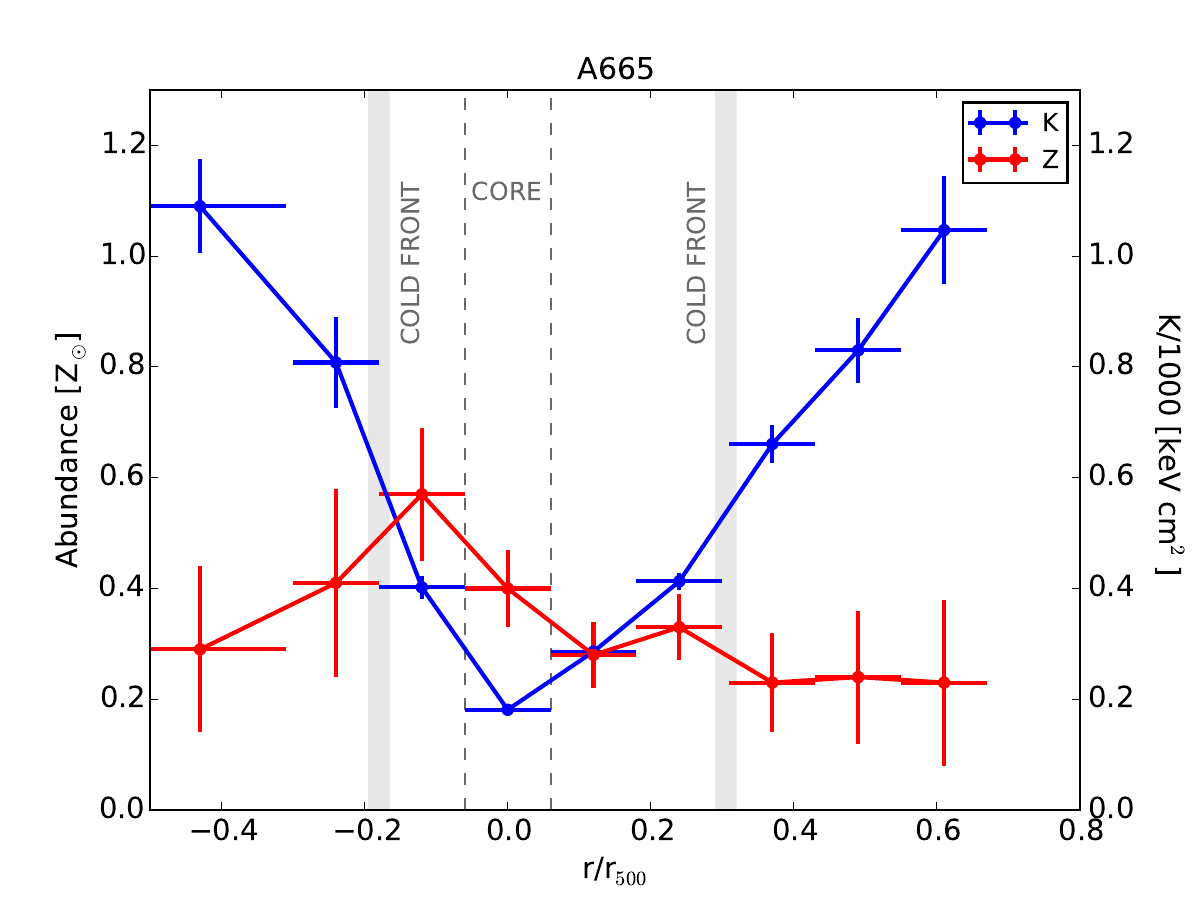} 
    \includegraphics[width=0.45\textwidth]{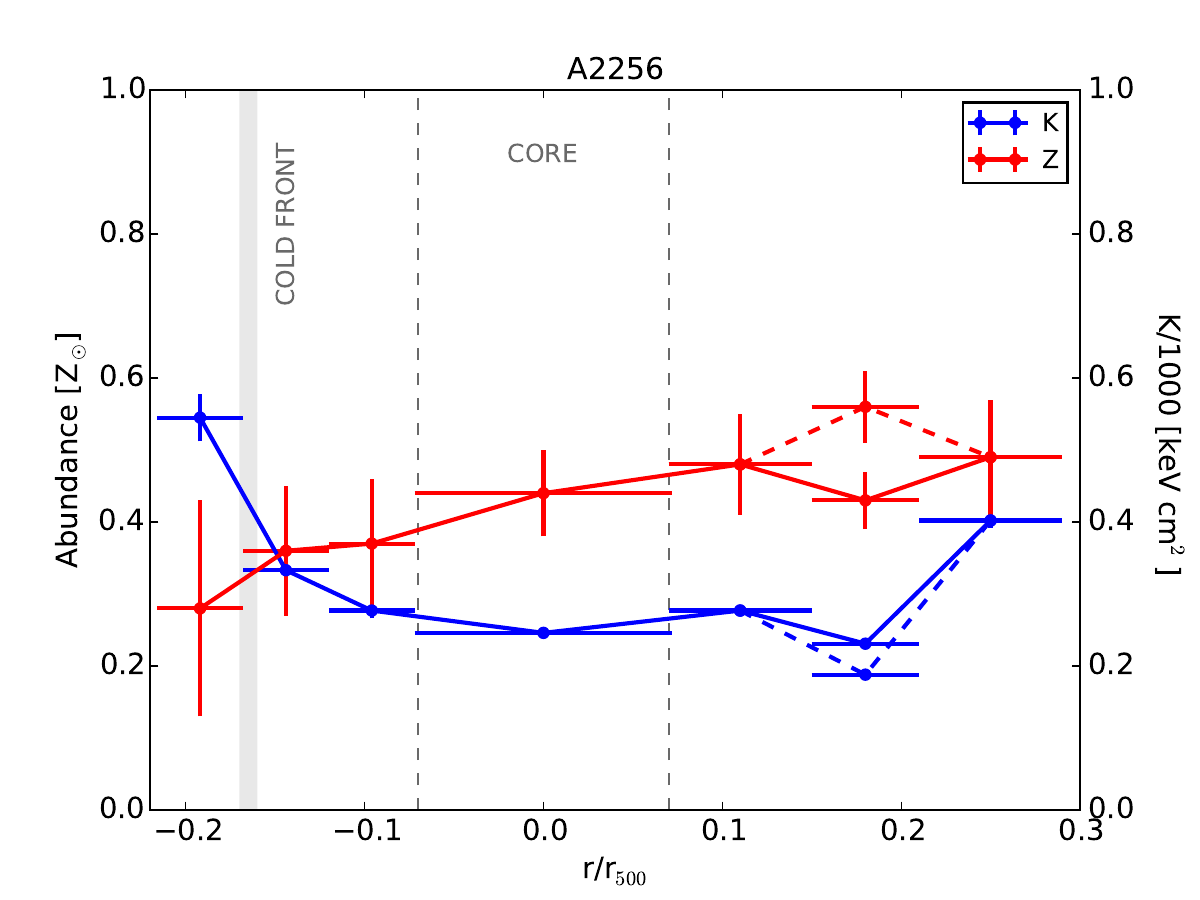} \\
\caption{Scaled pseudo-entropy ($K$/1000) and abundance distributions along the merging axis for CIZA2242, 1RXSJ0603, A3376, A3667, A665 and A2256. Dashed lines show the values of the yellow region after the A3667 cold front (see Fig. \ref{fig:fig7}) and the CORE2 values of A2256 (see Fig. \ref{fig:fig11}).}
\label{fig:figd21}
\end{figure*}
 
     The lower abundance value in the core for disturbed systems is suggested by \cite{Lovisari2019} to be caused by major mergers, where in some cases the core could be disrupted. As a result, the abundance peak in the center of the (CC) progenitors is spread out and distributed across larger radii. At first glance, the similarity of the Fe distribution observed between NCC clusters and these merging clusters tends to support this scenario. However, the steeper Fe peaks observed in the core of 1RXSJ0603 and in the vicinity of the core of A665 (perhaps offset from the BCG because of important sloshing motions) suggest that remnants of previous CC clusters can partly persist to (or, at least, not be entirely disrupted by) even the most major mergers in some cases (see Sect. 5.2). While this moderate central abundance increase could be associated with either (relatively old) core passage mergers, or on the contrary, early stage mergers, we note that our sample contains both cases. In fact, most of them are massive clusters, which have undergone a major merger with a core passage more than 0.6~Gyr ago. However, A2256 is a still on going merger, before core passage, although it is thought that the main cluster center is already affected by an older merger. Future larger samples, containing an equal proportion of early and late stage mergers, will help to relate the (re)-distribution of metals with merger history.

            \subsection{Abundance vs. pseudo-entropy relation} 
            \label{entropy}
           In order to study the thermal history of the ICM, we have calculated the   pseudo-entropy, $K \equiv kT \times n^{-2/3}_{e}$, as a function of the region distance for each merging cluster. These  pseudo-entropy distributions along the merging axis are shown together with the Fe abundance profiles in Fig. \ref{fig:figd21}. The temperature of each region, $kT$, is directly obtained from the spectral analysis, and the electron density, $n_e$, can be inferred from the emission measure, $Norm$, as $Norm$~$\propto$~1.2$n^{2}_{e} V$, where $V$ is the emitting volume projected onto the line-of-sight (LOS). To estimate $n_e$, we assume a spherical LOS in which only the sphere between the maximum and minimum radius contributes to the emission \citep{Henry2004,Madhavi2005}. We assume the emission volume $V$ to be $V = 2SL/3$ for the circular and elliptical sectors, and $V = \pi SL/2$ for rectangular ones. In each case, $L = 2\sqrt{(R_{max}^2-R_{min}^2)}$ and $S$ is the area of the sectors in the plane of the sky.

            \begin{figure*}[ht!]
               \centering
   \begin{minipage}[h]{0.45\textwidth}
    \centering
    \includegraphics[width=1\textwidth]{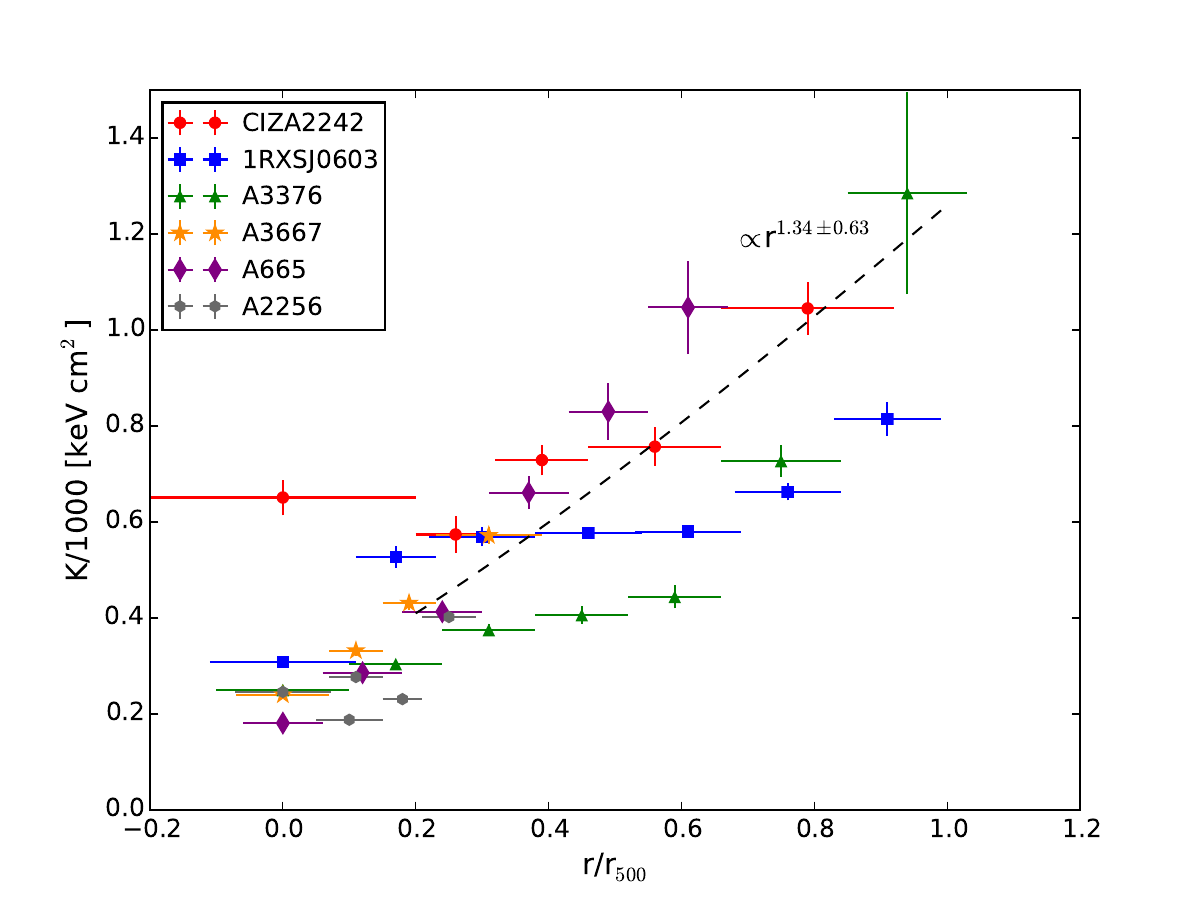}
  \end{minipage}
 \begin{minipage}[h]{0.45\textwidth}
    \centering
    \includegraphics[width=1\textwidth]{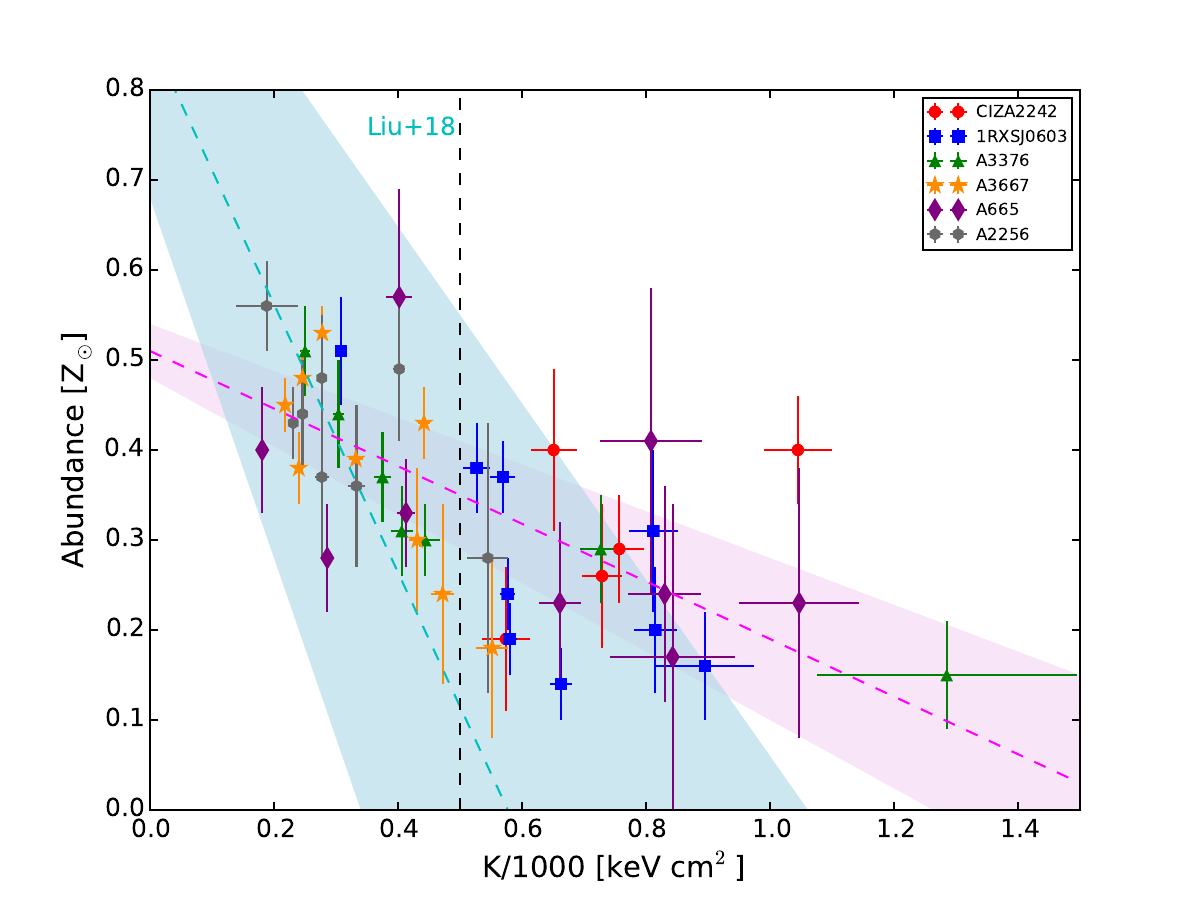}
 \end{minipage}
    \caption{\textit{Left panel}: Pseudo-entropy  distribution scaled by $r_{500}$.  \textit{Right panel}: The distribution of Fe abundance versus the scaled pseudo-entropy of all the measured regions of the clusters: CIZA2242, 1RXSJ0603, A3376, A3667, A665 and A2256. The magenta dashed line is the best-fit linear model and the shaded area shows the 1$\sigma$ uncertainties of the best-fit parameter. The cyan line and shadow are the results of \cite{Liu2018} for relaxed galaxy clusters.}
    \label{fig:figd22}

\end{figure*}
      
            Previous studies \citep{DeGrandi2004,Leccardi2010,Ghizzardi2014,Liu2018} have demonstrated, especially for relaxed (CC) clusters, the existence of a negative correlation between the Fe abundance and the gas entropy.  Low-entropy cores, where the cooling of the ICM might take place, are in fact associated with central abundance peaks. However, disturbed (NCC) clusters are not well characterised by this behaviour as mentioned by \cite{Leccardi2010} and \cite{Rossetti2010}.

       Pseudo-entropy quantifies the internal energy variation due to thermal processes such as, e.g. cooling, turbulent dissipation or shock heating. Regions under the influence of the latter, in particular, can be easily identified because they exhibit a positive gradient proportional to $r^{1.1}$ \citep{TOzzi2001}. In particular for our merging cluster selection, we observe a similar trend in the post-shock regions, see Fig. \ref{fig:figd21} and Fig. \ref{fig:figd22} (\textit{left}), although some systems (e.g. 1RXSJ0603, A3376) significantly deviate from this relation. If we fit simultaneously all our pseudo-entropy  distributions with a power-law ($\sim r^{\alpha}$), we obtain a slope of $\alpha$~=~1.34~$\pm$~0.63 (Fig. \ref{fig:figd22} \textit{left}). In the case of 1RXSJ0603, we see a flat profile between $\sim$0.2--0.4 $r_{500}$, coincident with the COREN location, possibly indicating a sign of the subcluster core remnant after the merger event. A similar flattening is observed in A3376, although its origin may be rather related to the central low-entropy gas stripped during the merger, thereby forming the cometary shape ICM.

       Most of NCC clusters are thought to have undergone a CC phase, before the (cool) core being disrupted completely or partially by major mergers \citep{Leccardi2010,Rossetti2010,ZuHone2011}. The on-going interactions heat and increase the entropy in the core. Therefore, NCC tend to have high-entropy cores and smaller entropy gradients. Some of these clusters present regions with lower entropy that have associated an abundance excess, in most of the cases associated with the BCG or a giant elliptical galaxy, and known as "CC-remnants" \citep{Rossetti2010}. It is suggested that these regions are the reminiscent of CC after the merging activity. Another consequence of merger events is the decrease of the abundance excess in the core, which can be erased by the shock-heating and gas mixing processes. 

  Figure \ref{fig:figd21} shows simultaneously the pseudo-entropy and Fe abundance  distributions scaled by $r_{500}$ for our six merging galaxy clusters. In the case of CIZA2242 and A3376, the entropy is slightly flat in the core and the profile increases smoothly (though with a larger gradient) out to the recently shocked region. Interestingly, 1RXSJ0603 exhibits a relative low-entropy core associated with the abundance peak. One natural interpretation is that   this region is the remainder of cool core after the merging activity, which agrees with \cite{ZuHone2011} simulations taking into account the mass ratio of 3:1 and it is found in 12 clusters of \cite{Rossetti2010}. An alternative scenario could be that 1RXSJ0603, as a well evolved merger ($\sim$2 Gyr since the shock producing the radio relic; e.g. \cite{Bruggen2012a}, \cite{Stroe2015a}), has started to recover its potential well and to re-build its entropy (and metal) stratification after the merger. For A3667 and A2256, we identify possible CC-remnants characterised by somewhat lower entropy and higher metallicity values \citep{Rossetti2010}. These remnants clearly appear after we include the measurements of the truncated yellow region after the A3667 cold front (see Fig. \ref{fig:fig7}) and the CORE2 values of A2256 (see Fig. \ref{fig:fig11}), in dashed lines in Fig. \ref{fig:figd21}. A2256 also contains a second minimum corresponding to the CORE1. As explained in \citet{Leccardi2010}, two low-entropy clumps can be found if we are in the presence of an early stage merger, as in this case, or an off-axis merger event. A665 also shows a relative low-entropy drop in the core, but in this case the abundance peak is displaced towards the cold front, as explained in Section 5.1, possibly due to the effect of the internal dynamics of the cold front during the infall of the subcluster. The above examples, and in particular the case of 1RXSJ0603, strongly suggest that in some cases,  signatures  of cool cores can remain even after the most major ICM merger events.

\begin{table}[ht!]
 \begin{center}
  \caption{Best-fit parameters of abundance vs pseudo-entropy distributions.}
   \label{tab:tabd3}
\begin{threeparttable}
\begin{tabular}{ccc}
\hline
\hline
 \noalign{\smallskip}
 K (keV cm$^2$) &$\alpha$ (Z$_{\odot}$ keV$^{-1}$ cm$^{-2}$) & Z$_0$ ($Z_\odot$) \\
 
  \noalign{\smallskip}
 \hline
 \noalign{\smallskip}
200--2300&0.32$~\pm~$0.06&0.51~$\pm$~0.03\\ 
  \noalign{\smallskip}
200--500&0.61$~\pm~$0.17 &0.61~$\pm$~0.05\\
  \noalign{\smallskip}
500--2300&0.07$~\pm~$0.07 &0.32~$\pm$~0.05 \\
 \noalign{\smallskip}
10--1000\tnote{a}&1.49$~\pm~$0.51 &0.86~$\pm$~0.18 \\
 \noalign{\smallskip}
  \hline
 \end{tabular}
  \tiny
  \begin{tablenotes}
        \item[a] \cite{Liu2018} distribution best-fit parameters.
  \end{tablenotes}
    \end{threeparttable} 
  \end{center}
 \end{table}

      Furthermore, we try to determine a correlation between Fe abundance and pseudo-entropy for our merging clusters. Figure \ref{fig:figd22} (\textit{right}) shows the distribution of $Z$-$K$. Generally speaking, our measured entropies are clearly higher than in the distribution of \cite{Liu2018}, with our lowest value being $\sim$200 keV cm$^2$ and the maximum abundance value being $\sim$0.5 $Z_{\odot}$. In fact, these values are characteristic of merging clusters \citep{Leccardi2010,ZuHone2011}. We fit the distribution with a linear function $Z \sim Z_0 - \alpha \, K$/1000. Our best-fit parameters are $Z_0 = 0.51 \pm 0.03$ and $\alpha = 0.32 \pm 0.06$. This correlation shows a flatter slope than obtained by \cite{Liu2018} for relaxed (CC) galaxy clusters ($Z_0 = 0.86 \pm 0.18$ and $\alpha = 1.49 \pm 0.51$). Our results suggest that the abundance and pseudo-entropy relation are not always tightly related. Instead, it is likely that, when a merger occurs, the entropy gets boosted in post-shock regions more rapidly than metals diffuse out. Interestingly, when we consider the values of pseudo-entropy only below $\sim$500 keV cm$^2$ -- which were presumably less affected by shock-heating, or which start to relax again (see vertical black dashed line in Fig. \ref{fig:figd22}), the slope of the relation considerably increases (Table \ref{tab:tabd3}).   This might suggest that in the lowest entropy regions ($\sim$200--500 keV cm$^2$) of  these merging clusters (NCC) -- usually associated with the core remnant (see Fig. \ref{fig:figd21}), the metal-rich gas (originally still coming from the BCG) gets mixed and redistributed at a moderate rate compared to the entropy changes during the merger event \citep{ZuHone2011}. Alternatively, chemical histories may differ between relaxed and merging systems.  Although beyond the scope of this paper, in some cases we cannot rule out the possibility of an extra Fe enrichment, produced by a boost of star formation or SNIa explosions, triggered by major mergers (e.g. \cite{Stroe2015a,Sob15}).

     Finally, we do not find evidence of a correlation in the higher (>500 keV cm$^2$) entropy range -- i.e. beyond $\sim$0.4--0.6 $r_{500}$. In these outer regions, the Fe abundance remains rather uniform around 0.2--0.3 Z$_\odot$, with no notable trend with cluster mass, radius or azimuth (Fig. \ref{fig:figd1}). Combined with the previous observations of a uniform $\sim$0.3 Z$_\odot$ Fe abundance in the outskirts of (mostly) CC clusters \citep{Fujita2008,Werner2013,T2016,Urban2017,Ezer2017}, we conclude that Fe distribution in cluster outskirts  of these merging clusters depends very weakly (or not at all) on the dynamical state of the cluster. This has important consequences regarding the chemical history of large-scale structures. In fact, although these previous measurements in relaxed systems strongly suggested that the most of the ICM enrichment took place in early stages of cluster formation \citep[i.e. at $z$ > 2--3; for recent reviews, see][]{Mernier2018b,Biffi2018b}, the alternative scenario of late enrichment coupled with a very efficient mixing of metals in the ICM, could not be totally excluded. Among the entire cluster population, merging clusters are systems in which metals had the least time to mix and diffuse out since their last merger event. Consequently, if the bulk of metals were mixed efficiently in the entire cluster volume (after being injected into the ICM at late stages), one would expect to find important variations of the Fe abundance at large radii in such dynamically disturbed systems, with values considerably higher than 0.2--0.3 Z$_\odot$ in some specific outer regions. Instead, the remarkable similarity of the Fe ndistribution in the outskirts of CC and these six merging clusters, and the above indication that in inner regions metal mixing lags somewhat behind entropy changes, constitute valuable additional evidence in favour of the "early enrichment" scenario.
        
        \section{Summary}
         In this work, we have analysed for the first time the Fe  distribution along the merging axis of six merging galaxy clusters ($0.05 < z < 0.23$) observed with \textit{XMM-Newton}/EPIC. We have obtained and specifically studied the  temperature, abundance and pseudo-entropy  distributions across the ICM of these disturbed systems. Specifically, we have provided a first estimate of the averaged Fe distribution out to $r_{500}$ -- as well as the relation between Fe abundance and pseudo-entropy -- in six merging clusters. The main results can be summarised as follows.
        
           \begin{itemize}
   	\item[$\bullet$] The Fe distribution is far from being universal, as it strongly varies from one merging cluster to another. While in some cases (e.g. CIZA2242, A2256) no Fe peak nor pseudo-entropy drop is detected in central regions, some of our observations (e.g. 1RXSJ0603, A665) strongly suggest that  remnants of previous (metal-rich) cool-cores can persist after major mergers.

	\item[$\bullet$]  Our averaged Fe distribution in  these six merging clusters is in excellent agreement with that in (non-merging) "non-cool-core" clusters presented by \cite{Lovisari2019}. The average profile clearly shows a moderate abundance central excess, thus present not only in CC systems but even in many spectacular mergers. Compared to the former, however, the central ($\lesssim 0.3 r_{500}$) Fe profile in these merging clusters is significantly less steep. This can be interpreted as the metal-rich gas of the CC progenitors being mixed and spread across moderate radii by merging processes. 		

	\item[$\bullet$] The minimum pseudo-entropy value obtained for these six merging clusters is $\sim$200 keV cm$^2$. This value is characteristic of typical unrelaxed systems, whose internal activity is dominated by turbulence, large-scale motions and/or AGN feedback. Whereas some clusters exhibit a  pseudo-entropy distribution that tends to follow the behaviour $\propto$~$r^{1.1}$ described by \cite{TOzzi2001}, some systems (e.g. 1RXSJ0603, A3376) deviate from this relation due to stripped cold gas from disrupted cores.

\item[$\bullet$] We find  a mild negative correlation between abundance and pseudo-entropy for the inner core region ($K$~=~200-500 keV cm$^2$), shallower than the anti-correlation obtained by \cite{Liu2018} for relaxed clusters. This might suggest that mergers affect the gas entropy more rapidly than they re-distribute metals (via gas mixing).  

	\item[$\bullet$]   In the outermost, high-entropy regions ($K$~=~500-2300 keV cm$^2$) studied here, we have no evidence of a correlation between metallicity and pseudo-entropy and the abundance is consistent with the uniform abundance value of 0.2--0.3 $Z_{\odot}$. This demonstrates that the lack of dependence of Fe abundance at such high entropies (and large radii) applies not only to the mass, radius and azimuthal characteristics of clusters, but also to the dynamical stage. In other words, our results provide important additional evidence towards the scenario of an early enrichment of the inter galactic medium ($z$~>2--3). 
	
	   \end{itemize}


\begin{acknowledgements}
 The authors thank Dr. R. van Weeren for the LOFAR radio data, Dr. R. Kale,  Dr. T. Shimwell and Dr. V. Vacca for providing the VLA radio data and Dr. A. Botteon for the SUMMSS data. S. Kara and. E. N. Ercan would like to thank TUBITAK for financial support through a 1002 project under the code 118F023. SRON is supported financially by NWO, the Netherlands Organization for Scientific Research. This work is based on observations obtained with XMM-Newton, an ESA science mission with instruments and contributions directly funded by ESA member states and the USA (NASA).
\end{acknowledgements}


--------------------------------------------------

  \begin{appendix}
  
  \section{Averaged abundance distribution along BCG displacement direction}
\begin{figure}[ht!]
               
   \begin{minipage}[h]{1\textwidth}
   
    \includegraphics[width=0.45\textwidth]{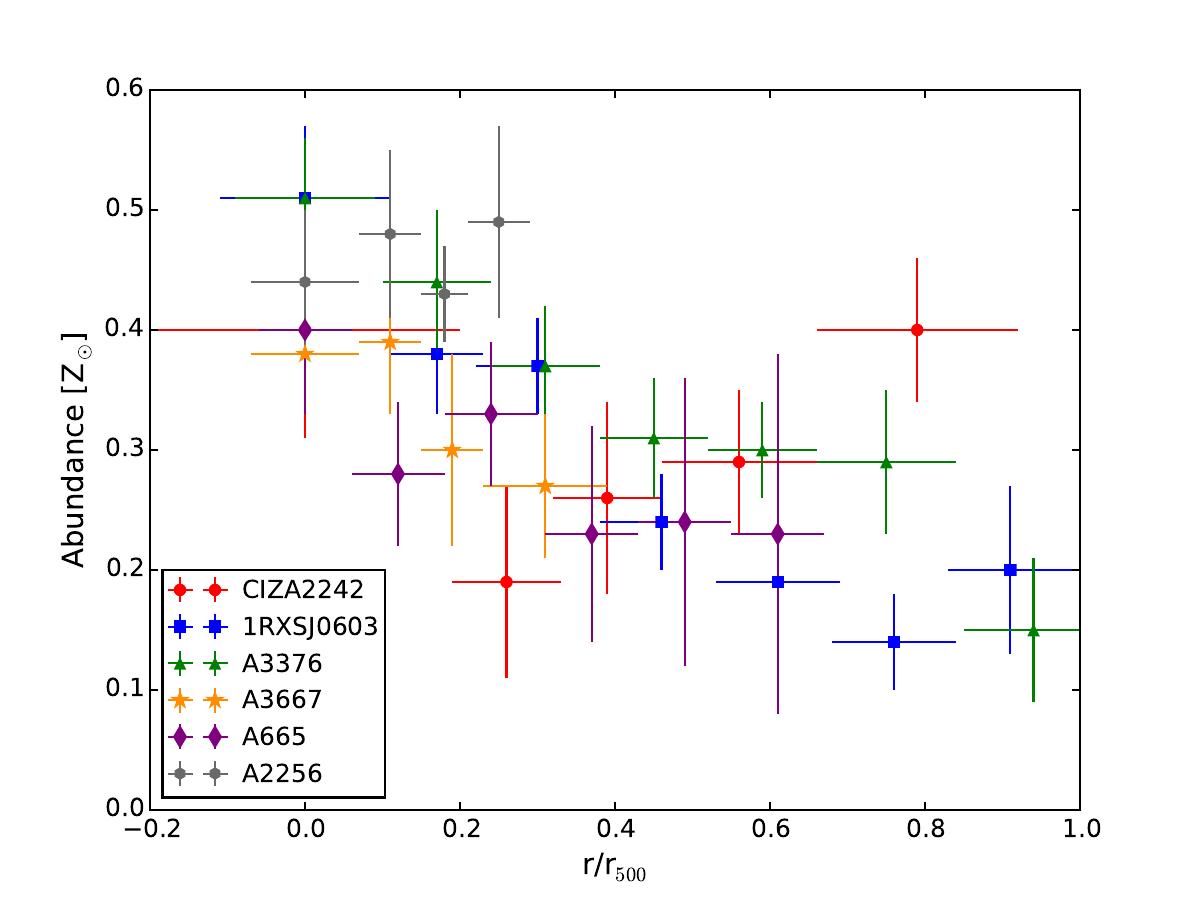}
  \end{minipage}
 \begin{minipage}[h]{1\textwidth}
    \includegraphics[width=0.45\textwidth]{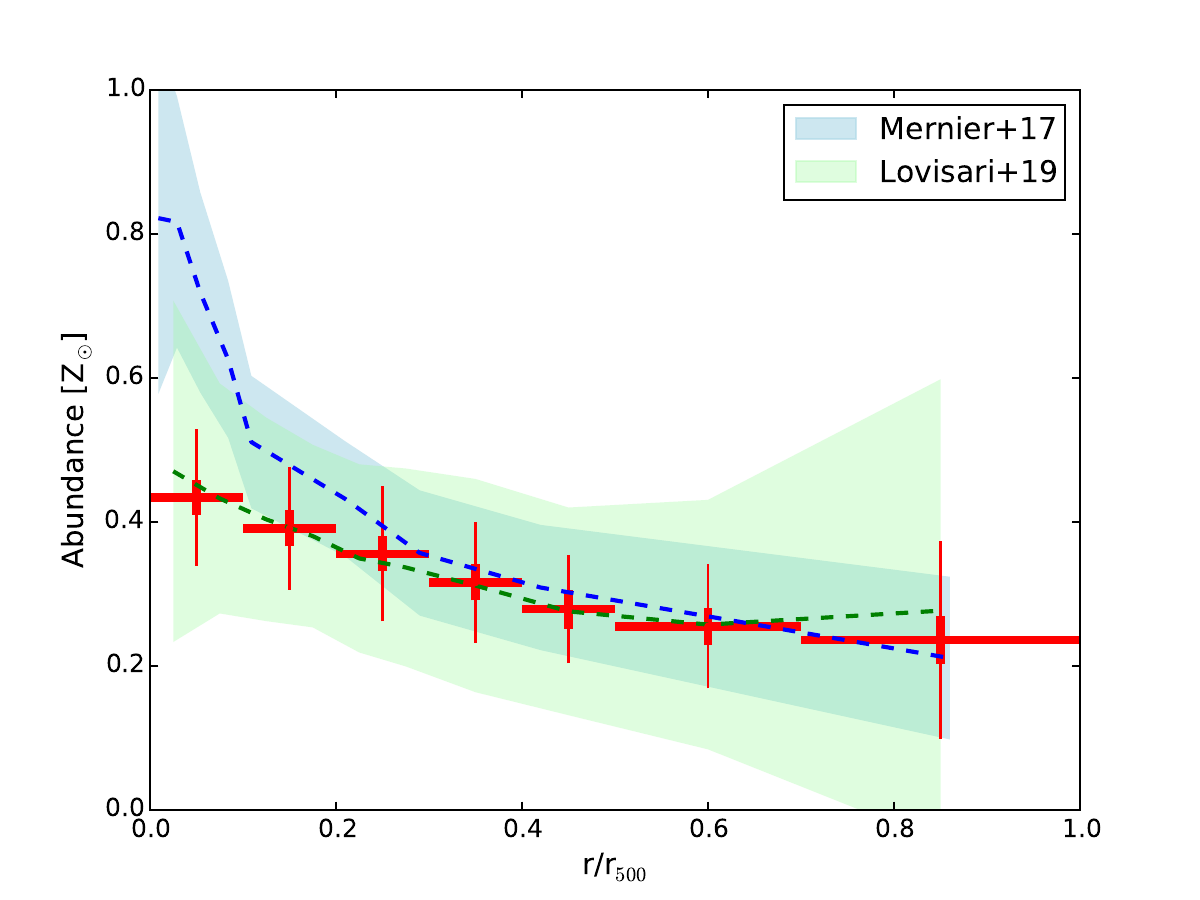}
 \end{minipage}
    \caption{\textit{Top panel}: Abundance distribution along the BCG displacement direction scaled by $r_{500}$.  \textit{Bottom panel}: Averaged abundance distribution scaled by $r_{500}$. Data points in red show the averaged value and the statistical error as a thick errorbar plus the scatter as a thinner errorbar for each bin of Table \ref{tab:tabd2}. The blue shaded area shows the average profile  for clusters (>1.7 keV) , including the statistical error and the scatter, derived by \cite{Mernier2017}. The green shaded area represents the mean values for disturbed systems, together with the scatter, obtained by \cite{Lovisari2019}. The blue and green dashed line follow the average and mean abundance value of \cite{Mernier2017} and \cite{Lovisari2019}, respectively.}
    \label{fig:figac1}

\end{figure}

 In addition to stacking our measurements over the entire merger axis as explained in Section \ref{ave}, we can alternatively choose to trace the ICM only from the BCG region (i.e. typically including the cluster X-ray peak), down to the freshly collided gas, i.e. along its cometary structure (defined in Figs. \ref{fig:fig2}, \ref{fig:fig4}, \ref{fig:fig6}, \ref{fig:fig8}, \ref{fig:fig10} and \ref{fig:fig12} as positive values on the x-axis). These regions better trace the disturbed and mixed plasma while avoiding special features such as cold fronts ahead of the merger (e.g. A3667, Fig. \ref{fig:fig8}); therefore the effects of the merging activity in the ICM are expected to be more significant. For CIZA2242, 1RXSJ0603 and A3376 this covers the complete elongation or cometary tail present in the ICM. For A3367 and A665 this includes the sectors 5--8 and 4--9, respectively. In the case of A2256, we consider the sectors 4--7, where the ICM of the main subclusters is predominantly affected by the presence of the infalling subclsuter. The individual Fe abundance distribution of the six merging galaxy clusters included in this analysis are plotted in Fig. \ref{fig:figac1}  (\textit{top}), the averaged values obtained with the stacking method described in Section \ref{ave} are shown in Fig. \ref{fig:figac1}  (\textit{bottom}) and Table  \ref{tab:tabac2}.
  
   \begin{table}[ht!]
 \begin{center}
  \caption{Averaged abundance distribution.}
   \label{tab:tabac2}
\begin{threeparttable}
 \begin{tabular}{lcccc}
\hline
\hline
 \noalign{\smallskip}
 Radius& $Z$& $\sigma_\text{stat}$& $\sigma_\text{scatter}$\\
 (r/r$_{500}$)& ($Z_\odot$)& &\\
  \noalign{\smallskip}
 \hline
 \noalign{\smallskip}
0.00--0.10&0.434&0.024&0.071\\ 
  \noalign{\smallskip}
 0.10--0.20&0.391&0.025&0.060\\
 \noalign{\smallskip}
  0.20--0.30&0.356&0.024&0.070\\
 \noalign{\smallskip}
   0.30--0.40&0.316&0.025&0.059\\
 \noalign{\smallskip}
   0.40--0.50&0.279&0.027&0.048\\
 \noalign{\smallskip}
   0.50--0.70&0.255&0.026&0.060\\
 \noalign{\smallskip}
   0.70--1.00&0.236&0.033&0.104\\
 \noalign{\smallskip}
  \hline

 \end{tabular}
  \tiny

  \end{threeparttable}
  \end{center}
 \end{table}
  
  Interestingly, the Fe averaged distribution is now in excellent agreement with the Fe average profile of NCC systems by \cite{Lovisari2019}. The major differences with respect to the distribution shown in Section \ref{ave} appear in the  0.1--0.2 $r_{500}$ (8$\%$ lower) and 0.2--0.3 $r_{500}$ (10$\%$ lower) radii. These lower values are obtained because the displaced Fe peak outside cluster center (see Section \ref{ave} and \ref{entropy} for more details) of A3667 and A665 are not included.
  
\clearpage
       \section{EPIC background modelling for the cluster sample}
       The following tables summarise the EPIC background model parameters used of each one of the galaxy clusters using the method described in Section 3.2. The norms are given in standard SPEX definition, which are intrinsically dependent of the distance to the source.
       
\begin{table}[ht!]
 \begin{center}
  \caption{Background components derived from the CIZA2242 observation.}
   \label{tab:taba1}
\begin{threeparttable}
 \begin{tabular}{lccc}
\hline
\hline
 \noalign{\smallskip}
 &\textit{Norm}\tnote{a}& \textit{kT} (keV)& $\Gamma$\\
  \noalign{\smallskip}
 \hline
 \noalign{\smallskip}
MWH&57.55$~\pm~0.28$&0.27 (fixed) &--\\ 
  \noalign{\smallskip}
LHB&258.33$~\pm~0.68$ &0.08 (fixed) & --\\
  \noalign{\smallskip}
CXB&6.75 (fixed)&--& 1.41 (fixed)\\
\noalign{\smallskip}
HF&  6.24$~\pm~2.56$ &0.7 (fixed)& --\\
 \noalign{\smallskip}
  \hline
   \noalign{\smallskip}
   SP MOS1&13.25$~\pm~1.98$&--&1.40$~\pm~0.19$\\ 
    \noalign{\smallskip}
   SP MOS2&18.10$~\pm~1.77$&--&1.40$~\pm~0.19$\\ 
    \noalign{\smallskip}
    SP PN&58.86$~\pm~11.51$&--&1.20$~\pm~0.15$\\ 
    \noalign{\smallskip}
  \hline
 \end{tabular}
 \tiny
 \begin{tablenotes}
    \item[a] For LHB and MWH norm in units of 10$^{69}$ m$^{-3}$ arcmin$^{-2}$ \\
             For CXB norm in units of 10$^{49}$ph s$^{-1}$keV$^{-1}$ arcmin$^{-2}$\\
             For SP norm in units of 10$^{44}$ph s$^{-1}$keV$^{-1}$ arcmin$^{-2}$
\end{tablenotes}
  \end{threeparttable}\label{tab:tabb4}
  \end{center}
 \end{table}

\begin{table}[ht!]
 \begin{center}
  \caption{Background components derived from the 1RXSJ0603 observation.}
   \label{tab:taba2}
\begin{threeparttable}
 \begin{tabular}{lccc}
\hline
\hline
 \noalign{\smallskip}
 &\textit{Norm}\tnote{a}& \textit{kT} (keV)& $\Gamma$\\
  \noalign{\smallskip}
 \hline
 \noalign{\smallskip}
MWH&2.58$~\pm~0.07$&$0.21~\pm~0.02$&--\\ 
  \noalign{\smallskip}
LHB&13.40$~\pm~0.06$ &0.08 (fixed) & --\\
  \noalign{\smallskip}
CXB&1.57 (fixed)&--& 1.41 (fixed)\\
 \noalign{\smallskip}
  \hline
   \noalign{\smallskip}
   SP MOS1&5.22$~\pm~0.46$&--&1.40$~\pm~0.01$\\ 
    \noalign{\smallskip}
   SP MOS2&6.34$~\pm~1.01$&--&1.40$~\pm~0.01$\\ 
    \noalign{\smallskip}
    SP PN&42.71$~\pm~0.90$&--&1.40$~\pm~0.04$\\ 
    \noalign{\smallskip}
  \hline
 \end{tabular}
 \tiny
 \begin{tablenotes}
    \item[a] For LHB and MWH norm in units of 10$^{70}$ m$^{-3}$ arcmin$^{-2}$ \\
             For CXB norm in units of 10$^{50}$ph s$^{-1}$keV$^{-1}$ arcmin$^{-2}$\\
             For SP norm in units of 10$^{44}$ph s$^{-1}$keV$^{-1}$ arcmin$^{-2}$
\end{tablenotes}
  \end{threeparttable}\label{tab:tabb4}
  \end{center}
 \end{table}

 \begin{table}[ht!]
 \begin{center}
  \caption{Best-fit parameter values of the total background estimated for A3376 E observation.}
 \label{tab:taba3a}
\begin{threeparttable}
 \begin{tabular}{lccc}
\hline
\hline
 \noalign{\smallskip}
 &\textit{Norm}\tnote{a}& \textit{kT} (keV)& $\Gamma$\\
  \noalign{\smallskip}
 \hline
 \noalign{\smallskip}
MWH&2.31 (fixed)&0.27 (fixed)&--\\ 
  \noalign{\smallskip}
LHB&86.00 (fixed) & 0.08 (fixed) & --\\
  \noalign{\smallskip}
CXB&3.90 (fixed)&--& 1.41 (fixed)\\
 \noalign{\smallskip}
  \hline
   \noalign{\smallskip}
   SP MOS1&2.18$~\pm~0.37$&--&1.40$~\pm~0.04$\\ 
    \noalign{\smallskip}
   SP MOS2&1.30$~\pm~0.37$&--&1.40$~\pm~0.04$\\ 
       \noalign{\smallskip}
   SP PN&2.77$~\pm~0.66$&--&1.09$~\pm~0.12$\\ 
    \noalign{\smallskip}
  \hline

 \end{tabular}
 \tiny
 \begin{tablenotes}
    \item[a] For LHB and MWH norm in units of 10$^{68}$ m$^{-3}$ arcmin$^{-2}$ \\
             For CXB norm in units of 10$^{48}$ph s$^{-1}$keV$^{-1}$ arcmin$^{-2}$\\
             For SP norm in units of 10$^{45}$ph s$^{-1}$keV$^{-1}$ arcmin$^{-2}$
\end{tablenotes}
  \end{threeparttable}\label{tab:tabb4}
  \end{center}
 \end{table}
 
\begin{table}[ht!]
 \begin{center}
  \caption{Best-fit parameter values of the total background estimated for A3376 W observation.}
 \label{tab:taba3b}
\begin{threeparttable}
 \begin{tabular}{lccc}
\hline
\hline
 \noalign{\smallskip}
 &\textit{Norm}\tnote{a}& \textit{kT} (keV)& $\Gamma$\\
  \noalign{\smallskip}
 \hline
 \noalign{\smallskip}
MWH&2.31 (fixed)&0.27 (fixed)&--\\ 
  \noalign{\smallskip}
LHB&86.00 (fixed) & 0.08 (fixed) & --\\
  \noalign{\smallskip}
CXB&3.90 (fixed)&--& 1.41 (fixed)\\
 \noalign{\smallskip}
  \hline
   \noalign{\smallskip}
   SP MOS1&2.87$~\pm~0.40$&--&1.40$~\pm~0.01$\\ 
    \noalign{\smallskip}
   SP MOS2&23.65$~\pm~0.45$&--&1.40$~\pm~0.01$\\ 
       \noalign{\smallskip}
   SP PN&7.57$~\pm~0.57$&--&1.21$~\pm~0.04$\\ 
    \noalign{\smallskip}
  \hline

 \end{tabular}
 \tiny
 \begin{tablenotes}
    \item[a] For LHB and MWH norm in units of 10$^{68}$ m$^{-3}$ arcmin$^{-2}$ \\
             For CXB norm in units of 10$^{48}$ph s$^{-1}$keV$^{-1}$ arcmin$^{-2}$\\
             For SP norm in units of 10$^{45}$ph s$^{-1}$keV$^{-1}$ arcmin$^{-2}$
\end{tablenotes}
  \end{threeparttable}\label{tab:tabb4}
  \end{center}
 \end{table}

 \begin{table}[ht!]
 \begin{center}
  \caption{Background components derived from the A3667 observation.}
   \label{tab:taba4}
\begin{threeparttable}
 \begin{tabular}{lccc}
\hline
\hline
 \noalign{\smallskip}
 &\textit{Norm}\tnote{a}& \textit{kT} (keV)& $\Gamma$\\
  \noalign{\smallskip}
 \hline
 \noalign{\smallskip}
MWH&1.49$~\pm~0.07$&$0.26~\pm~0.01$&--\\ 
  \noalign{\smallskip}
LHB&26.21$~\pm~1.56$ &0.08 (fixed) & --\\
  \noalign{\smallskip}
CXB&0.74 (fixed)&--& 1.41 (fixed)\\
 \noalign{\smallskip}
  \hline
   \noalign{\smallskip}
   SP MOS1&12.26$~\pm~1.82$&--&1.40$~\pm~0.09$\\ 
    \noalign{\smallskip}
   SP MOS2&7.55$~\pm~3.12$&--&1.40$~\pm~0.09$\\ 
    \noalign{\smallskip}
    SP PN&0.40$~\pm~0.06$&--&0.35$~\pm~0.07$\\ 
    \noalign{\smallskip}
 
  \hline

 \end{tabular}
 \tiny
 \begin{tablenotes}
    \item[a] For LHB and MWH norm in units of 10$^{69}$ m$^{-3}$ arcmin$^{-2}$ \\
             For CXB norm in units of 10$^{49}$ph s$^{-1}$keV$^{-1}$ arcmin$^{-2}$\\
             For SP norm in units of 10$^{44}$ph s$^{-1}$keV$^{-1}$ arcmin$^{-2}$
\end{tablenotes}
  \end{threeparttable}\label{tab:tabb4}
  \end{center}
 \end{table}

 \begin{table}[ht!]
 \begin{center}
  \caption{Background components derived from the A665 observation.}
   \label{tab:taba5}
\begin{threeparttable}
 \begin{tabular}{lccc}
\hline
\hline
 \noalign{\smallskip}
 &\textit{Norm}\tnote{a}& \textit{kT} (keV)& $\Gamma$\\
  \noalign{\smallskip}
 \hline
 \noalign{\smallskip}
MWH&0.89$~\pm~0.06$&$0.19~\pm~0.03$&--\\ 
  \noalign{\smallskip}
LHB&15.00$~\pm~3.54$ & 0.08 (fixed) & --\\
  \noalign{\smallskip}
CXB&0.41 (fixed)&--& 1.41 (fixed)\\
 \noalign{\smallskip}
  \hline
   \noalign{\smallskip}
   SP MOS1&22.12$~\pm~1.17$&--&1.40$~\pm~0.04$\\ 
    \noalign{\smallskip}
   SP MOS2&9.88$~\pm~0.48$&--&1.40$~\pm~0.04$\\ 

  \hline

 \end{tabular}
 \tiny
 \begin{tablenotes}
    \item[a] For LHB and MWH norm in units of 10$^{70}$ m$^{-3}$ arcmin$^{-2}$ \\
             For CXB norm in units of 10$^{50}$ph s$^{-1}$keV$^{-1}$ arcmin$^{-2}$\\
             For SP norm in units of 10$^{44}$ph s$^{-1}$keV$^{-1}$ arcmin$^{-2}$
\end{tablenotes}
  \end{threeparttable} 
  \end{center}
 \end{table}

\begin{table*}[!ht]
\begin{minipage}[b]{0.45\textwidth}\centering
\caption{Background components derived from the A2256 pointing 1 observation.}
   \label{tab:taba6}
\begin{threeparttable}
 \begin{tabular}{lccc}
\hline
\hline
 \noalign{\smallskip}
 &\textit{Norm}\tnote{a}& \textit{kT} (keV)& $\Gamma$\\
  \noalign{\smallskip}
 \hline
 \noalign{\smallskip}
MWH&4.77$~\pm~0.18$&$0.31~\pm~0.04$&--\\ 
  \noalign{\smallskip}
LHB&54.50$~\pm~1.70$ & 0.08 (fixed) & --\\
  \noalign{\smallskip}
CXB&8.17 (fixed)&--& 1.41 (fixed)\\
 \noalign{\smallskip}
  \hline
   \noalign{\smallskip}
   SP MOS1&9.66$~\pm~0.61$&--&0.91$~\pm~0.02$\\ 
    \noalign{\smallskip}
   SP MOS2&9.99$~\pm~0.72$&--&0.91$~\pm~0.02$\\ 
   \noalign{\smallskip}
   SP PN&20.94$~\pm~3.51$&--&1.08$~\pm~0.08$\\
  \hline

 \end{tabular}
 \tiny
 \begin{tablenotes}
    \item[a] For LHB and MWH norm in units of 10$^{68}$ m$^{-3}$ arcmin$^{-2}$ \\
             For CXB norm in units of 10$^{48}$ph s$^{-1}$keV$^{-1}$ arcmin$^{-2}$\\
             For SP norm in units of 10$^{45}$ph s$^{-1}$keV$^{-1}$ arcmin$^{-2}$
\end{tablenotes}
  \end{threeparttable} 
\end{minipage}
\hspace{0.5cm}
\begin{minipage}[b]{0.45\textwidth}
\centering
\caption{Background components derived from the A2256 pointing 2 observation.}
   \label{tab:taba6}
\begin{threeparttable}
 \begin{tabular}{lccc}
\hline
\hline
 \noalign{\smallskip}
 &\textit{Norm}\tnote{a}& \textit{kT} (keV)& $\Gamma$\\
  \noalign{\smallskip}
 \hline
 \noalign{\smallskip}
MWH&4.77$~\pm~0.18$&$0.31~\pm~0.04$&--\\ 
  \noalign{\smallskip}
LHB&54.50$~\pm~1.70$ & 0.08 (fixed) & --\\
  \noalign{\smallskip}
CXB&8.17 (fixed)&--& 1.41 (fixed)\\
 \noalign{\smallskip}
  \hline
   \noalign{\smallskip}
   SP MOS1&3.23$~\pm~0.52$&--&1.40$~\pm~0.02$\\ 
    \noalign{\smallskip}
   SP MOS2&3.13$~\pm~0.46$&--&1.40$~\pm~0.02$\\ 
   \noalign{\smallskip}
   SP PN&5.02$~\pm~0.88$&--&1.40$~\pm~0.03$\\
  \hline

 \end{tabular}
 \tiny
 \begin{tablenotes}
    \item[a] For LHB and MWH norm in units of 10$^{68}$ m$^{-3}$ arcmin$^{-2}$ \\
             For CXB norm in units of 10$^{48}$ph s$^{-1}$keV$^{-1}$ arcmin$^{-2}$\\
             For SP norm in units of 10$^{45}$ph s$^{-1}$keV$^{-1}$ arcmin$^{-2}$
\end{tablenotes}
  \end{threeparttable} 
\end{minipage}
\end{table*}
 
\newpage
\section{Temperature and abundance analysis for the cluster sample} 
 The following tables summarise the best-fit parameters for the temperature and abundance  distribution along the merging axis. 
\begin{table*}[!hbt]
 \begin{center}  
 \caption{Best-fit parameters for CIZA2242 regions shown in Fig. \ref{fig:fig1}.}
   \label{tab:tabb1}
 \begin{tabular}{ccccccc}
\hline
\hline
\noalign{\smallskip}
Region &Radius& \textit{kT} &$Z$ &\textit{Norm}&N$_{\rm{H}}$&C-stat/d.o.f.\\ 
& (\arcmin)&(keV)&($Z_\odot$)&(10$^{72}$ m$^{-3}$ arcmin$^{-2}$)&(10$^{21}$ cm$^{-2}$)&\\ 
 \noalign{\smallskip}
 \hline
\noalign{\smallskip}
 1 &0.0~$\pm~1.5$&9.62~$\pm~0.54$&0.40~$\pm~0.09$&1.64~$\pm~0.04$&3.60~$\pm~0.10$&603/574\\
 \noalign{\smallskip}
 2&2.0~$\pm~0.5$&8.21$~\pm~0.55$&0.19~$\pm~0.09$&1.98~$\pm~0.06$&3.79~$\pm~0.13$&553/539\\
 \noalign{\smallskip}
 3 &3.0~$\pm~0.5$&9.14~$\pm~0.40$&0.26~$\pm~0.08$&1.63~$\pm~0.04$&3.85~$\pm~0.11$&570/560\\
   \noalign{\smallskip}
 4&4.3~$\pm~0.8$&8.07~$\pm~0.42$&0.29~$\pm~0.06$&1.46~$\pm~0.05$&3.85~$\pm~0.07$&624/606\\
 \noalign{\smallskip}
  5&6.0~$\pm~1.0$&8.68~$\pm~0.45$&0.40~$\pm~0.06$&0.95~$\pm~0.03$&4.27~$\pm~0.09$&653/619\\
  \noalign{\smallskip}
  5a (70--97$\degree$) &6.0~$\pm~1.0$&7.41$~\pm~$0.56&0.58$~\pm~$0.11&0.79~$\pm~0.03$&4.10~$\pm~0.13$&561/546\\ 
  \noalign{\smallskip}
5b (97--117$\degree$)&6.0~$\pm~1.0$&8.69$~\pm~$0.67& 0.29$~\pm~$0.09&1.44~$\pm~0.04$&4.32~$\pm~0.14$&525/559  \\
  \noalign{\smallskip}
5c (117--140$\degree$)& 6.0~$\pm~1.0$&9.61$~\pm~$1.03& 0.18$~\pm~$0.10&0.83~$\pm~0.03$&5.57~$\pm~0.07$&566/551\\
\noalign{\smallskip}
  6&9.0~$\pm~2.0$&10.29~$\pm~0.79$&0.29~$\pm~0.10$&0.25~$\pm~0.01$&5.13~$\pm~0.07$&662/643\\
    \noalign{\smallskip}
  BCGN&6.0~$\pm~1.5$&8.16~$\pm~0.41$&0.26~$\pm~0.06$&1.37~$\pm~0.03$&4.33~$\pm~0.10$&645/591\\
 \noalign{\smallskip}
 \hline
 \end{tabular}
  \end{center}
 \end{table*}

\begin{table*}[!hbt]
 \begin{center}  
 \caption{Best-fit parameters for 1RXSJ0603 regions shown in Fig. \ref{fig:fig3}.}
   \label{tab:tabb2}
 \begin{tabular}{cccccc}
\hline
\hline
\noalign{\smallskip}
Region &Radius& \textit{kT} &$Z$ &\textit{Norm}&C-stat/d.o.f.\\ 
& (\arcmin)&(keV)&($Z_\odot$)&(10$^{72}$ m$^{-3}$ arcmin$^{-2}$)&\\ 
 \noalign{\smallskip}
 \hline
\noalign{\smallskip}
 1 &-1.9~$\pm~0.4$&10.07~$\pm~0.46$&0.31~$\pm~0.09$&1.62~$\pm~0.09$&524/590\\
 \noalign{\smallskip}
 2&-1.1$\pm~0.4$&9.80$~\pm~0.41$&0.38~$\pm~0.05$&4.16~$\pm~0.07$&606/625\\
 \noalign{\smallskip}
 3 (CORES) &0.0~$\pm~0.7$&8.43~$\pm~0.27$&0.51~$\pm~0.06$&7.02~$\pm~0.13$&613/572\\
   \noalign{\smallskip}
 4&2.0~$\pm~0.5$&8.92~$\pm~0.31$&0.37~$\pm~0.04$&8.85~$\pm~0.13$&640/647\\
 \noalign{\smallskip}
  5&3.0~$\pm~0.5$&8.57~$\pm~0.18$&0.24~$\pm~0.04$&9.20~$\pm~0.08$&717/652\\
  \noalign{\smallskip}
  6&4.0~$\pm~0.5$&7.86~$\pm~0.15$&0.19~$\pm~0.04$&8.10~$\pm~0.10$&783/652\\
 \noalign{\smallskip}
  7&5.0~$\pm~0.5$&7.42~$\pm~0.20$&0.14~$\pm~0.04$&5.08~$\pm~0.05$&729/631\\
 \noalign{\smallskip}
   8&6.0~$\pm~0.5$&6.92~$\pm~0.29$&0.20~$\pm~0.07$&2.43~$\pm~0.03$&619/602\\
 \noalign{\smallskip}
   9&10.0~$\pm~2.0$&2.93~$\pm~0.22$&0.16~$\pm~0.06$&0.36~$\pm~0.02$&974/661\\
 \noalign{\smallskip}
 COREN&3.0~$\pm~0.4$&8.85~$\pm~0.36$&0.37~$\pm~0.08$&5.05~$\pm~0.01$&606/562\\
 \noalign{\smallskip}
 \hline
 \end{tabular}
  \end{center}
 \end{table*}

\begin{table*}[!hbt]
 \begin{center}  
 \caption{Best-fit parameters for A3376 regions shown in Fig. \ref{fig:fig5}.}
   \label{tab:tabb3}
 \begin{tabular}{cccccc}
 
\hline
\hline
\noalign{\smallskip}
Region &Radius& \textit{kT} &$Z$ &\textit{Norm}&C-stat/d.o.f.\\ 
& (\arcmin)&(keV)&($Z_\odot$)&(10$^{70}$ m$^{-3}$ arcmin$^{-2}$)&\\ 
 \noalign{\smallskip}
 \hline
\noalign{\smallskip}
 1 (BCG2) &0.0~$\pm~2.0$&4.17~$\pm~0.13$&0.51~$\pm~0.05$&6.88~$\pm~0.12$&553/607\\
 \noalign{\smallskip}
 2&3.5~$\pm~1.5$&3.89$~\pm~0.12$&0.44~$\pm~0.06$&7.14~$\pm~0.18$&642/579\\
 \noalign{\smallskip}
 3&6.5~$\pm~1.5$&4.00~$\pm~0.14$&0.37~$\pm~0.05$&5.61~$\pm~0.13$&647/603\\
 \noalign{\smallskip}
  \hline
   \noalign{\smallskip}
 4&9.5~$\pm~1.5$&3.79~$\pm~0.10$&0.31~$\pm~0.05$&4.54~$\pm~0.68$&1250/1185\\
 \noalign{\smallskip}
  5&12.5~$\pm~1.5$&3.88~$\pm~0.16$&0.30~$\pm~0.04$&4.29~$\pm~0.58$&1276/1195\\
 \noalign{\smallskip}
  \hline
  \noalign{\smallskip}
  6&16.0~$\pm~2.0$&4.21~$\pm~0.19$&0.29~$\pm~0.06$&1.63~$\pm~0.03$&617/643\\
 \noalign{\smallskip}
  7&20.0~$\pm~2.0$&5.36~$\pm~0.84$&0.15~$\pm~0.06$&0.68~$\pm~0.10$&657/599\\
 \noalign{\smallskip}
  BCG1&18.2~$\pm~2.0$&4.49~$\pm~0.35$&0.21~$\pm~0.12$&1.35~$\pm~0.01$&512/544\\
 \noalign{\smallskip}
 \hline
 \end{tabular}
  \end{center}
 \end{table*}

\begin{table*}[!hbt]
 \begin{center}  
 \caption{Best-fit parameters for A3667 regions shown in Fig. \ref{fig:fig7}.}
   \label{tab:tabb4}
 \begin{tabular}{cccccc}
 
\hline
\hline
\noalign{\smallskip}
Region &Radius& \textit{kT} &$Z$ &\textit{Norm}&C-stat/d.o.f.\\ 
& (\arcmin)&(keV)&($Z_\odot$)&(10$^{71}$ m$^{-3}$ arcmin$^{-2}$)&\\ 
 \noalign{\smallskip}
 \hline
\noalign{\smallskip}
 1&-7.6~$\pm~0.9$&5.46~$\pm~0.11$&0.43~$\pm~0.04$&1.24~$\pm~0.01$&802/664\\
 \noalign{\smallskip}
 2&-5.9~$\pm~0.9$&4.56$~\pm~0.06$&0.52~$\pm~0.03$&2.57~$\pm~0.02$&843/682\\
 \noalign{\smallskip}
 3&-4.1~$\pm~0.9$&5.04~$\pm~0.07$&0.48~$\pm~0.03$&4.16~$\pm~0.02$&761/685\\
 \noalign{\smallskip}
 4&-2.4~$\pm~0.9$&5.54~$\pm~0.09$&0.45~$\pm~0.03$&6.12~$\pm~0.47$&756/651\\
 \noalign{\smallskip}
  5&0.0~$\pm~1.5$&6.79~$\pm~0.17$&0.38~$\pm~0.04$&4.29~$\pm~0.05$&760/652\\
 \noalign{\smallskip}
  6&2.4~$\pm~0.9$&7.05~$\pm~0.23$&0.39~$\pm~0.06$&3.54~$\pm~0.05$&675/618\\
 \noalign{\smallskip}
  7&4.1~$\pm~0.9$&7.29~$\pm~0.24$&0.30~$\pm~0.08$&2.35~$\pm~0.04$&729/631\\
 \noalign{\smallskip}
  8&6.8~$\pm~1.8$&6.79~$\pm~0.20$&0.27~$\pm~0.06$&1.48~$\pm~0.02$&733/673\\
 \noalign{\smallskip}
 \hline

 \end{tabular}
  \end{center}
 \end{table*}

\begin{table*}[!hbt]
 \begin{center}  
 \caption{Best-fit parameters for A665 regions shown in Fig. \ref{fig:fig9}.}
   \label{tab:tabb4}
 \begin{tabular}{cccccc}
 
\hline
\hline
\noalign{\smallskip}
Region &Radius& \textit{kT} &$Z$ &\textit{Norm}&C-stat/d.o.f.\\ 
& (\arcmin)&(keV)&($Z_\odot$)&(10$^{72}$ m$^{-3}$ arcmin$^{-2}$)&\\ 
 \noalign{\smallskip}
 \hline
\noalign{\smallskip}
 1&-3.5~$\pm~1.0$&6.73~$\pm~0.52$&0.29~$\pm~0.15$&0.26~$\pm~0.01$&393/400\\
 \noalign{\smallskip}
  2&-2.0~$\pm~0.5$&8.43~$\pm~0.85$&0.41~$\pm~0.17$&0.67~$\pm~0.02$&373/362\\
 \noalign{\smallskip}
  3&-1.0~$\pm~0.5$&7.30~$\pm~0.38$&0.57~$\pm~0.12$&2.51~$\pm~0.05$&413/372\\
 \noalign{\smallskip}
   4&0.0~$\pm~0.5$&7.80~$\pm~0.37$&0.40~$\pm~0.07$&11.79~$\pm~0.21$&431/421\\
 \noalign{\smallskip}
  5&1.0~$\pm~0.5$&7.39~$\pm~0.27$&0.28~$\pm~0.06$&7.17~$\pm~0.09$&452/405\\
 \noalign{\smallskip}
 6&2.0~$\pm~0.5$&7.36~$\pm~0.27$&0.33~$\pm~0.06$&3.35~$\pm~0.03$&476/405\\
 \noalign{\smallskip}
 7&3.0~$\pm~0.5$&8.72~$\pm~0.44$&0.23~$\pm~0.09$&1.66~$\pm~0.03$&412/395\\
 \noalign{\smallskip}
8&4.0~$\pm~0.5$&8.32$~\pm~0.59$&0.24~$\pm~0.12$&0.84~$\pm~0.02$&393/386\\
 \noalign{\smallskip}
 9&5.0~$\pm~0.5$&8.56~$\pm~0.79$&0.23~$\pm~0.15$&0.51~$\pm~0.01$&375/374\\
 \noalign{\smallskip}
 \hline

 \end{tabular}
  \end{center}
 \end{table*}
 
\begin{table*}[!hbt]
 \begin{center}  
 \caption{Best-fit parameters for A2256 regions shown in Fig. \ref{fig:fig11}.}
   \label{tab:tabb5}
 \begin{tabular}{cccccc}
 
\hline
\hline
\noalign{\smallskip}
Region &Radius& \textit{kT} &$Z$ &\textit{Norm}&C-stat/d.o.f.\\ 
& (\arcmin)&(keV)&($Z_\odot$)&(10$^{71}$ m$^{-3}$ arcmin$^{-2}$)&\\ 
 \noalign{\smallskip}
 \hline
\noalign{\smallskip}
1&-4.0~$\pm~0.5$&9.03~$\pm~0.54$&0.28~$\pm~0.15$&1.86~$\pm~0.06$&1203/1095\\
 \noalign{\smallskip}
   2&-3.0~$\pm~0.5$&6.97~$\pm~0.27$&0.36~$\pm~0.09$&3.25~$\pm~0.08$&1161/1101\\
 \noalign{\smallskip}
  3&-2.0~$\pm~0.5$&6.56~$\pm~0.25$&0.37~$\pm~0.09$&3.83~$\pm~0.09$&1149/1081\\
 \noalign{\smallskip}
 4 (CORE1)&0.0~$\pm~1.5$&6.89~$\pm~0.15$&0.44~$\pm~0.06$&4.78~$\pm~0.07$&1218/1192\\
 \noalign{\smallskip}
 4&2.3~$\pm~0.8$&6.60~$\pm~0.20$&0.48~$\pm~0.07$&5.33~$\pm~0.09$&1077/1073\\
 \noalign{\smallskip}
 5&3.8~$\pm~0.8$&5.09$~\pm~0.13$&0.43~$\pm~0.04$&5.41~$\pm~0.09$&1308/1145\\
 \noalign{\smallskip}
   6&5.3~$\pm~0.8$&4.24~$\pm~0.14$&0.49~$\pm~0.08$&0.70~$\pm~0.02$&1145/1024\\
 \noalign{\smallskip}
   CORE2&-3.3~$\pm~0.5$&4.70~$\pm~0.12$&0.56~$\pm~0.05$&5.79~$\pm~0.09$&1259/1096\\
 \noalign{\smallskip}
 \hline

 \end{tabular}
  \end{center}
 \end{table*}

  \end{appendix}

\end{document}